\documentclass[journal=jctcce,manuscript=article]{achemso}
\setkeys{acs}{articletitle = true}
\setkeys{acs}{chaptertitle = true}
\setkeys{acs}{abbreviations = false}
\setkeys{acs}{maxauthors=150}
\setkeys{acs}{etalmode=truncate}

\usepackage{graphicx}
\usepackage{multirow}
\usepackage{siunitx}
\usepackage{xcolor}
\usepackage{braket}
\usepackage{amsmath}
\usepackage{bm}
\usepackage{float}

\usepackage[hidelinks]{hyperref}
\hypersetup{colorlinks=false}

\usepackage{numprint}
\usepackage[utf8]{inputenc}
\usepackage[T1]{fontenc}
\usepackage{mathptmx}
\usepackage[english]{babel}
\usepackage{subfigure}
\usepackage{comment}
\usepackage{siunitx}

\usepackage{xr}
\DeclareMathAlphabet{\mathcal}{OMS}{cmsy}{m}{n}
\title{Reduced density matrix and cumulant approximations of quantum linear response} 

\author{Theo Juncker von Buchwald}
\email{tjvbu@kemi.dtu.dk}
\affiliation{Department of Chemistry, Technical University of Denmark, Kemitorvet Building 207, DK-2800 Kongens Lyngby, Denmark.}
\author{Erik Rosendahl Kjellgren}
\email{kjellgren@sdu.dk}
\affiliation{Department of Physics, Chemistry and Pharmacy,
University of Southern Denmark, Campusvej 55, 5230 Odense, Denmark.}
\author{Jacob Kongsted}
\affiliation{Department of Physics, Chemistry and Pharmacy,
 University of Southern Denmark, Campusvej 55, 5230 Odense, Denmark.}
\author{Stephan P. A. Sauer}
\affiliation{Department of Chemistry, University of Copenhagen, DK-2100 Copenhagen \O, Denmark.}
\author{Sonia Coriani}
\affiliation{Department of Chemistry, Technical University of Denmark, Kemitorvet Building 207, DK-2800 Kongens Lyngby, Denmark.}
\author{Karl Michael Ziems}
\email{K.M.Ziems@soton.ac.uk}
\affiliation{School of Chemistry, University of Southampton, Highfield, Southampton SO17 1BJ, United Kingdom}
\affiliation{Department of Chemistry, Technical University of Denmark, Kemitorvet Building 207, DK-2800 Kongens Lyngby, Denmark.}

\date{August 10, 2025}

\begin{document}

\begin{abstract}
Linear response (LR) is an important tool in the computational chemist's toolbox. It is therefore no surprise that the emergence of quantum computers has led to a quantum version, quantum LR (qLR). However, the current quantum era of near-term intermediary scale quantum (NISQ) computers is dominated by noise, short decoherence times, and slow measurement speed. It is therefore of interest to find approximations that greatly reduce the quantum workload while only slightly impacting the quality of a method. In an effort to achieve this, we approximate the naive qLR with singles and doubles (qLRSD) method by either directly approximating the reduced density matrices (RDMs) or indirectly through their respective reduced density cumulants (RDCs). We present an analysis of the measurement costs behind qLR with RDMs, and report qLR results for model Hydrogen ladder systems; for varying active space sizes of OCS, SeH$_2$, and H$_2$S; and for symmetrically stretched H$_2$O and BeH$_2$.

Discouragingly, while approximations to the 4-body RDMs and RDCs seem to produce good results for systems at the equilibrium geometry and for some types of core excitations, they both tend to fail when the system exhibits strong correlation. All approximations to the 3-body RDMs and/or RDCs severely affect the results and cannot be applied.
\end{abstract}

\maketitle

\section{Introduction}
\label{sec:intro}

Predicting molecular properties like excitation energies, oscillator strengths, and rotational strengths plays a crucial role in interpreting spectroscopic data. This capability also serves as a foundation in various scientific fields, including photochemistry, photophysics, photocatalysis, and the design of photoactivatable products for novel therapeutic tools. To this end, the molecular response framework~\cite{olsen1985linear,Christiansen1998,helgaker2012recent,Pawlowski2015} is an obvious choice to compute molecular properties, as it has been long established for conventional hardware and has, in the past years, been reformulated for quantum hardware.~\cite{Kumar2023,ziems2023options,jensen2024quantum,von2024reduced}

On conventional hardware, linear response has been formulated and implemented for a wide range of electronic structure methods such as Hartree Fock theory,~\cite{olsen1985linear,Norman1995} multi-configurational self-consistent field (MCSCF) theory,~\cite{olsen1985linear,Hettema1992,Jonsson1996} coupled cluster theory,~\cite{koch1990,Christiansen1998,Pawlowski2015} Møller-Plesset perturbation theory,~\cite{jod23,jod05-jpca109-11618,spas188} and time-dependent density functional theory.~\cite{Salek2002,Parker2018}

On the quantum side, linear response has been adapted to the quantum linear response (qLR) framework,~\cite{Kumar2023,ziems2023options,von2024reduced,reinholdt2024subspace,reinholdt2025self,ziems2025understanding} the variational quantum response (VQR) method,~\cite{huang2022variational} and other general LR frameworks.~\cite{cai2020quantum,kharazi2024efficient} Since its initial proposal,~\cite{Kumar2023} qLR has been further extended upon with, e.g., the development of an orbital-optimized variant for active spaces,~\cite{ziems2023options} a reduced density matrix implementation,~\cite{ziems2023options,von2024reduced} a Davidson solver approach,~\cite{reinholdt2024subspace} and the introduction of polarizable embedding environments.~\cite{reinholdt2025self} Recently, multiple strategies to reduce the hardware requirements were utilized in a successful application of qLR on quantum hardware,~\cite{ziems2025understanding} along with a detailed analysis of the algorithm.~\cite{kjellgren2024divergences,kjellgren2025redundant} Other approaches to calculate excited states and response properties using quantum hardware such as quantum equation of motion,~\cite{ollitrault2020quantum,asthana2023quantum,kim2023two,jensen2024quantum} state-average VQE variants,~\cite{nakanishi2019subspace,Parrish2019-fi,yalouz2021state,grimsley2025challenging} and variational quantum deflation,~\cite{higgott2019variational,chan2021molecular} have also been introduced.

Not just the fault-tolerant evaluation of linear response functions is currently out of reach, but even going beyond a few qubits is impossible on current quantum devices due to noise rates, decoherence, and sampling speed. Therefore, approximations to qLR have been considered and carefully studied against their accuracy. 
In the classical regime, MCSCF is used to significantly reduce the computational requirements by only treating the correlation of the active space, thereby reducing the size of the problem. In the paper by \citeauthor{ziems2023options},~\cite{ziems2023options} an MCSCF approach to qLR was employed, where an active space was used to reduce the quantum hardware requirements. The active space approach was combined with orbital optimization and the truncation of excitation rank within the active space. In the same work, the authors introduced eight parameterization schemes, of which three were labeled as applicable in the near-term. Of these, in a previous paper, we formulated the naive parameterization, restricted to singles and doubles excitation in the active space, in a reduced density matrix (RDM) formalism needing up to the four-body RDM (4-RDM) and reducing the classical aspects of the parameterization with the use of rank reduction.~\cite{von2024reduced} This allowed for the calculation of response properties for medium sized molecules with moderately sized basis sets.

The RDM formalism, however, still comes at a relatively steep cost, since the 4-RDM scales with the number of orbitals in the active space to the eight power, $(N_A^8)$.  It is therefore of interest to find approximations in an effort to further reduce the size of the problem. One possible approach would be to directly approximate the RDMs. Another approach would be to approximate the corresponding reduced density cumulants (RDCs)~\cite{mazziotti1998approximate,kutzelnigg1999cumulant,Harris2002} and reconstruct the RDMs. Approximations using RDCs are a well-known strategy within conventional quantum chemistry methods such as NEVPT2,~\cite{zgid2009study} CASPT2,~\cite{nakatani2017density} DMRG,~\cite{saitow2013multireference,kurashige2014complete} and DSRG.~\cite{li2023intruder} Inspired by these works, we here explore the possibility to use reduced density matrix and reduced density cumulant approximations in our formulation of naive quantum linear response to reduce the number of quantum measurements as well as the classical cost.

This paper is organized as follows. In Section~\ref{sec:theo} we shortly review the active space approximation and the orbital-optimized unitary coupled cluster (oo-UCC) method, and then introduce the qLR framework. In Section~\ref{sec:theo}, RDMs and RDCs as well as their approximations are introduced. In Section~\ref{sec:comp}, we provide the computational details, followed by Section~\ref{sec:results} where we start by providing a detailed analysis of the amount of quantum measurements depending on RDMs, active space, and mapping, as well as analyzing how this translate to the actual qLR algorithm. Next, we present results for our approximations, which are firstly applied to Hydrogen ladder model systems of increasing size in a minimal basis, followed by various molecular systems in different active spaces. Lastly, we investigate the effects of strong correlation on the approximations. Concluding remarks are given at the end.

\section{Theory}
\label{sec:theo}

Throughout this paper, unless otherwise stated, the following notation will be used: $p$, $q$, $r$, $s$, $t$, $u$, $m$, and $n$ for general orbital indices; \textit{a, b, c,} and \textit{d} for virtual (secondary) orbital indices; \textit{i, j, k,} and \textit{l} are inactive (doubly occupied) orbital indices; and $\nu_a$ and $\nu_i$ refer to active-space orbitals that are, respectively, virtual and doubly occupied in the Hartree-Fock reference state. 

\subsection{Active space}
\label{ssec:active_space}

In the active space approximation the full space is divided into three subspaces: an inactive, an active, and a virtual (secondary) space. The orbitals in the inactive space are considered to be doubly occupied, the orbitals in the virtual (secondary) space are empty, and the orbitals in the active space are then ordinarily treated with a full configuration interaction (FCI) expansion. Expectation values of an operator may be found as a product of the different spaces
\begin{equation}
\label{eq:AS_seperation}
\braket{0(\boldsymbol{\theta}) | \hat O | 0(\boldsymbol{\theta})} = \braket{I | \hat O_I | I} \braket{A(\theta) | \hat O_A | A(\theta)} \braket{V | \hat O_V | V}~, 
\end{equation}
with the separation of the reference wave function into an inactive, $\ket{I}$, an active, $\ket{A(\boldsymbol{\theta})}$, and a virtual (secondary), $\ket{V}$, part
\begin{equation}
\label{wf:partitioned}
\left|0\left(\boldsymbol{\theta}\right)\right> = \left|I\right>\otimes \left|A\left(\boldsymbol{\theta}\right)\right> \otimes \left|V\right>~,
\end{equation}
and the operator decomposition
\begin{equation}
    \hat{O} = \hat{O}_I\otimes \hat{O}_A\otimes \hat{O}_V~.
\end{equation}
Note the dependence of the active space wavefunction component on the wavefunction parameters~$\boldsymbol{\theta}$.

The selection of the active space is commonly done by choosing orbitals and number of electrons for the active space part. This gives the notation $(N_e, N_A)$ where $N_e$ is the number of electrons included in the active space and $N_A$ is the number of spatial orbitals in the active space. This strategy is famously used for the classical method CASSCF (complete active space self-consistent field),~\cite{Siegbahn1980,roos1980complete,Siegbahn1981} where the active space is treated using a complete CI expansion. Recently, the active space strategy has also been applied in quantum computing for chemistry, where only the active space is treated on the quantum device whereas the inactive space and virtual space are treated classically.~\cite{takeshita2020increasing} A popular parametrization of the active space is unitary coupled cluster (UCC), wherein a further approximation is the truncation of the cluster operator.~\cite{bartlett1989alternative,Mizukami2020,Sokolov2020} The expectation value of an operator acting on the active space (i.e.~being evaluated on quantum hardware) is treated by translating the operator to a sum of Pauli strings
\begin{equation}
    \braket{A\left(\boldsymbol{\theta}\right) | \hat{O}_A | A\left(\boldsymbol{\theta}\right)} = \sum_{l} c_l \braket{A\left(\boldsymbol{\theta}\right) | \hat{P}_l | A\left(\boldsymbol{\theta}\right)}.
\end{equation}
Each Pauli string, $\hat{P}_l$, has a corresponding coefficient, $c_l$, both of which are dependent on the mapping chosen. Different operators may share Pauli strings, therefore, the results of a given Pauli string measurement are saved in memory to be used for different operators.\cite{ziems2025understanding} The measurement overhead can be further reduced by utilizing commuting groups,~\cite{patel2025quantum} such as qubit-wise commutativity (QWC).~\cite{verteletskyi2020measurement}

\subsection{Orbital optimized unitary coupled cluster}
\label{ssec:oo-ucc}

An extension of UCC is the orbital optimized unitary coupled cluster (oo-UCC),~\cite{bartlett1989alternative,Mizukami2020,Sokolov2020} where UCC is performed in the active space and the orbital optimization is performed between the inactive and active spaces, the active and virtual (secondary) spaces, and the inactive and virtual (secondary) spaces. 

The UCC wave function is given as an exponential parameterization of a reference wave function $\ket{0}$
\begin{equation}
\label{UCC}
\ket{\text{UCC}(\boldsymbol{\theta})} = \textrm{e}^{\hat{T}(\boldsymbol{\theta}) 
- \hat{T}^\dagger
(\boldsymbol{\theta})}\ket{0}~,
\end{equation}
where $\hat{T}(\boldsymbol{\theta})$ is the cluster operator which generates excited determinants
\begin{align}
    \hat{T}(\boldsymbol{\theta}) = & \hat{T}_1(\boldsymbol{\theta}) + \hat{T}_2(\boldsymbol{\theta}) + \ldots \\
    \hat{T}_1(\boldsymbol{\theta}) = & \sum_{\nu_a\nu_i}\theta_{\nu_a\nu_i}\hat{E}_{\nu_a\nu_i} \\
    \hat{T}_2(\boldsymbol{\theta}) = & \frac{1}{2} \sum_{\substack{\nu_a\nu_i\\\nu_b\nu_j}}\theta_{\nu_a\nu_i\nu_b\nu_j}\hat{E}_{\nu_a\nu_i}\hat{E}_{\nu_b\nu_j} \\
    \vdots\nonumber &
\end{align}
and $\hat{E}_{pq}$ is the singlet one-electron excitation operator, defined by the creation $\hat{a}^\dagger_{p\sigma}$ and annihilation $\hat{a}_{q\sigma}$ operators, $\hat{E}_{pq} = \hat{a}^\dagger_{p\alpha}\hat{a}_{q\alpha} + \hat{a}^\dagger_{p\beta}\hat{a}_{q\beta}$. The adjoint cluster operator $\hat{T}^\dagger(\boldsymbol{\theta})$ generates all de-excitations and ensures unitarity. Like in CI and in regular CC, the cluster operator $\hat{T}(\boldsymbol{\theta})$ (and $\hat{T}^\dagger(\boldsymbol{\theta})$) can be truncated. As an example, UCC singles and doubles (UCCSD) only includes the singles and doubles cluster operators. Unlike CI, the UCC method remains size-consistent when truncating the cluster operator.~\cite{anand2022quantum}

The orbital optimization is done by exponentially parameterizing the UCC wave function with the orbital rotation operator $\hat{\kappa}(\boldsymbol{\kappa})$
\begin{align}
    \ket{\text{oo-UCC}(\boldsymbol{\kappa},\boldsymbol{\theta})}~=~& \textrm{e}^{-\hat{\kappa}(\boldsymbol{\kappa})}\ket{\textrm{UCC}(\theta)} \\
    \hat{\kappa}({\boldsymbol{\kappa}}) = & \sum_{p>q}\kappa_{pq}\hat{E}^-_{pq} \\
    \hat{E}^-_{pq} = & \hat{E}_{pq} - \hat{E}_{qp}
\end{align}
where $\hat{E}^-_{pq}$ only acts between the inactive, active, and virtual spaces. The $\boldsymbol{\theta}$ and $\boldsymbol{\kappa}$ parameters are found by minimizing the energy using the orbital optimized variation quantum eigensolver (oo-VQE)~\cite{Mizukami2020,Sokolov2020} 
\begin{equation}
    E(\boldsymbol{\kappa},\boldsymbol{\theta}) = \min_{\boldsymbol{\kappa},\boldsymbol{\theta}}\langle\text{UCC}(\boldsymbol{\theta})|\hat{H}(\boldsymbol{\kappa})|\text{UCC}(\boldsymbol{\theta})\rangle
\end{equation}
where the Hamiltonian has been similarity transformed by the orbital rotation parameters. 

\subsection{Linear response}
\label{ssec:LR}

As anticipated in Section~\ref{sec:intro}, properties such as excitation energies, oscillator strengths, and rotational strengths may be calculated through the linear response framework. For an MCSCF reference wave function, $\ket{0}$, this is achieved by an exponential unitary transformation~\cite{olsen1985linear}
\begin{equation}
    \ket{\tilde{0}(t)} = \text{e}^{\hat\kappa(t)}\text{e}^{\hat S(t)}\ket{0}
\end{equation}
parameterized by the time-dependent orbital operator $\hat{\kappa}(t)$ and active space rotation operator $\hat{S}(t)$. They contain, respectively, the orbital rotation operators $\hat q_\mu$ and the active space excitation operators $\hat G_n$
\begin{align}
    \hat\kappa(t) & = \sum_\mu \left( \kappa_\mu(t) \hat q_\mu + \kappa_\mu^*(t)\hat q_\mu^\dagger \right) \label{eq:kappa_t} \\
    \hat S(t) & = \sum_n \left( S_n(t) \hat G_n + S_n^*(t)\hat G_n^\dagger \right). \label{eq:s_t}
\end{align}

Introducing a perturbative (and Fourier) expansion of the parameters, excitation energies are then found by solving the generalized eigenvalue quation~\cite{olsen1985linear,helgaker2012recent}
\begin{equation}\textbf{E}^{[2]}\boldsymbol{\beta}_k = \omega_k \textbf{S}^{[2]}\boldsymbol{\beta}_k~, \label{eq:LR_exc}
\end{equation}
where $\boldsymbol{\beta}_k$ is the excitation vector and $\omega_k$ the corresponding excitation energy for excited state $k$. The Hessian $\textbf{E}^{[2]}$ and metric $\textbf{S}^{[2]}$ matrices are given as
\begin{align}
 \textbf{E}^{[2]} &= \begin{pmatrix}
    {\boldsymbol{A}} & {\boldsymbol{B}} \\
      {\boldsymbol{B}}^* & {\boldsymbol{A}}^*           
     \end{pmatrix}, \quad 
     \textbf{S}^{[2]} = \begin{pmatrix}
    \boldsymbol{\Sigma} & \boldsymbol{\Delta} \\
     -\boldsymbol{\Delta} ^* &  -\boldsymbol{\Sigma}^*           
     \end{pmatrix} \quad
\end{align}
with the submatrices
\begin{align}
    {\bm{A}} &= \begin{pmatrix} \label{eq:LR_A}
\left<0\left|\left[\hat{q}_\mu^\dagger,\left[\hat{H},\hat{q}_{\nu}\right]\right]\right|0\right>
& \left<0\left|\left[\hat{q}^\dagger_\mu,\left[\hat{H},\hat{G}_{m}\right]\right]\right|0\right> \\
\left<0\left|\left[\hat{G}_{n}^\dagger,\left[\hat{H},\hat{q}_{\nu}\right]\right]\right|0\right>
& \left<0\left|\left[\hat{G}_{n}^\dagger,\left[\hat{H},\hat{G}_{m}\right]\right]\right|0\right>
\end{pmatrix} \\
    \boldsymbol{B} &= \begin{pmatrix} \label{eq:LR_B}
\left<0\left|\left[\hat{q}_\mu^\dagger,\left[\hat{H},\hat{q}_{\nu}^\dagger\right]\right]\right|0\right>
& \left<0\left|\left[\hat{q}^\dagger_\mu,\left[\hat{H},\hat{G}^\dagger_{m}\right]\right]\right|0\right> \\
\left<0\left|\left[\hat{G}_{n}^\dagger,\left[\hat{H},\hat{q}_{\nu}^\dagger\right]\right]\right|0\right>
& \left<0\left|\left[\hat{G}_{n}^\dagger,\left[\hat{H},\hat{G}_{m}^\dagger\right]\right]\right|0\right>
\end{pmatrix} \\
    \boldsymbol{\Sigma} &= \begin{pmatrix} \label{eq:LR_sigma}
\left<0\left|\left[\hat{q}_\mu^\dagger,\hat{q}_{\nu}\right]\right|0\right>
& \left<0\left|\left[\hat{q}_{\mu}^\dagger,\hat{G}_{m}\right]\right|0\right> \\
\left<0\left|\left[\hat{G}_{n}^\dagger,\hat{q}_{\nu}\right]\right|0\right>
& \left<0\left|\left[\hat{G}_{n}^\dagger,\hat{G}_{m}\right]\right|0\right>
\end{pmatrix} \\
    \boldsymbol{\Delta} &= \begin{pmatrix} \label{eq:LR_delta}
\left<0\left|\left[\hat{q}_\mu^\dagger,\hat{q}_{\nu}^\dagger\right]\right|0\right>
& \left<0\left|\left[\hat{q}_{\mu}^\dagger,\hat{G}_{m}^\dagger\right]\right|0\right> \\
\left<0\left|\left[\hat{G}_{n}^\dagger,\hat{q}_{\nu}^\dagger\right]\right|0\right>
& \left<0\left|\left[\hat{G}_{n}^\dagger,\hat{G}_{m}^\dagger\right]\right|0\right>
\end{pmatrix}.
\end{align}
and the excitation vector as
\begin{align}
    {\boldsymbol{\beta}}_k = \begin{pmatrix}
    {\boldsymbol{Z}}_k \\
      {\boldsymbol{Y}}_k^*            
     \end{pmatrix} = \begin{pmatrix}
    \boldsymbol{\kappa}_k \\
      \boldsymbol{S}_k \\
      \boldsymbol{\kappa}_{-k}^* \\
      \boldsymbol{S}_{-k}^* \\
     \end{pmatrix}
\end{align}
where the excitation block $\boldsymbol{Z}_k$ corresponds to $\omega_k >0$, and the de-excitation block $\boldsymbol{Y}_k^*$ corresponds to $\omega_k<0$. The elements of the two subblocks are, respectively, the state-specific orbital rotation parameters $\boldsymbol{\kappa}_k$ and active-space excitation  parameters $\boldsymbol{S}_k$ that define the first-order expansion of eqs. \eqref{eq:kappa_t} and \eqref{eq:s_t}. 

The oscillator strength related to the transition from the ground state to the excited state $k$ can be calculated as
\begin{equation}
    f_k = \frac{2}{3}\omega_k\sum_\gamma|\langle0|[\hat{\mu}_\gamma,\hat{\tilde{O}}_k]|0\rangle|^2
\end{equation}
using the electric dipole moment operator
\begin{equation}
    \hat{\mu}_\gamma = -\sum_{pq}\langle p|\Vec{r}_\gamma|q\rangle\hat{E}_{pq},
\end{equation}
and the normalized excitation operator
\begin{equation}
    \hat{\tilde{O}}_k = \frac{\hat{O}_k}{\sqrt{\langle0|[\hat{O}_k,\hat{O}_k^\dagger]|0\rangle}}
\end{equation}
\begin{equation}
    \hat{O}_k = \sum_{L\in\{\mu,n\}}(Z_{k,L}\hat{X}_L^\dagger + Y_{k,L}\hat{X}_L),
\end{equation}
where $Z_{k,L}$ and $Y_{k,L}$ are weights in $\boldsymbol{\beta}_k$ of the corresponding operator $\hat{X}_L \in \{\hat{q}_{\mu}, \hat{G}_n\}$.

In the paper by~\citeauthor{ziems2023options},\cite{ziems2023options} eight LR Ans{\"a}tze were introduced through specific choices of the operators $\hat{q}$ and $\hat{G}$. We focus here on the naive parameterization and truncate the linear response excitation rank to spin-adapted singles and doubles excitations~\cite{paldus1977application,piecuch1989orthogonally} (which we will refer to as qLRSD from now on),
\begin{align}
    \hat q_{pq} = \frac{1}{\sqrt{2}}\hat{E}_{pq}~, \quad\textrm{with} \quad pq=\{\nu i, ai,a\nu \}  
    \label{eq:q}
\end{align}
and
\begin{align}
    \hat{G} \in \Bigg\{\frac{1}{\sqrt{2}}\hat{E}_{v_av_i},\quad &\frac{1}{2\sqrt{\left(1+\delta_{v_av_b}\right)\left(1+\delta_{v_iv_j}\right)}}\left(\hat{E}_{v_av_i}\hat{E}_{v_bv_j} + \hat{E}_{v_av_j}\hat{E}_{v_bv_i}\right), \label{eq:G} \\\nonumber
    &\frac{1}{2\sqrt{3}}\left(\hat{E}_{v_av_i}\hat{E}_{v_bv_j} - \hat{E}_{v_av_j}\hat{E}_{v_bv_i}\right)\Bigg\}.
\end{align} 
Note that if the active space spans the full space, like in FCI, there is no inactive and virtual (secondary) space and, as such, $\text{e}^{\hat{\kappa}(t)}$ disappears. 

\subsection{Reduced density matrices}
\label{ssec:RDM}

In our former study,~\cite{von2024reduced} we reformulated the naive qLRSD parametrization of Ref.~\citenum{ziems2023options} using reduced density matrices which reduced the classical requirements by rank reduction, allowing larger basis sets to be employed. The Hessian, metric, and property gradient were then constructed using RDMs, where the active space RDM contributions were evaluated on (emulated) quantum device (see eqs. (25) -- (29) and (S5) -- (S28) in Ref. \citenum{von2024reduced}).

When restricting the excitation rank in the active space in naive qLR to singles and doubles excitations, only the RDMs up to the 4-RDM are needed to construct the Hessian.~\cite{von2024reduced} In second quantization, the full 1-, 2-, 3-, and 4-RDM are given as
\begin{align}
    \Gamma^{[1]}_{pq} = & \sum_{\tau\in\{\alpha,\beta\}} \left<0\left|\hat a^\dagger_{p\tau}\hat a_{q\tau}\right|0\right> = \left<0\left|\hat{E}_{pq}\right|0\right> \\
    \Gamma^{[2]}_{pqrs} = & \sum_{\tau\gamma\in\{\alpha,\beta\}} \left<0\left|\hat a^\dagger_{p\tau}\hat a^\dagger_{r\gamma}\hat a_{s\gamma}\hat a_{q\tau}\right|0\right> = \left<0\left|\hat{e}_{pqrs}\right|0\right> \\\nonumber
    = & \left<0\left|\hat{E}_{pq}E_{rs}\right|0\right> - \delta_{qr}\Gamma^{[1]}_{ps} \\
    \Gamma^{[3]}_{pqrstu} = & \sum_{\tau\gamma\delta\in\{\alpha,\beta\}} \left<0\left|\hat a^\dagger_{p\tau}\hat a^\dagger_{r\gamma}\hat a^\dagger_{t\delta}\hat a_{u\delta}\hat a_{s\gamma}\hat a_{q\tau}\right|0\right> \\\nonumber
    = & \left<0\left|\hat{E}_{pq}\hat{E}_{rs}\hat{E}_{tu}\right|0\right> - \delta_{ts}\Gamma^{[2]}_{pqru} - \delta_{rq}\Gamma^{[2]}_{pstu} - \delta_{tq}\Gamma^{[2]}_{purs} - \delta_{ts}\delta_{rq}\Gamma^{[1]}_{pu} \\
    \Gamma^{[4]}_{pqrstumn} = & \sum_{\tau\gamma\delta\sigma\in\{\alpha,\beta\}} \left<0\left|\hat a^\dagger_{p\tau}\hat a^\dagger_{r\gamma}\hat a^\dagger_{t\delta}\hat a^\dagger_{m\sigma}\hat a_{n\sigma}\hat a_{u\delta}\hat a_{s\gamma}\hat a_{q\tau}\right|0\right> \\\nonumber
    = & \left<0\left|\hat{E}_{pq}\hat{E}_{rs}\hat{E}_{tu}\hat{E}_{mn}\right|0\right> - \delta_{rq}\Gamma^{[3]}_{pstumn} - \delta_{tq}\Gamma^{[3]}_{pursmn} - \delta_{mq}\Gamma^{[3]}_{pnrstu} - \delta_{mu}\Gamma^{[3]}_{pqrstn} 
    \\\nonumber
    & - \delta_{ts}\Gamma^{[3]}_{pqrumn} - \delta_{ms}\Gamma^{[3]}_{pqrntu} - \delta_{mu}\delta_{rq}\Gamma^{[2]}_{pstn} - \delta_{mu}\delta_{tq}\Gamma^{[2]}_{pnrs} - \delta_{ts}\delta_{mu}\Gamma^{[2]}_{pqrn} \\\nonumber
    & - \delta_{ts}\delta_{rq}\Gamma^{[2]}_{pumn} - \delta_{ts}\delta_{mq}\Gamma^{[2]}_{pnru} - \delta_{ms}\delta_{rq}\Gamma^{[2]}_{pntu} - \delta_{ms}\delta_{tq}\Gamma^{[2]}_{purn} - \delta_{mu}\delta_{ts}\delta_{rq}\Gamma^{[1]}_{pn}.
\end{align}
The computationally expensive step is measuring the 3- and 4-RDM in the active space, which scale as $\mathcal{O}\left(N_A^6\right)$ and $\mathcal{O}\left(N_A^8\right)$, respectively. Therefore, it is favorable to find approximations for the RDMs. A straightforward way is to ignore (parts of) the 3 -and 4-RDM. As summarized in Table~\ref{tab:rdm_rdc_approximations}, we here investigate four direct approximations to the RDMs, namely setting the complete (3- and) 4-RMD to zero or just their off-diagonals.

\begin{table}[H]
    \centering
    \begin{tabular}{ll}
        \hline\hline
        Approximation & Description \\\hline
        4-zRDM & Set the entire 4-RDM to zero \\
        3-zRDM \& 4-zRDM & Set the entire 3-RDM and 4-RDM to zero \\
        4-dRDM & Set all off-diagonal elements of the 4-RDM to zero \\
        3-dRDM \& 4-dRDM & Set all off-diagonal elements of the 3-RDM and 4-RDM to zero \\\hline
        4-zRDC & Set the entire 4-RDC to zero \\
        3-zRDC \& 4-zRDC & Set the entire 3-RDC and 4-RDC to zero \\
        4-dRDC & Set all off-diagonal elements of the 4-RDC to zero \\
        3-dRDC \& 4-dRDC & Set all off-diagonal elements of the 3-RDC and 4-RDC to zero \\\hline\hline
    \end{tabular}
    \caption{RDM and RDC approximations.}
    \label{tab:rdm_rdc_approximations}
\end{table}

\subsection{Reduced density cumulants}
\label{ssec:RDC}

An alternative to introducing direct approximations in the RDMs is to reconstruct them from reduced density cumulants (RDCs).~\cite{Harris2002,mazziotti1998approximate} The 1-RDM through 4-RDM can be constructed by RDCs as
\begin{align} 
    \boldsymbol{\Gamma}^{[1]} &= \boldsymbol{\Delta}_1 \label{eq:1RDM} \\
    \boldsymbol{\Gamma}^{[2]} &= \boldsymbol{\Delta}_1 \wedge \boldsymbol{\Delta}_1 + \boldsymbol{\Delta}_2 \label{eq:2RDM} \\
    \boldsymbol{\Gamma}^{[3]} &= \boldsymbol{\Delta}_1 \wedge \boldsymbol{\Delta}_1 \wedge \boldsymbol{\Delta}_1 + 3 \boldsymbol{\Delta}_2 \wedge \boldsymbol{\Delta}_1 + \boldsymbol{\Delta}_3 \label{eq:3RDM} \\
    \boldsymbol{\Gamma}^{[4]} &= \boldsymbol{\Delta}_1 \wedge \boldsymbol{\Delta}_1 \wedge \boldsymbol{\Delta}_1 \wedge \boldsymbol{\Delta}_1 + 6 \boldsymbol{\Delta}_2 \wedge \boldsymbol{\Delta}_1 \wedge \boldsymbol{\Delta}_1 + 3 \boldsymbol{\Delta}_2 \wedge \boldsymbol{\Delta}_2 + 4 \boldsymbol{\Delta}_3 \wedge \boldsymbol{\Delta}_1 + \boldsymbol{\Delta}_4 \label{eq:4RDM}
\end{align}
where $\boldsymbol{\Delta}_n$ is the $n$-RDC and $\boldsymbol{\Delta}_n \wedge \boldsymbol{\Delta}_m$ is the Grassmann product of the $n$-RDC and $m$-RDC given by
\begin{equation}
    (\Delta_n \wedge \Delta_m)^{r_1\ldots r_k}_{s_1\ldots s_k} = \frac{1}{(k!)^2} \sum_{\mathcal{P}_k\mathcal{Q}_k} \epsilon_{\mathcal{P}_k}\epsilon_{\mathcal{Q}_k} \Delta^{p_1\ldots p_n}_{q_1\ldots q_n}\Delta^{p_{n+1}\ldots p_{k}}_{q_{n+1}\ldots q_k}
\end{equation}
where $\mathcal{P}$ is the set of permutations over the upper indices, $\mathcal{Q}$ is the set of permutations over the lower indices, $\mathcal{P}_k$ and $\mathcal{Q}_k$ are elements of the sets of permutations, $\epsilon_{\mathcal{P}_k}$ and $\epsilon_{\mathcal{Q}_k}$ are the parities of permutations $\mathcal{P}_k$ and $\mathcal{Q}_k$, and $k = n + m$ is the total number of upper or lower indices. $\mathcal{P}$ and $\mathcal{Q}$ both contain $k!\:$ permutations and, as such, the summation runs over $(k!)^2$ elements. The Grassmann product is commutative
\begin{equation}
    \bm{A} \wedge \bm{B} = \bm{B} \wedge \bm{A}
\end{equation}
associative
\begin{equation}
    \bm{A} \wedge ( \bm{B} \wedge \bm{C} ) = ( \bm{A} \wedge \bm{B} ) \wedge \bm{C}
\end{equation}
and antisymmetric in the upper and lower indices
\begin{align}
    & (\Delta_n \wedge \Delta_m)^{r_1\ldots r_\mu\ldots r_\nu\ldots r_k}_{s_1 \ldots s_\gamma \ldots s_\delta \ldots s_k} = - (\Delta_n \wedge \Delta_m)^{r_1 \ldots r_\nu \ldots r_\mu \ldots r_k}_{s_1 \ldots s_\gamma \ldots s_\delta \ldots s_k} \\
    & = - (\Delta_n \wedge \Delta_m)^{r_1 \ldots r_\mu \ldots r_\nu \ldots r_k}_{s_1\ldots s_\delta \ldots s_\gamma \ldots s_k} = (\Delta_n \wedge \Delta_m)^{r_1 \ldots r_\nu \ldots r_\mu \ldots r_k}_{s_1 \ldots s_\delta \ldots s_\gamma \ldots s_k}~~.
\end{align}
The expression for the $k$-RDC is found by isolation in the $k$-RDM expression in eqs. \eqref{eq:1RDM}--\eqref{eq:4RDM} 
\begin{align}
    \boldsymbol{\Delta}_1 & = \boldsymbol{\Gamma}^{[1]} \label{eq:1RDC} \\
    \boldsymbol{\Delta}_2 & = \boldsymbol{\Gamma}^{[2]} - \boldsymbol{\Delta}_1 \wedge \boldsymbol{\Delta}_1 \label{eq:2RDC} \\
    \boldsymbol{\Delta}_3 & = \boldsymbol{\Gamma}^{[3]} - \boldsymbol{\Delta}_1 \wedge \boldsymbol{\Delta}_1 \wedge \boldsymbol{\Delta}_1 - 3 \boldsymbol{\Delta}_2 \wedge \boldsymbol{\Delta}_1 \label{eq:3RDC} \\
    \boldsymbol{\Delta}_4 & = \boldsymbol{\Gamma}^{[4]} - \boldsymbol{\Delta}_1 \wedge \boldsymbol{\Delta}_1 \wedge \boldsymbol{\Delta}_1 \wedge \boldsymbol{\Delta}_1 - 6 \boldsymbol{\Delta}_2 \wedge \boldsymbol{\Delta}_1 \wedge \boldsymbol{\Delta}_1 - 3 \boldsymbol{\Delta}_2 \wedge \boldsymbol{\Delta}_2 - 4 \boldsymbol{\Delta}_3 \wedge \boldsymbol{\Delta}_1.
\end{align}

Reconstructing the RDMs from the RDCs provides more possibilities for approximations, as summarized in Table~\ref{tab:rdm_rdc_approximations}. In the approximations called zRDC, the RDC of a given order is set to zero and the RDM of same order is partially reconstructed from the lower order RDCs. For instance, in the 4-zRDC approximation, we use the exact 1-, 2-, and 3-RDMs and reconstruct the 4-RDM using  eq.~\eqref{eq:4RDM} with $\boldsymbol{\Delta}_4 = 0$, i.e.~only utilizing the exact 1-, 2-, and 3-RDCs. In the dRDC approximations, the exact diagonal of the RDM is kept and the off-diagonal elements of the RDM are partially reconstructed from the lower order RDCs as shown above. These approximations are employed to both 3- and 4-RDM, and to the 4-RDM alone.

Since the summation in the Grassmann product runs over $(k!)^2$ elements, the reconstruction of the $k$-RDM naively scales as $\mathcal{O}\left\{N_A^{2k}\cdot(k!)^2\right\}$, where $N_A^{2k}$ comes from the size of the $k$-RDM and $(k!)^2$ comes from the summation in the Grassmann product. This scaling can be reduced to $\mathcal{O}\left\{\left(\frac{N_A!}{(N_A-k)!}\right)^2\cdot\frac{1}{n!m!}\right\}$ by utilizing the antisymmetry of the Grassmann product and the symmetries of the RDCs:
\begin{itemize}
\item[-] First, due to the antisymmetry of the Grassmann product, if two or more indices in the upper or lower indices are equal, the element must be zero. This reduces $N_A^{2k}$ to $\left(\frac{N_A!}{(N_A-k)!}\right)^2$.
\item[-] Second, due to the antisymmetry of the Grassmann product, only elements that are unique in the upper and lower indices need to be calculated. Other elements may be found by permuting the indices to a known combination and multiplying with the parities of the permutations. As an example, take an element of the Grassmann product, $(\Delta_n \wedge \Delta_m)^{prtm}_{qsun}$, such that $n+m=4$, $p>r>t>m$, and $q>s>u>n$. From this element all permutation of the indices $p$, $r$, $t$, and $m$ as well as all permutations of the indices $q$, $s$, $u$, and $n$ can be found. This reduces the scaling by a factor of $(k!)^2$.
\item[-] 
Third, due to the symmetries of the RDCs, the number of permutations that need to be considered in the Grassmann product can be reduced. This reduces the scaling by a factor of $n!m!$.
\end{itemize}

\section{Computational Details}
\label{sec:comp}

First, we quantify the potential savings of our approximations in terms of number of measurements on a quantum device, by calculating the number of Pauli strings needed for each approximation. This is done, for increasing active space sizes, for both the Jordan-Wigner, Parity,~\cite{seeley2012bravyi} and the Bravyi-Kitaev~\cite{bravyi2002fermionic} mappings. Furthermore, we utilize QWC to reduce the number of Pauli strings. To this end, the Pauli strings needed for a given RDM are simply sorted in a reverse alphanumerical order and then grouped using QWC using first-fit bin packing. Note that the computational basis string is always included in the list.

For the analysis of the eight approximations (Table~\ref{tab:rdm_rdc_approximations}) to naive qLRSD in the RDM framework, we investigate their effects on increasing sizes of active spaces. For this purpose, we use ladders of dihydrogen molecules of various lengths in the minimal basis STO-3G,~\cite{hehre1969a,hehre1970a,pietro1980a} with a FCI wave function as our ansatz. The FCI ansatz parameters and the RDMs were obtained using \texttt{PySCF}.~\cite{pyscf,pyscf_2} The H-H bond distance in the dihydrogen molecules (the rung) is 2.0 Bohr and the distance between the dihydrogens (the distance between the rungs) is 1.5 Bohr. The qLR calculation is also performed in the full space (all electrons in all orbitals) with the excitation rank reduced to singles and doubles excitations (qLRSD) using our in-house \texttt{DensityMatrixDrivenModule}(\texttt{DMDM}).\cite{dmdm}  The integrals needed in \texttt{DMDM} are gathered through the \texttt{PySCF} interface to \texttt{LIBCINT}.~\cite{libcint}

Going beyond model systems, we also investigate the effects of the RDM and RDC approximations for chemically stable molecules with active spaces. For this purpose, we use the test systems shown in Table~\ref{tab:comp_details}, where the number of doubly occupied (inactive) and virtual orbitals in the Hartree-Fock reference is given along with the utilized active space. All calculations adopted the STO-3G basis set and all wave function optimizations were performed at the oo-UCCSD level using \texttt{SlowQuant},~\cite{slowquant} after which a qLRSD calculation was performed using \texttt{DMDM}.
\begin{table}[H]
    \centering
    \begin{tabular}{c|ccc}
        \hline\hline
        \rule{0pt}{13pt} Molecules & $N_I^\text{HF}$ & $N_V^\text{HF}$ & Active spaces \\[2pt]\hline
        H$_2$S & 9 & 2 & (4, 4); (8, 6) \\
        OCS & 15 & 4 & (4, 4); (6, 6) \\
        SeH$_2$ & 18 & 2 & (4, 4); (8, 6) \\
        \hline\hline
    \end{tabular}
    \caption{List of molecules considered, with the number of inactive (doubly occupied), $N^{\textrm{HF}}_{I}$, and virtual, $N^{\textrm{HF}}_{V}$, orbitals in the Hartree-Fock reference state, and the active spaces used. The notation 
    $(N_e, N_A)$ is used for the active space, where $N_e$ is the number of active electrons and $N_A$ is the number of active orbitals.}
    \label{tab:comp_details}
\end{table}

To test the RDM/RDC approximations for strongly correlated systems, we additionally symmetrically stretch the geometry of H$_2$O and BeH$_2$ to 1.0, 1.5, and 2.0 times the equilibrium bond length of 0.92 \AA\ and 1.35 \AA, respectively. This is done using the cc-pVDZ~\cite{dunning1989a} basis set in the \mbox{(6, 6)} active space for water, and the cc-pVDZ basis set in both a (4, 4) and a (6, 6) active space for BeH$_2$. Also in these cases, calculations are performed with an oo-UCC ansatz using \texttt{SlowQuant}, after which a qLRSD calculation is performed on top using \texttt{DMDM}.

\section{\label{sec:results}Results and Discussion}

\subsection{Pauli savings}
\label{ssec:pauli_savings}

Table~\ref{tab:pauli_savings} collects the number of unique Pauli strings that are needed to measure the RDMs when using, respectively, the Jordan-Wigner, the Parity, and the Bravyi-Kitaev transformations. Note that the number of Pauli strings for the RDMs is a system independent variable -- that is, it is the same for different molecules when the same size of active space is used. The system specific information is encapsulated in the state vector (Ansatz) the RDM operators act on as well as the integrals.

\begin{table}[H]
    \centering
    \caption{Number of additional Pauli strings to be measured for each order of RDMs in various active space sizes using the Jordan-Wigner, Bravyi-Kitaev, and Parity mapping. For the QWC reduction, the Pauli strings are ordered reverse alphanumerically with the computational basis string always included. The notation 
    $(N_e, N_A)$ is used for the active space, where $N_e$ is the number of active electrons and $N_A$ is the number of active orbitals.
    }
    \begin{tabular}{c|cccc}
        \hline\hline
        \multicolumn{5}{c}{Jordan-Wigner mapping} \\
        \hline\hline
        Active space & 1-RDM & 2-RDM & 3-RDM & 4-RDM  \\\hline
        (4, 4)  & 17    & 273     & 834     & 785      \\
        (4, 6)  & 37    & 1\,403  & 10\,838 & 37\,106  \\
        (6, 6)  & 37    & 1\,403  & 10\,838 & 37\,106  \\
        (8, 6)  & 37    & 1\,403  & 10\,838 & 37\,106  \\
        (10, 6) & 37    & 1\,403  & 10\,838 & 37\,106  \\
        (8, 8)  & 65    & 4\,498  & 67\,037 & 446\,726 \\
        \hline\hline
        \multicolumn{5}{c}{Parity mapping} \\
        \hline\hline
        Active space & 1-RDM & 2-RDM & 3-RDM & 4-RDM  \\\hline
        (4, 4)  & 15    & 234     & 336     & 144      \\
        (4, 6)  & 35    & 1\,369  & 9\,320  & 24\,787  \\
        (6, 6)  & 35    & 1\,369  & 9\,325  & 25\,099  \\
        (8, 6)  & 35    & 1\,369  & 9\,320  & 24\,787  \\
        (10, 6) & 35    & 1\,369  & 9\,325  & 25\,099  \\
        (8, 8)  & 63    & 4\,524  & 62\,258 & 387\,835 \\
        \hline\hline
        \multicolumn{5}{c}{Bravyi-Kitaev mapping} \\
        \hline\hline
        Active space & 1-RDM & 2-RDM & 3-RDM & 4-RDM  \\\hline
        (4, 4)  & 15    & 231     & 339     & 144      \\
        (4, 6)  & 59    & 1\,931  & 10\,473 & 37\,375  \\
        (6, 6)  & 59    & 1\,931  & 10\,473 & 37\,375  \\
        (8, 6)  & 59    & 1\,931  & 10\,473 & 37\,375  \\
        (10, 6) & 59    & 1\,931  & 10\,473 & 37\,375  \\
        (8, 8)  & 67    & 4\,827  & 53\,990 & 297\,466 \\
        \hline\hline
    \end{tabular}
    \label{tab:pauli_savings}
\end{table}

Each entry in the table corresponds to the additional number of Pauli strings needed to measure the $k$-RDM when compared to the $(k-1)$-RDM. The trivial trend of an increasing number of Pauli strings with increasing active space size is observed for all mappings. In the Jordan-Wigner and in the Bravyi-Kitaev mappings, the number of electrons in the active space has no effect on the number of Pauli strings, and so the mapping is only dependent on the number of orbitals in the active space. The Parity mapping shows a small dependency on the number of electrons in the active space with an oscillatory behavior between even and odd numbers of $\alpha/\beta$ electrons. For the active spaces considered here, the Jordan-Wigner mapping consistently requires more Pauli string measurements than the Parity and Bravyi-Kitaev mappings. The Parity mapping initially requires fewer Pauli string measurements than the Bravyi-Kitaev mapping, but this changes for the \mbox{(8, 8)} active space where Bravyi-Kitaev requires the fewest Pauli string measurements.

It is apparent from Table~\ref{tab:pauli_savings} that a substantial amount of measurements can be saved by approximating the 3- and 4-RDMs, with greater savings at larger active spaces, and thus, motivating our approximations (Table~\ref{tab:rdm_rdc_approximations}). The $k$-zRDM and $k$-zRDC approximations require zero additional measurements compared to the $(k-1)$-RDM, as both approximations are equivalent to not measuring the $k$-RDM. As the diagonals of the RDMs are in the computational basis, which is always included, no additional Pauli strings need to be measured for the 3-d and 4-dRDMs either. Due to this, the amount of measurements needed for each of the 3- \& 4- or 4-RDM approximations are equivalent and the primary difference between them comes down to classical costs.

Having seen that the 4-RDM is by far the largest contributor to the measurement costs, we next want to understand how many unique elements of the 4-RDM are used in the oo-naive qLR algorithm. Importantly, this is independent of the mapping used and only depends on the number of electrons and orbitals in the active space. The orbital optimization does not use elements of the 4-RDM, and so this discussion is equally valid for naive qLR and oo-naive qLR. 

In Table~\ref{tab:4rdm_percent} we show that: \textbf{(I)} as the number of orbitals in the active space increases, so does the percentage of the 4-RDM that is used, and \textbf{(II)} for a fixed number of orbitals, the number of elements used has a maximum when there are as many electrons as orbitals in the active space. Both of these trends stem from the number of $\hat{G}$ operators. This is due to an increase in allowed index combinations out of the Hartree-Fock reference. To this end, trend \textbf{(II)} is equivalent to maximizing the function $\frac{\text{V}^2\cdot\text{O}^2}{(\text{V}+\text{O})^4}~,$ where O and V are the number of orbitals in the active space that are occupied and virtual in the Hartree-Fock reference, respectively. For a given active space, $\text{V}+\text{O}$ is constant. The maximum of this function is found at $\text{V}=\text{O}$, i.e.\ an active space with an equal number of electrons and orbitals. This indicates that for strongly correlated systems, where the 4-RDM is more populated, the 4-RDM and 4-RDC approximations may produce significant errors.

\begin{table}[H]
    \centering
    \begin{tabular}{c|ccc}
        \hline\hline Active space & Elements used & Total elements & Percent used \\
        \hline
        (4, 4) & 1\,761 & 1\,996 & 88.23 \% \\
        (4, 6) & 35\,800 & 41\,406 & 86.46 \% \\
        (6, 6) & 37\,415 & 41\,406 & 90.36 \% \\
        (8, 6) & 35\,311 & 41\,406 & 85.28 \% \\
        (10, 6) & 26\,785 & 41\,406 & 64.69 \% \\
        (8, 8) & 349\,877 & 384\,112 & 91.09 \% \\
        (10, 10) & 2\,022\,907 & 2\,212\,750 & 91.42 \% \\
        (12, 12) & 8\,555\,818 &  9\,340\,332 & 91.60 \% \\
        \hline\hline
    \end{tabular}
    \caption{Number of symmetry unique elements used from the 4-RDM in the oo-naive qLR algorithm.}
    \label{tab:4rdm_percent}
\end{table}

\subsection{\texorpdfstring{H\textsubscript{2}}{} ladders}
\label{ssec:h-ladder}

\begin{figure}[H]
    \centering
    \includegraphics[width=.47\linewidth]{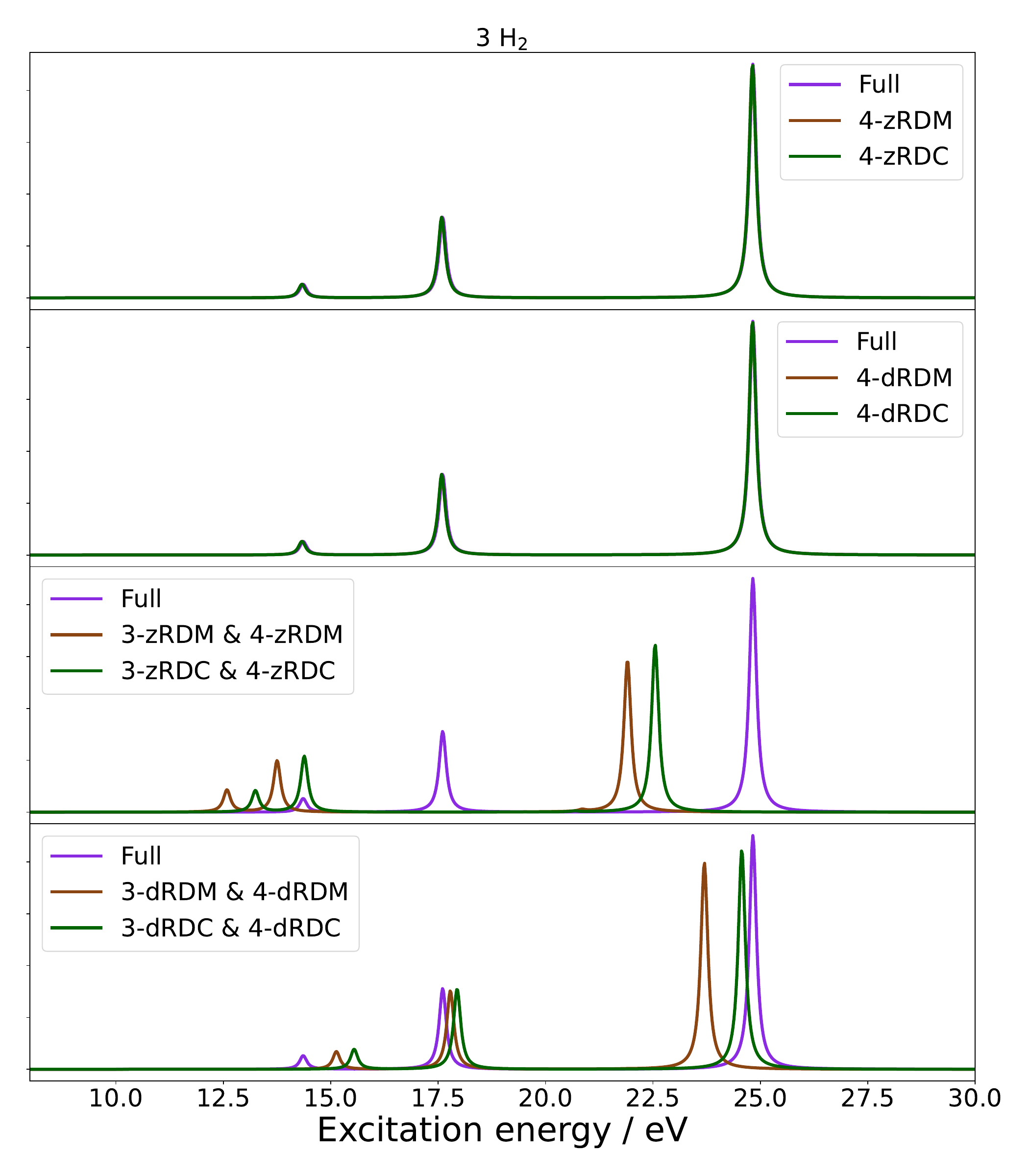}
    \includegraphics[width=.47\linewidth]{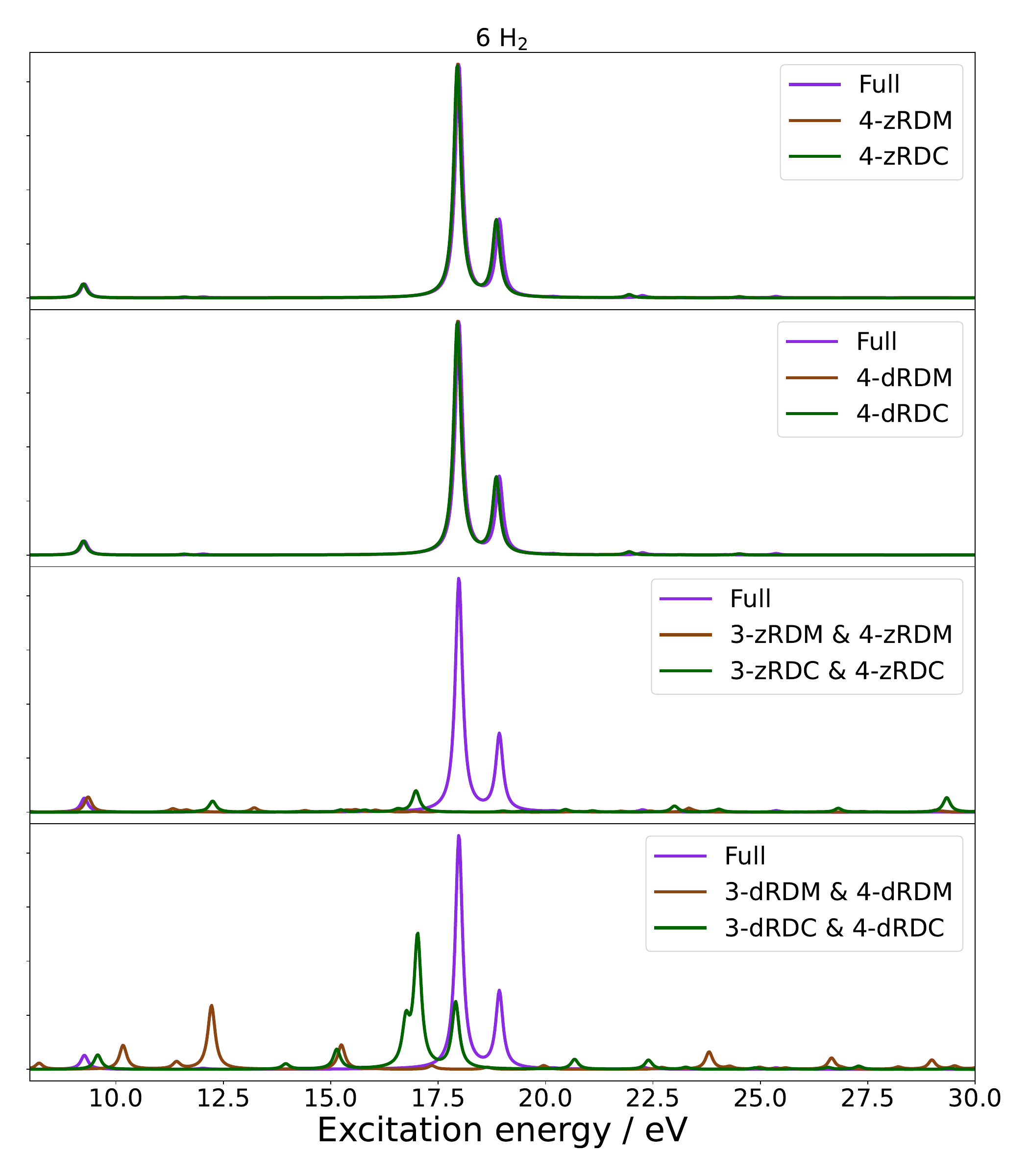}
    \caption{Absorption spectra of H$_2$ ladders containing three (left) and six (right) rugs. Each figure contains four panels comparing the naive qLRSD absorption spectrum with no approximation to the absorption spectrum of naive qLRSD using the 4-zRDM and 4-zRDC approximations (first panel), the absorption spectrum of naive qLRSD using the 4-dRDM and 4-dRDC approximations (second panel), absorption spectrum of naive qLRSD using the 3-zRDM \& 4-zRDM and 3-zRDC \& 4-zRDC approximations (third panel), and absorption spectrum of naive qLRSD using the 3-dRDM \& 4-dRDM and 3-dRDC \& 4-dRDC approximations (fourth panel).}
    \label{fig:H2}
\end{figure}

We start by assessing the impact of our eight approximations on the excitation energies and absorption spectra of H$_2$ ladders of various length. The approximations to the RDMs and RDCs has a possibility of introducing linear dependencies in the linear response equations. This leads to non-physical excitation energies with a value of $0$ Ha. In our investigations we remove these and refer to the remainder as "non-zero excitation energies". In Fig.~\ref{fig:H2}, the absorption spectra of 3 H$_2$ and 6 H$_2$ are shown (1 H$_2$, 2 H$_2$, 4 H$_2$, and 5 H$_2$ in Fig.~\ref{SI-fig:H2}), while Table~\ref{tab:H2} quantifies the mean absolute errors (MAE) compared to FCI for a maximum of 100 non-zero excitation energies (the oscillator strengths can be found in Table~\ref{SI-tab:H2}). 

It is immediately clear from Fig.~\ref{fig:H2} that approximations to the 4-RDM, whether done by approximating/neglecting the 4-RDM itself or its cumulant, has little qualitative impact on the spectra of the qLRSD method, while increasingly large errors are seen in Table~\ref{tab:H2} with increasing number of H$_2$. From the MAE in Table~\ref{tab:H2}, it is also clear that there is no difference between neglecting the entire 4-RDM (4-zRDM) or neglecting only its off-diagonal values (4-dRDM). This is further supported by counting the number of times that diagonal values of the 4-RDM are accessed, where for all lengths of the H$_2$ ladder the diagonal 4-RDM values are accessed zero times. It is also apparent from Table~\ref{tab:H2} that approximations to the 4-RDC perform similarly to direct approximations of the 4-RDM. From the results in Table~\ref{tab:H2} a general trend of the MAE rising with an increase of system size can be seen in line with the trend seen in Table~\ref{tab:4rdm_percent} of an increase of RDM usage with system size.  

When we look at approximations to the 3-RDM, we can see in the bottom two panels of Fig.~\ref{fig:H2} that any approximations to the 3-RDM,   whether by approximating the RDM itself or its cumulant,  significantly impact the spectrum. Therein, the 3- \& 4-dRDM approximations perform better than the 3- \& 4-zRDM approximations, and both 3-RDC approximations perfom better than the respective 3-RDM approximations. The latter effect is more pronounced in the 3- \& 4-dRDC approximation and therefore shows that the diagonal of the 3-RDC contributes appreciably to the excitation energies.

\begin{table}[H]
    \centering
    \begin{tabular}{c|cccc|cccc}
        \hline\hline & \multicolumn{4}{c|}{RDM approximations} & \multicolumn{4}{c}{RDC approximations} \\
        System & 4-z & 3- \& 4-z & 4-d & 3- \& 4-d & 4-z & 3- \& 4-z & 4-d & 3- \& 4-d  \\
        \hline
        1 H$_2$ & 0.0000 & 0.0000 & 0.0000 & 0.0000 & 0.0000 & 0.0000 & 0.0000 & 0.0000 \\
        2 H$_2$ & 0.2999 & 4.2836 & 0.2999 & 2.2334 & 0.2999 & 4.2835 & 0.2999 & 2.2291 \\
        3 H$_2$ & 0.4273 & 7.4679 & 0.4273 & 1.6097 & 0.4414 & 7.0933 & 0.4296 & 1.7333 \\
        4 H$_2$ & 0.8669 & 9.6414 & 0.8669 & 4.9432 & 0.8443 & 8.2002 & 0.8443 & 2.0725 \\
        5 H$_2$ & 0.6302 & 12.4120 & 0.6302 & 4.8132 & 0.6972 & 10.4634 & 0.6972 & 2.7017 \\
        6 H$_2$ & 0.8699 & 15.5930 & 0.8699 & 8.0726 & 0.8781 & 14.0385 & 0.8781 & 3.1770 \\
        \hline\hline
    \end{tabular}
    \caption{Mean absolute errors (MAE) in eV for the eight RDM and RDC approximations for a maximum 100 non-zero excitation energies for different lengths of the H$_2$ ladder.}
    \label{tab:H2}
\end{table}

\subsection{Other molecules}
\label{ssec:stable}

Next, we investigate whether the findings of the H$_2$ model ladders hold for a set of more realistic molecules. Special interest is on validating (or disproving) the previous finding that the 4-RDM can be neglected/approximated for the qLR algorithm. 

\begin{figure}[H]
    \centering
    \includegraphics[width=0.51\linewidth]{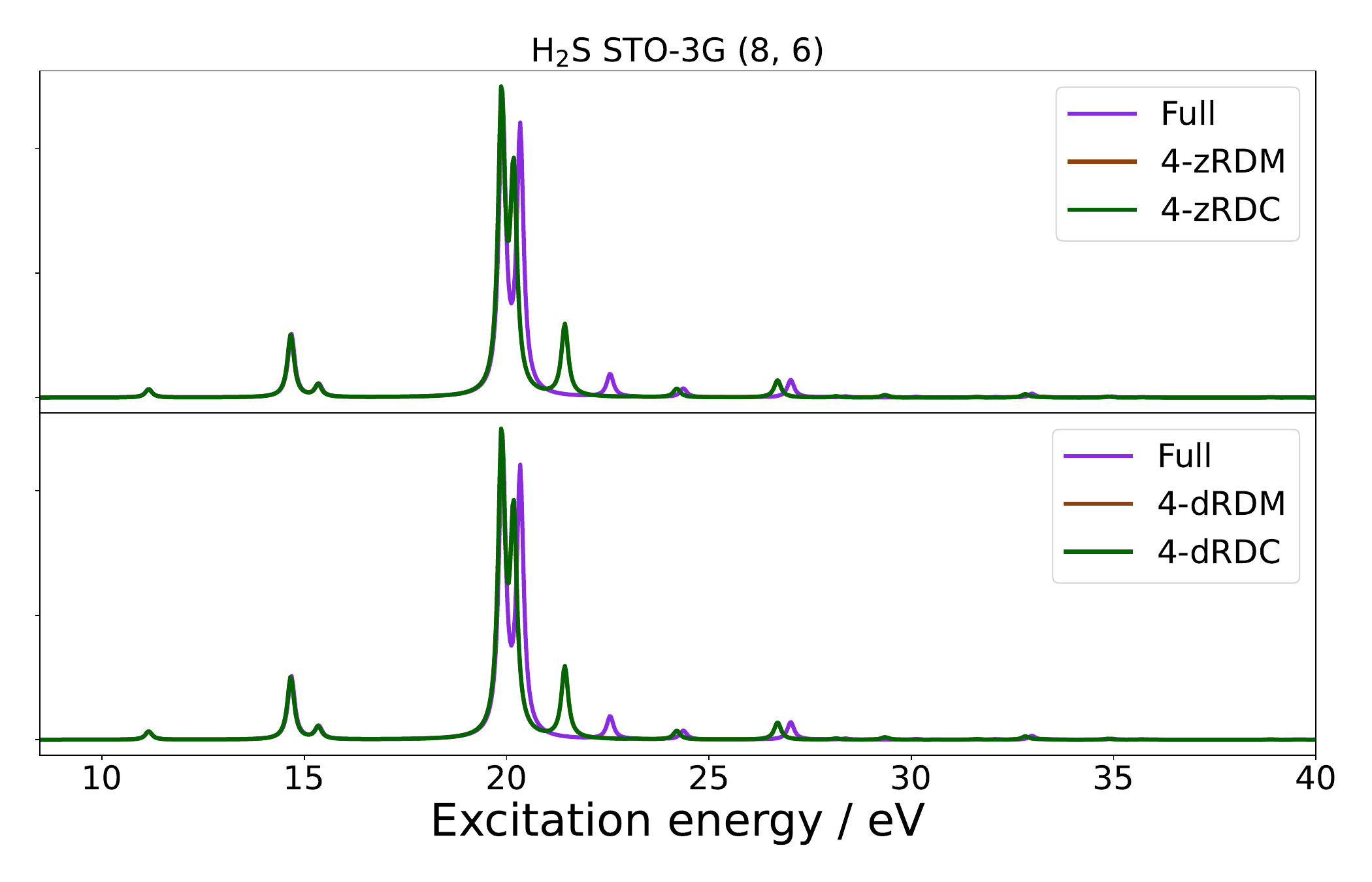}
    \includegraphics[width=0.51\linewidth]{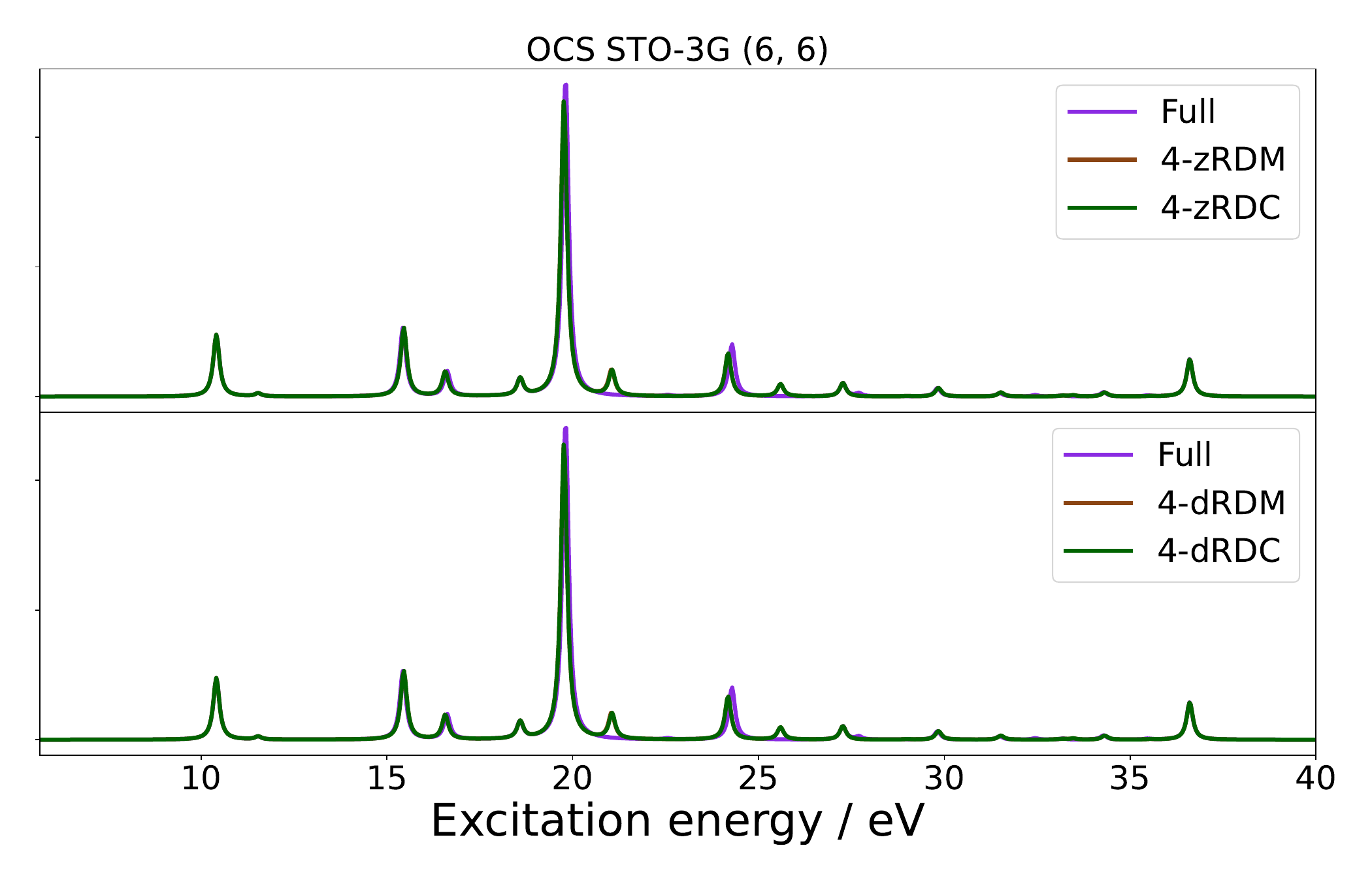}
    \includegraphics[width=0.51\linewidth]{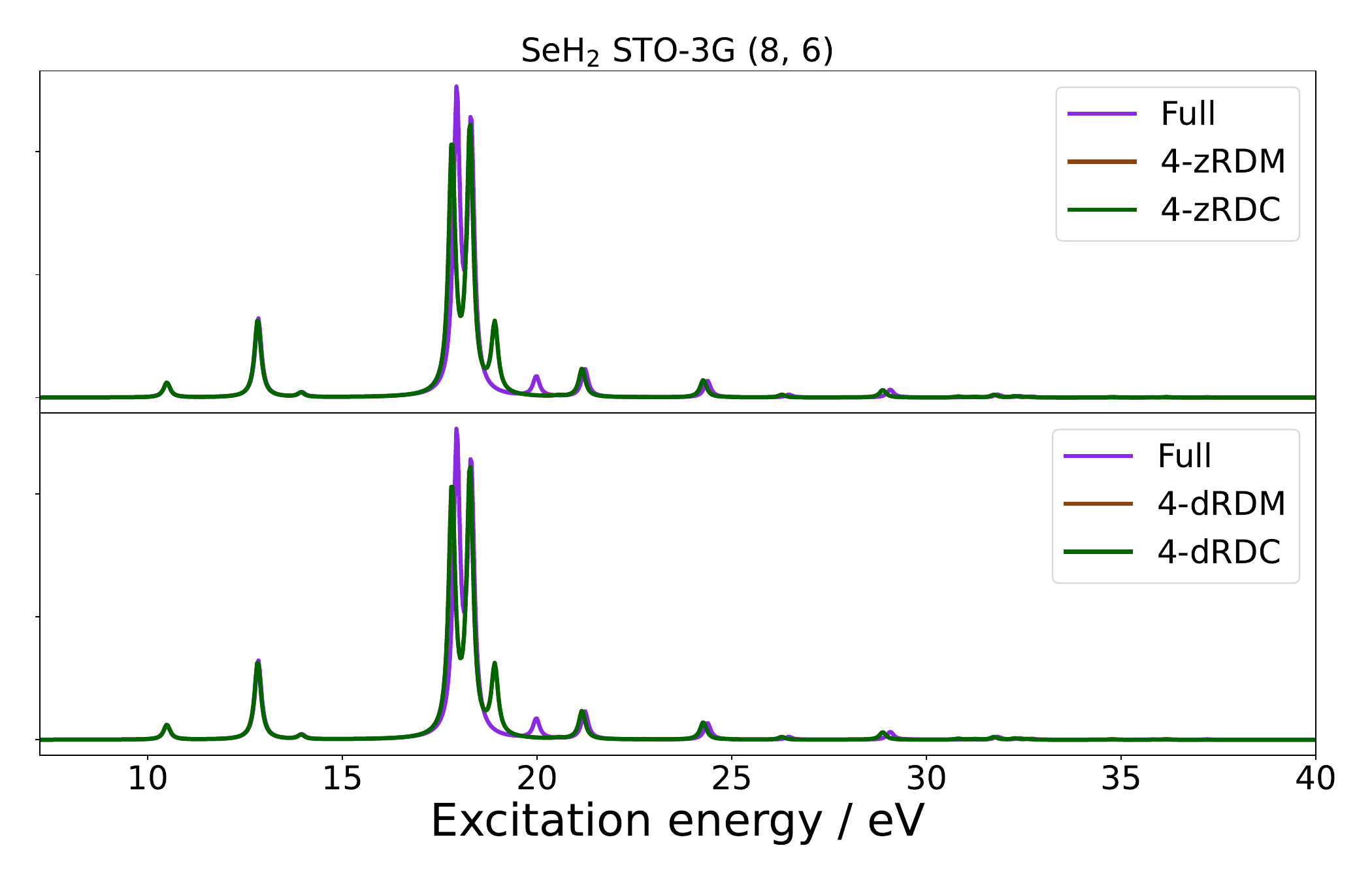}
    \caption{Absorption spectrum of H$_2$S (8, 6) [top], OCS (6, 6) [middle], and SeH$_2$ [bottom]. Each figure contains two panels comparing the naive qLRSD absorption spectrum with no approximation to the absorption spectrum of naive qLRSD using the 4-zRDM and 4-zRDC approximations (first panel) and the absorption spectrum of naive qLRSD using the 4-dRDM and 4-dRDC approximations (second panel).}
    \label{fig:molecules_large_active_space}
\end{figure}

In Fig.~\ref{fig:molecules_large_active_space} the absorption spectra of the molecules H$_2$S, OCS, and SeH$_2$ in the active spaces of (8, 6), (6, 6), and (8, 6) are shown; Table~\ref{tab:other_molecules} collects the MAE of the first 100 non-zero excitation energies of H$_2$S, OCS, and SeH$_2$ in all the calculated active spaces. Results for the (4, 4) active space calculations in addition to the 3-RDM and 3-RDC approximations may be found in the SI in Figs.~\ref{SI-fig:other_molecules_4-4} and~\ref{SI-fig:molecules_large_active_space}. For standard deviations and values regarding oscillator strengths we refer to Table~\ref{SI-tab:other_molecules}.

The results in Table~\ref{tab:other_molecules} confirm  that in general the contribution from the 4-RDM is very small, with a maximum MAE of $0.5964 \pm 0.6293$ eV for the excitation energies of OCS in a (6, 6) active space in the 4-zRDM approximation. While the trend of increasing errors with an increasing active space is apparent, the errors are smaller due to the contribution from the classically-treated inactive space. This reduces the overall contribution from the active space on the excitation energies and thereby the errors caused by the approximations. Once again, the errors increase when approximating the 3-RDM to such a degree that it is not applicable to approximate the 3-RDM or its cumulant. It is noted that the 3- \& 4-dRDC approximation is the best performing variant of the 3-RDM approximations.

\begin{table}[H]
    \begin{tabular}{c|cccc|cccc}
        \hline\hline & \multicolumn{4}{c|}{RDM approximations} & \multicolumn{4}{c}{RDC approximations} \\
        Molecule & 4-z & 3- \& 4-z & 4-d & 3- \& 4-d & 4-z & 3- \& 4-z & 4-d & 3- \& 4-d  \\
        \hline
        H$_2$S (4, 4) & 0.0748 & 1.0614 & 0.0748 & 0.6222 & 0.0748 & 1.0614 & 0.0748 & 0.6227 \\
        H$_2$S (8, 6) & 0.3286 & 1.4537 & 0.3286 & 1.1384 & 0.3295 & 1.3105 & 0.3295 & 1.1497 \\
        OCS (4, 4) & 0.0835 & 26.5346 & 0.0835 & 25.3310 & 0.0835 & 26.5340 & 0.0835 & 25.3311 \\
        OCS (6, 6) & 0.5964 & 10.1088 & 0.5964 & 5.9769 & 0.5934 & 13.3942 & 0.5934 & 3.8050 \\
        SeH$_2$ (4, 4) & 0.0368 & 0.4996 & 0.0368 & 0.3318 & 0.0368 & 0.4995 & 0.0368 & 0.3318 \\
        SeH$_2$ (8, 6) &  0.2327 & 1.1153 & 0.2327 & 0.6882 & 0.2341 & 1.0362 & 0.2341 & 0.8265 \\
        \hline\hline
    \end{tabular}
    \caption{Mean absolute errors (MAE) in eV for the eight RDM and RDC approximations for a maximum of 100 non-zero excitation energies. The notation 
    $(N_e, N_A)$ is used for the active space, where $N_e$ is the number of active electrons and $N_A$ is the number of active orbitals.}
    \label{tab:other_molecules}
\end{table}

\subsection{Strongly correlated systems}
\label{ssec:corr}

\begin{figure}[H]
    \centering
    \includegraphics[width=0.47\linewidth]{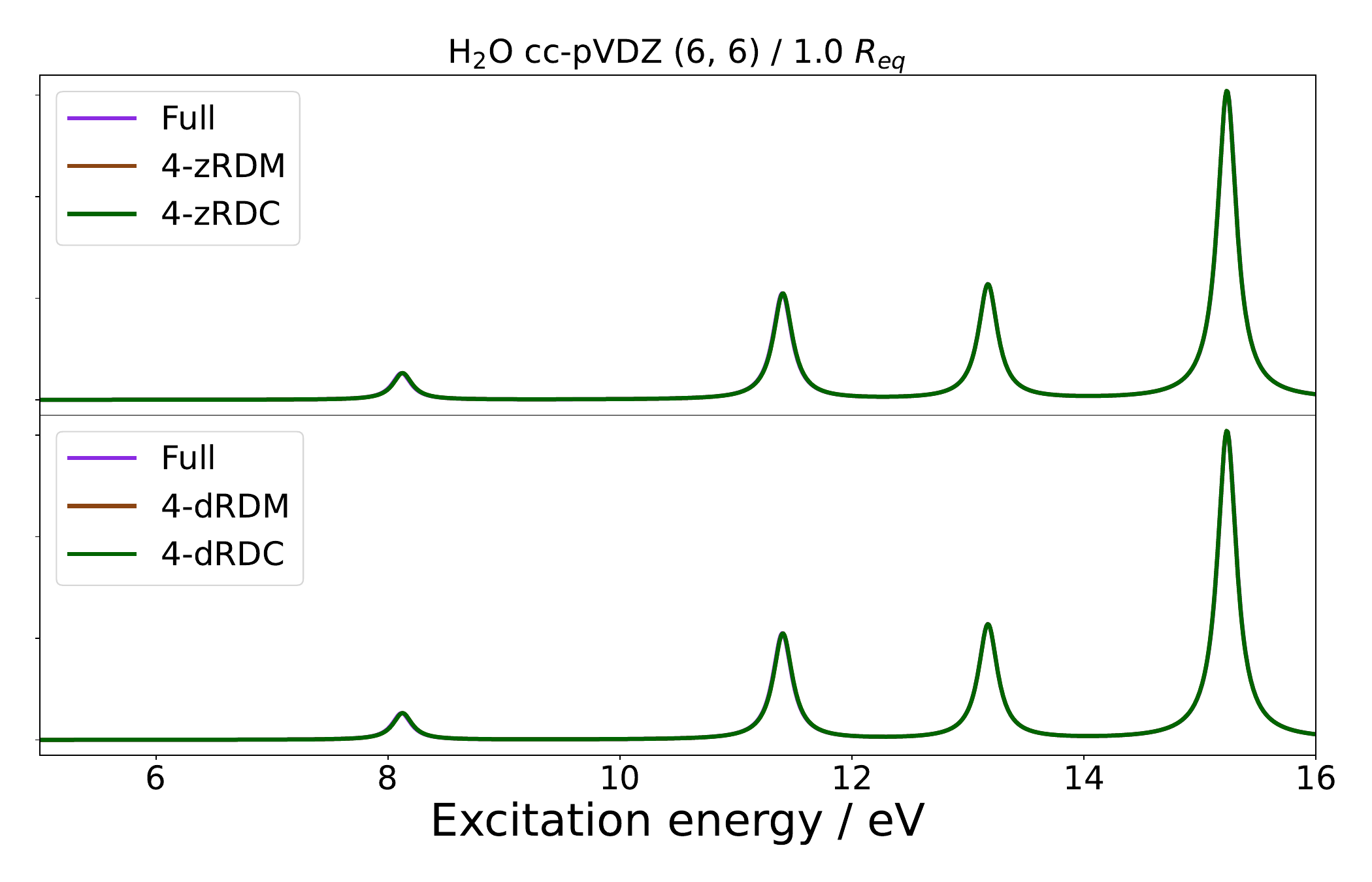}
    \includegraphics[width=0.47\linewidth]{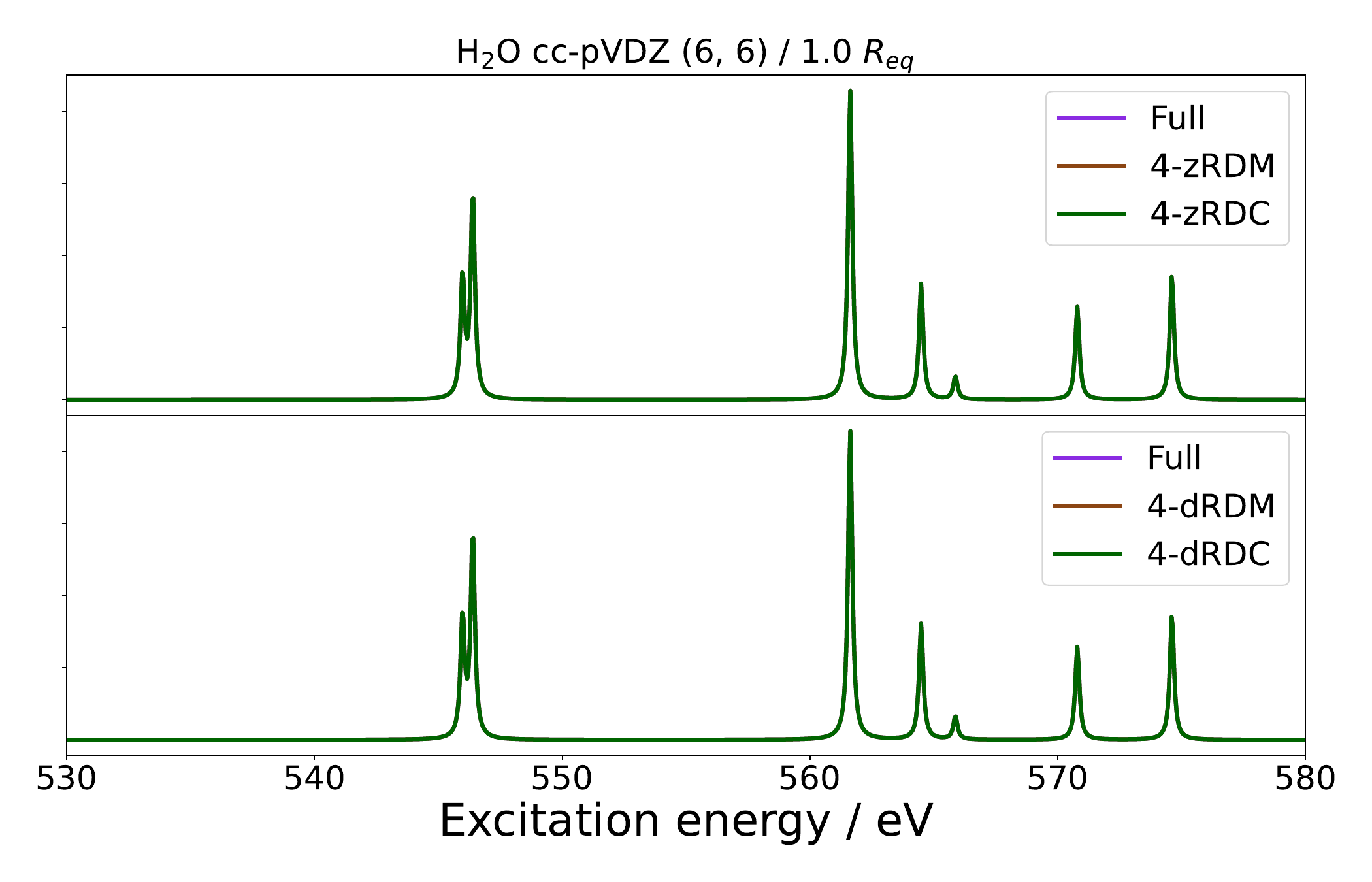}
    \includegraphics[width=0.47\linewidth]{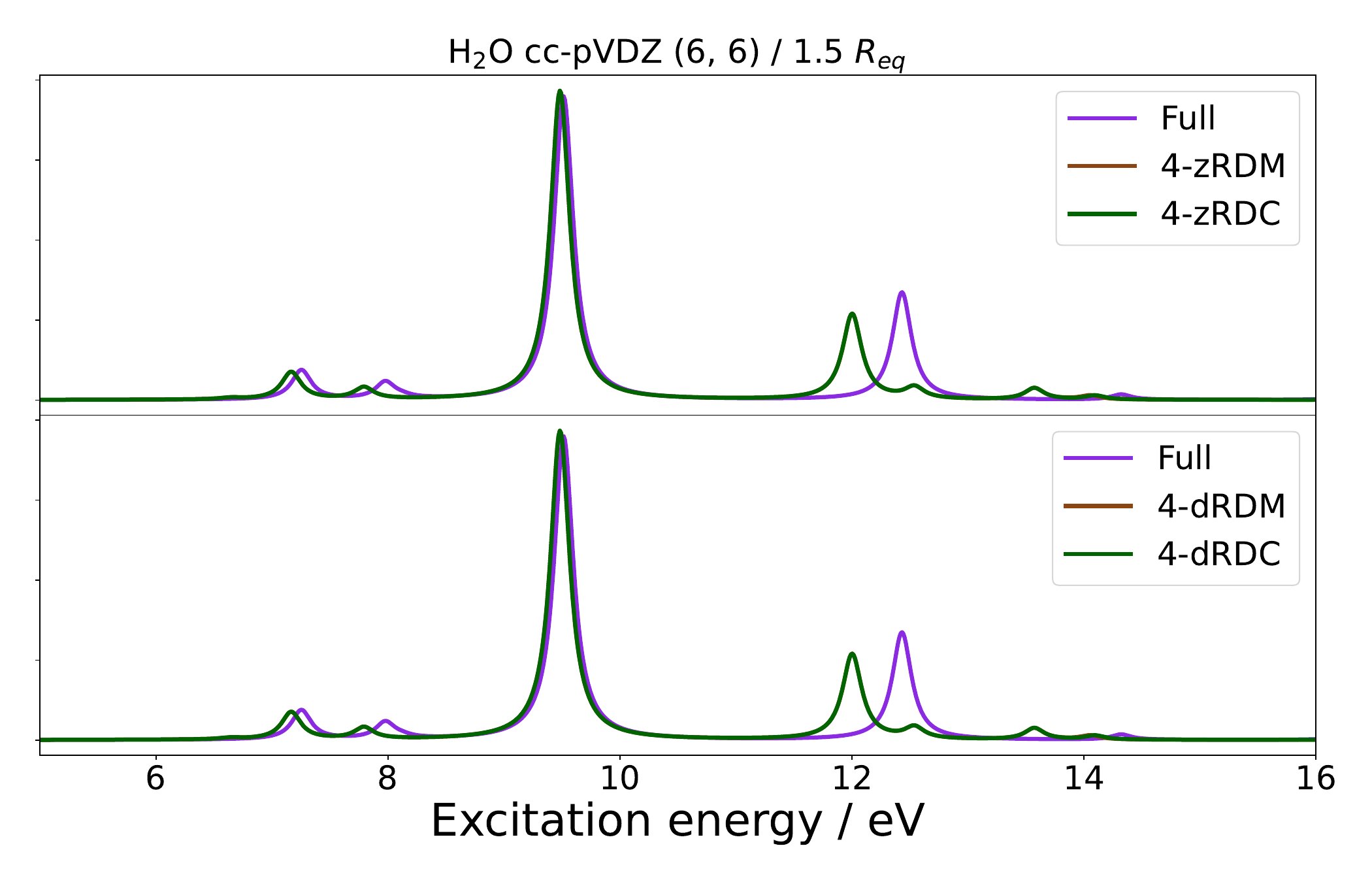}
    \includegraphics[width=0.47\linewidth]{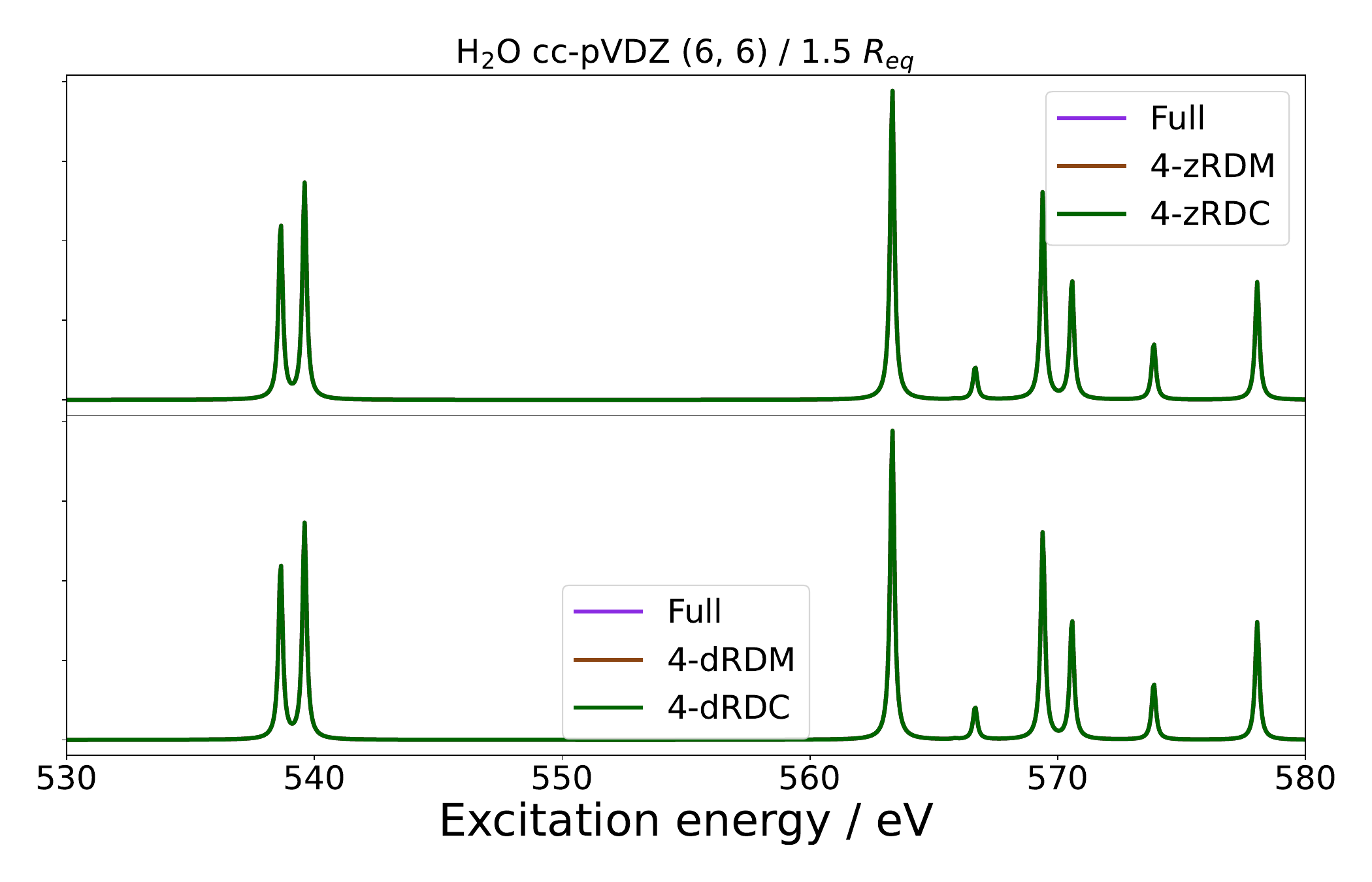}
    \includegraphics[width=0.47\linewidth]{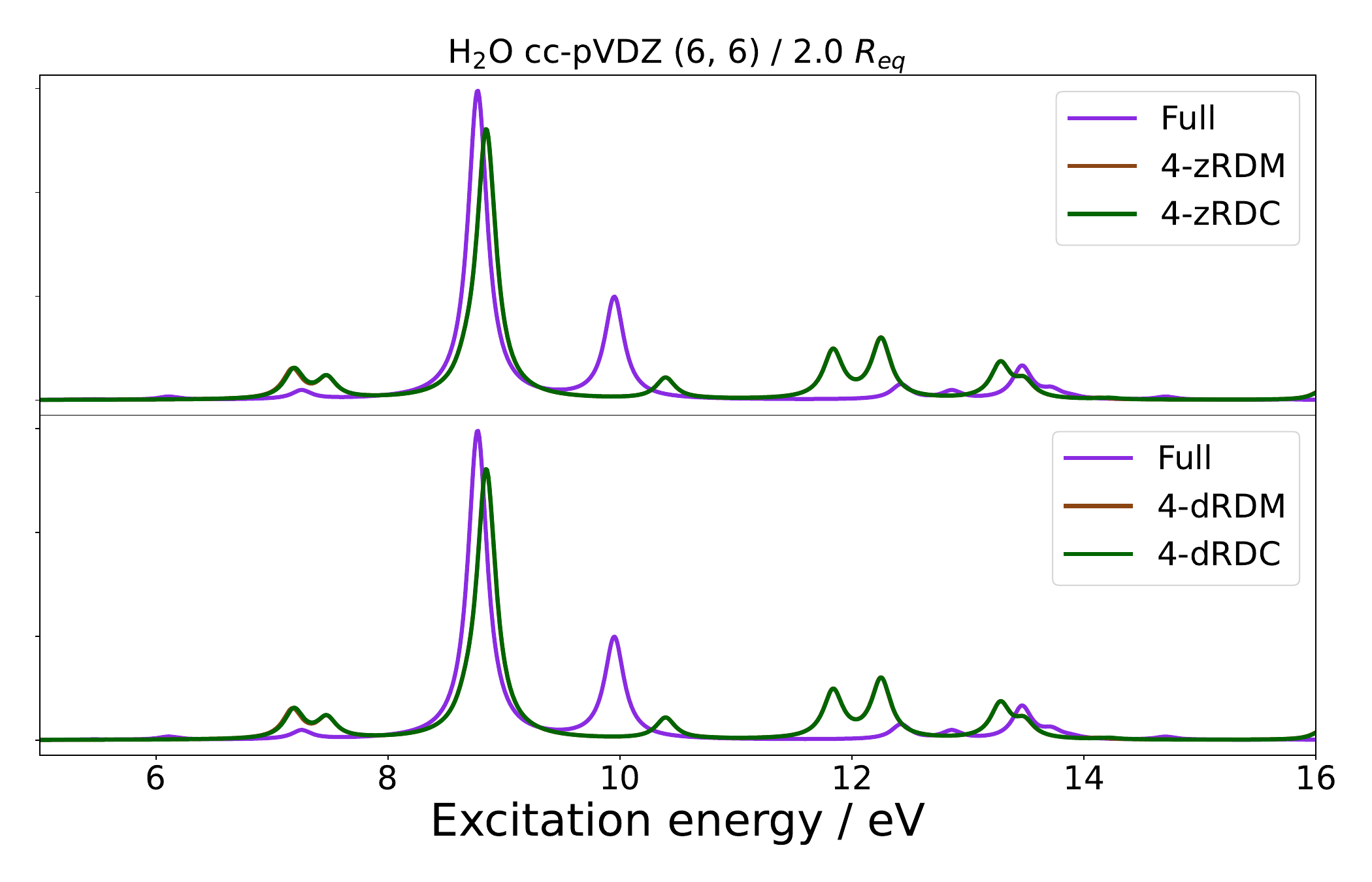}
    \includegraphics[width=0.47\linewidth]{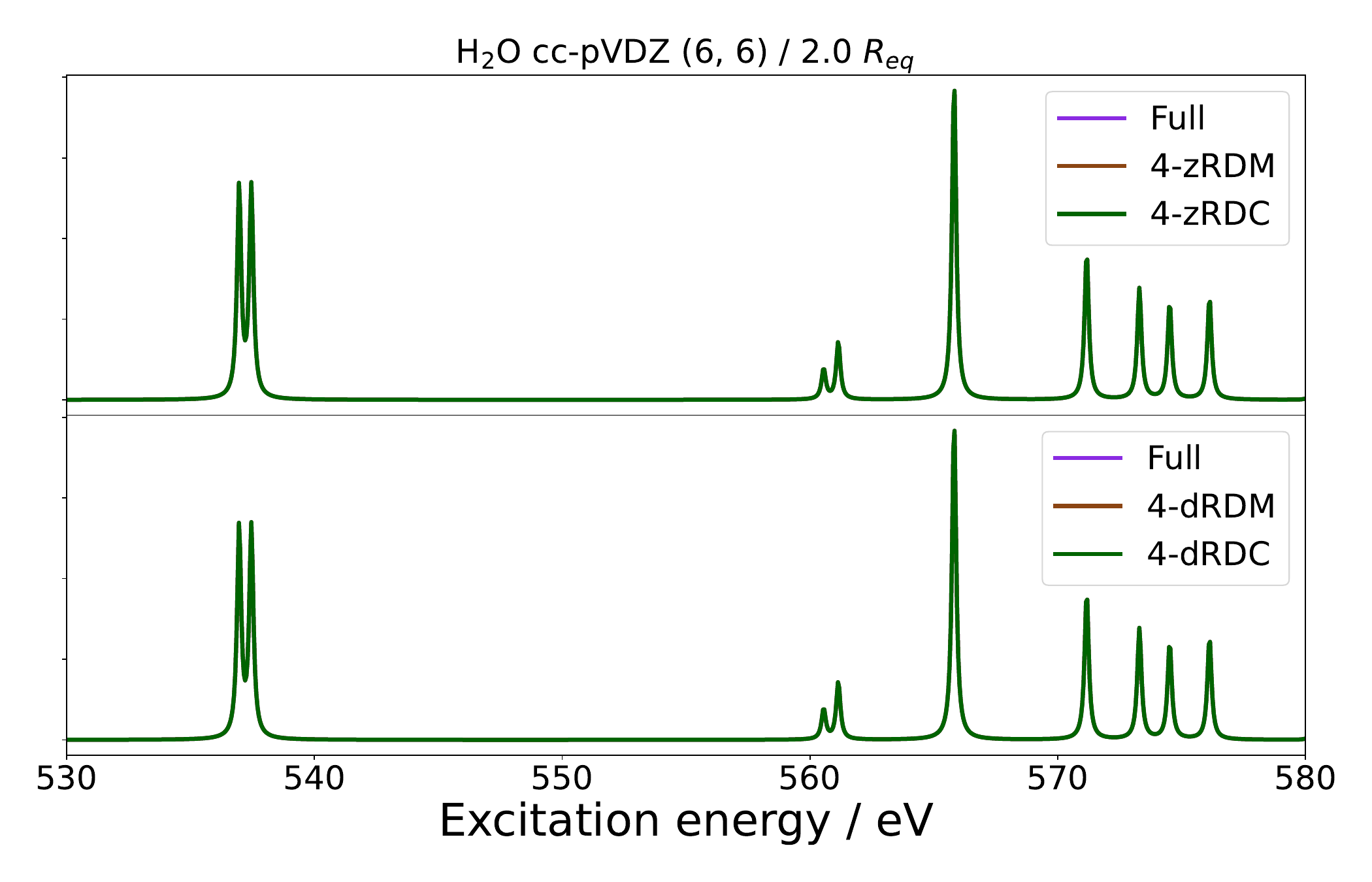}
    \caption{Absorption spectra of H$_2$O in a (6, 6) active space with the cc-pVDZ basis set at differing symmetric O-H stretches in the valence [left] and core [right] excitation regions. Each figure contains two panels comparing the naive qLRSD absorption spectrum with no approximation to the absorption spectrum of naive qLRSD using the 4-zRDM and 4-zRDC approximations (first panel) and the absorption spectrum of naive qLRSD using the 4-dRDM and 4-dRDC approximations (second panel).}
    \label{fig:stretched_water_valence_dz}
\end{figure}

\begin{figure}[H]
    \centering
    \includegraphics[width=0.47\linewidth]{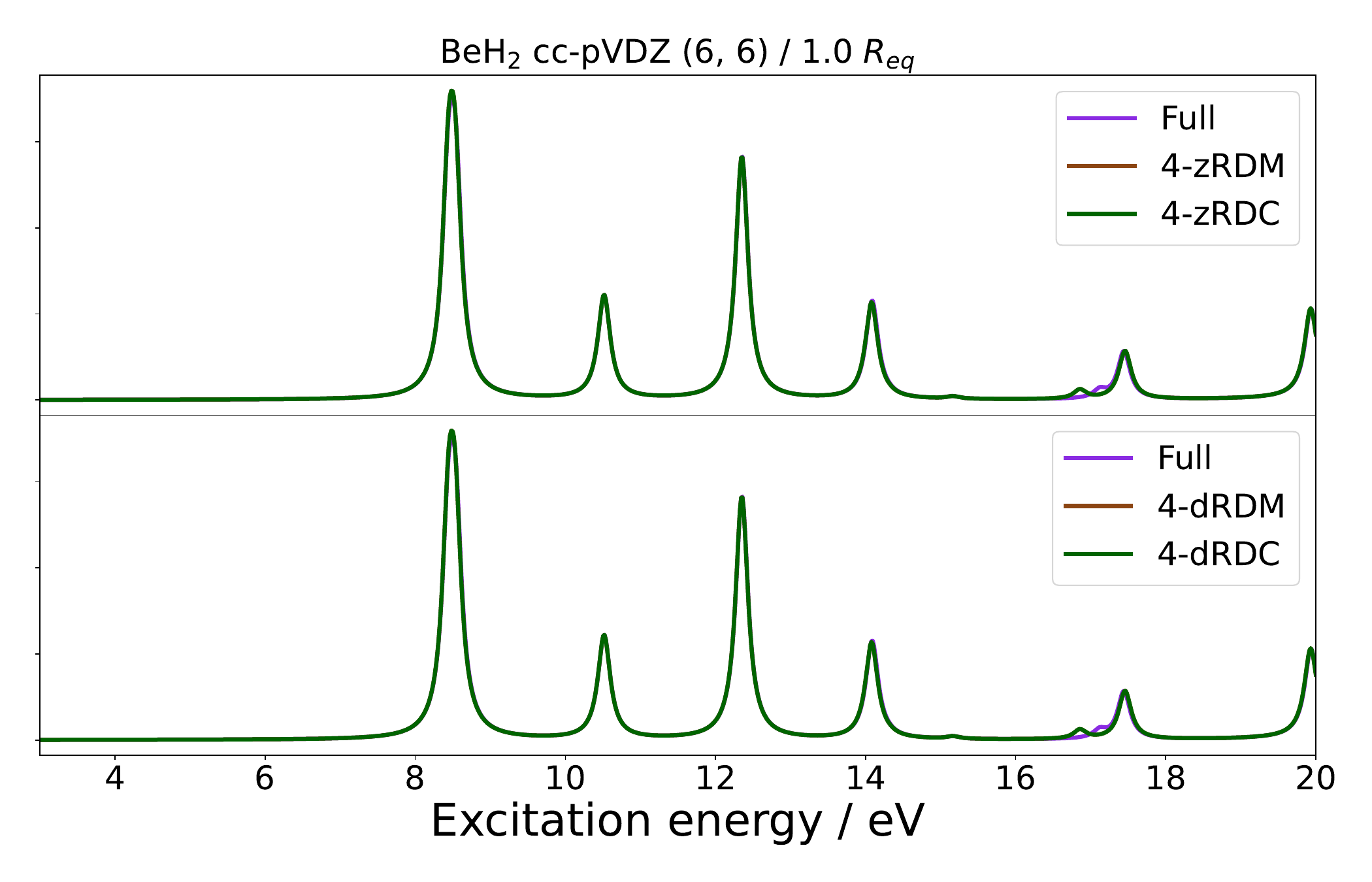}
    \includegraphics[width=0.47\linewidth]{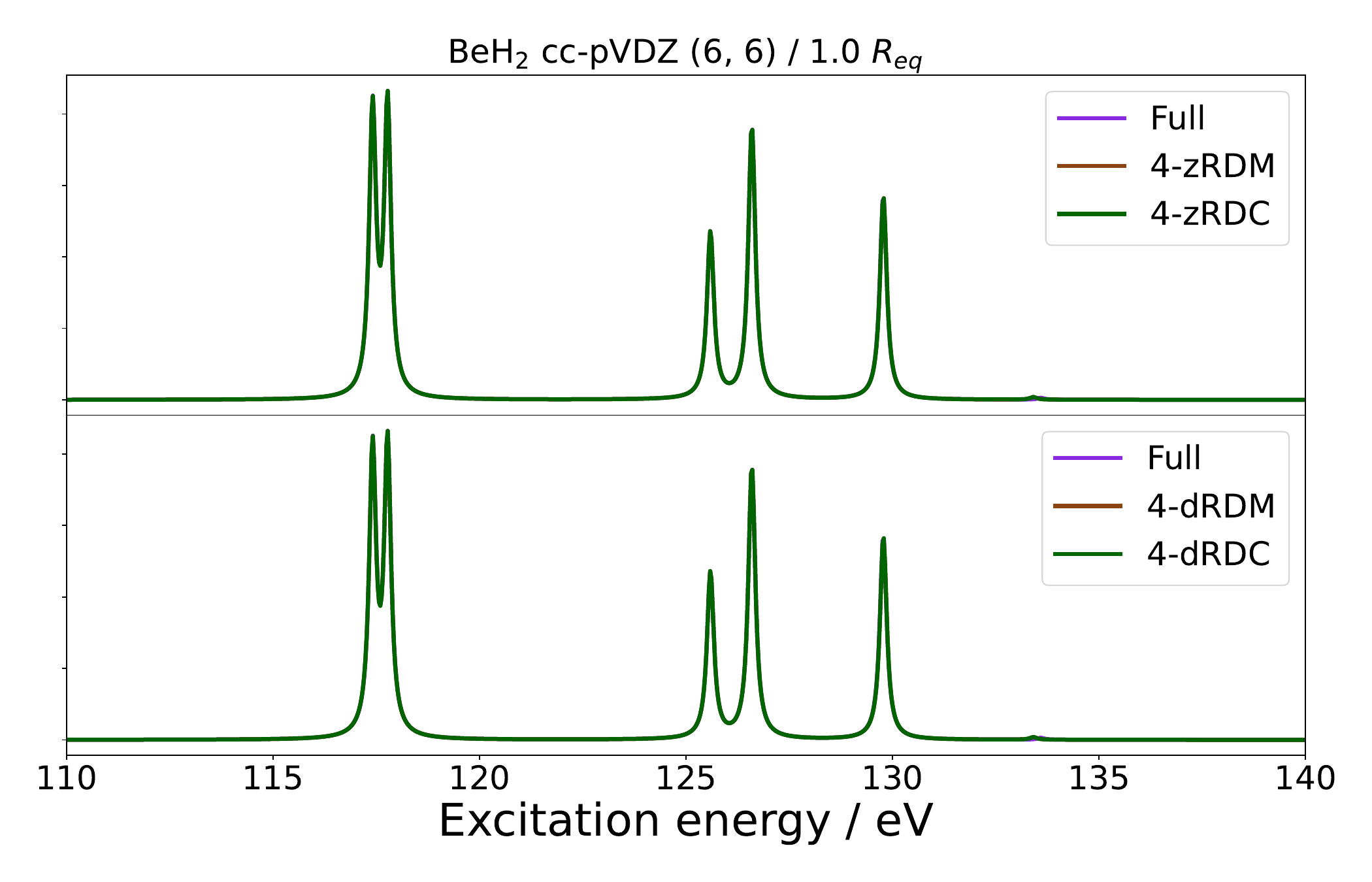}
    \includegraphics[width=0.47\linewidth]{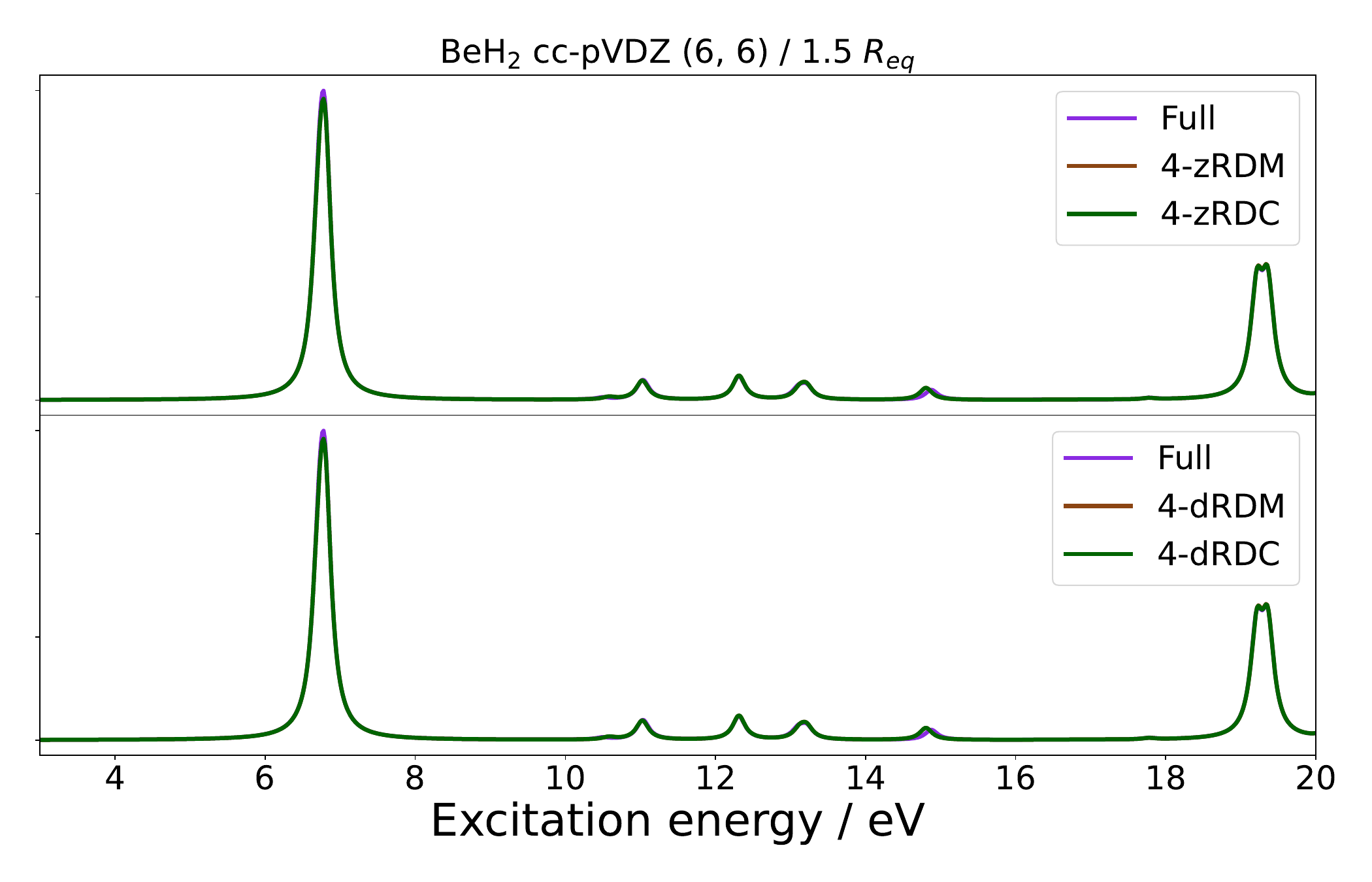}
    \includegraphics[width=0.47\linewidth]{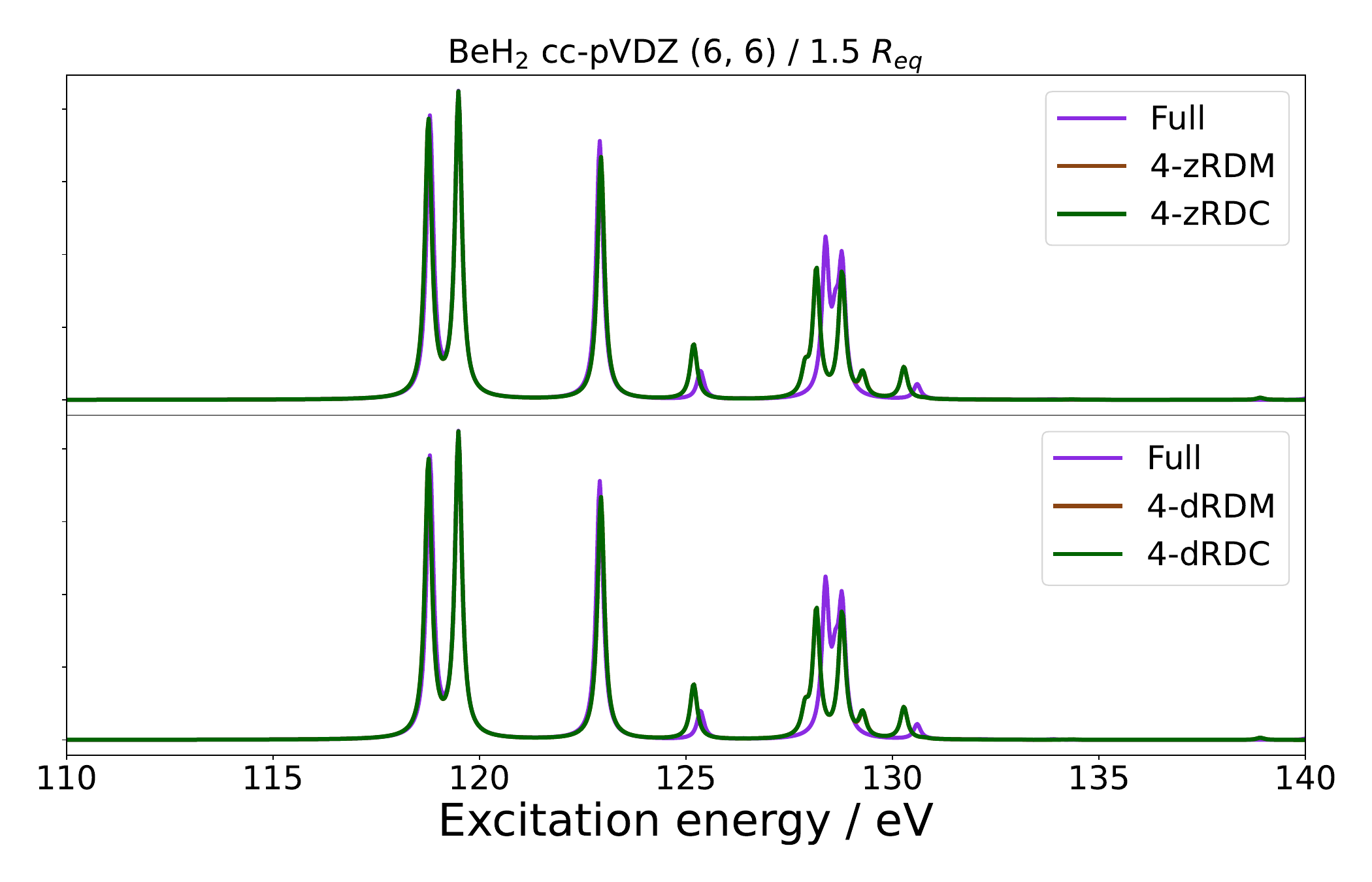}
    \includegraphics[width=0.47\linewidth]{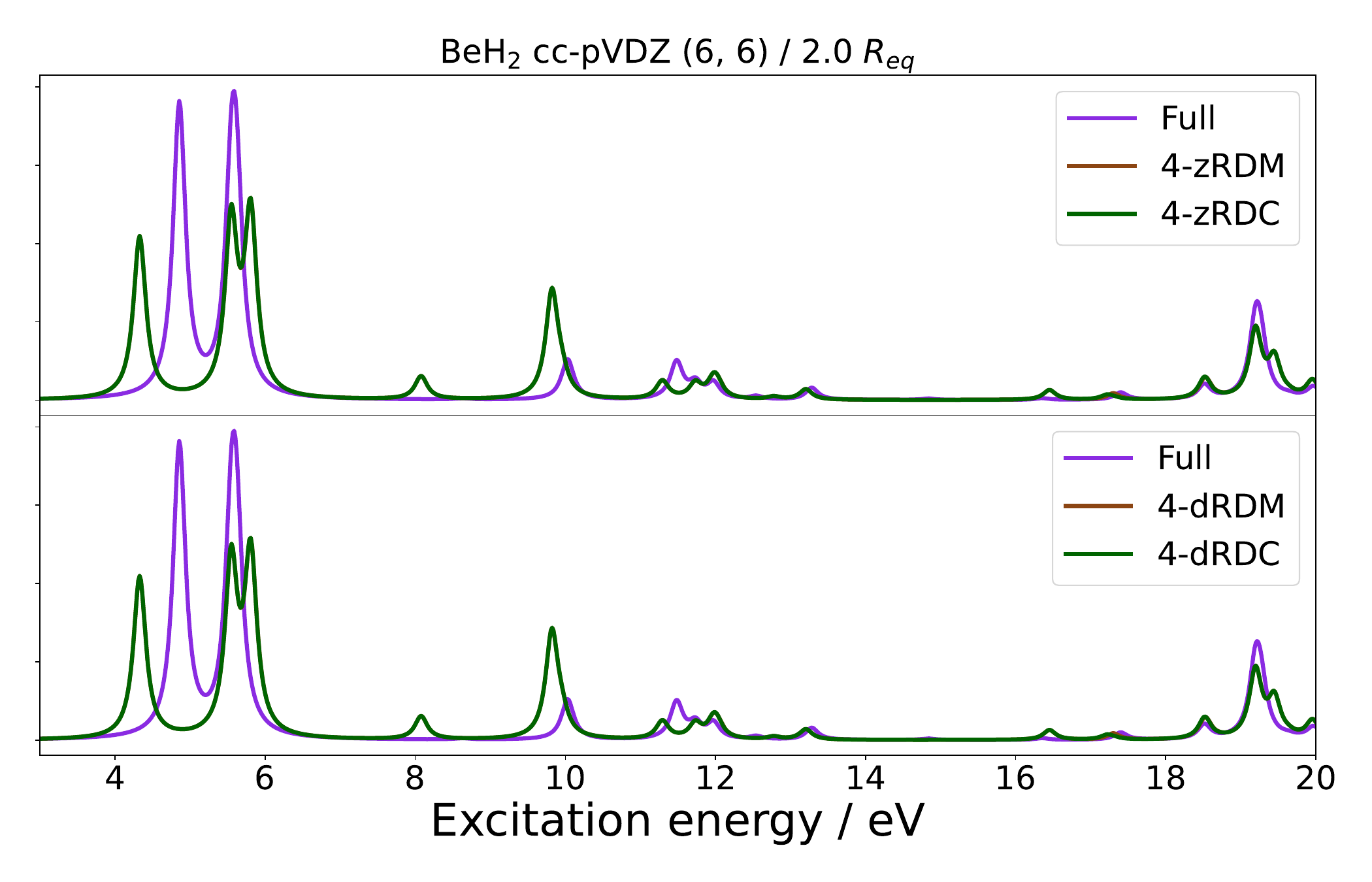}
    \includegraphics[width=0.47\linewidth]{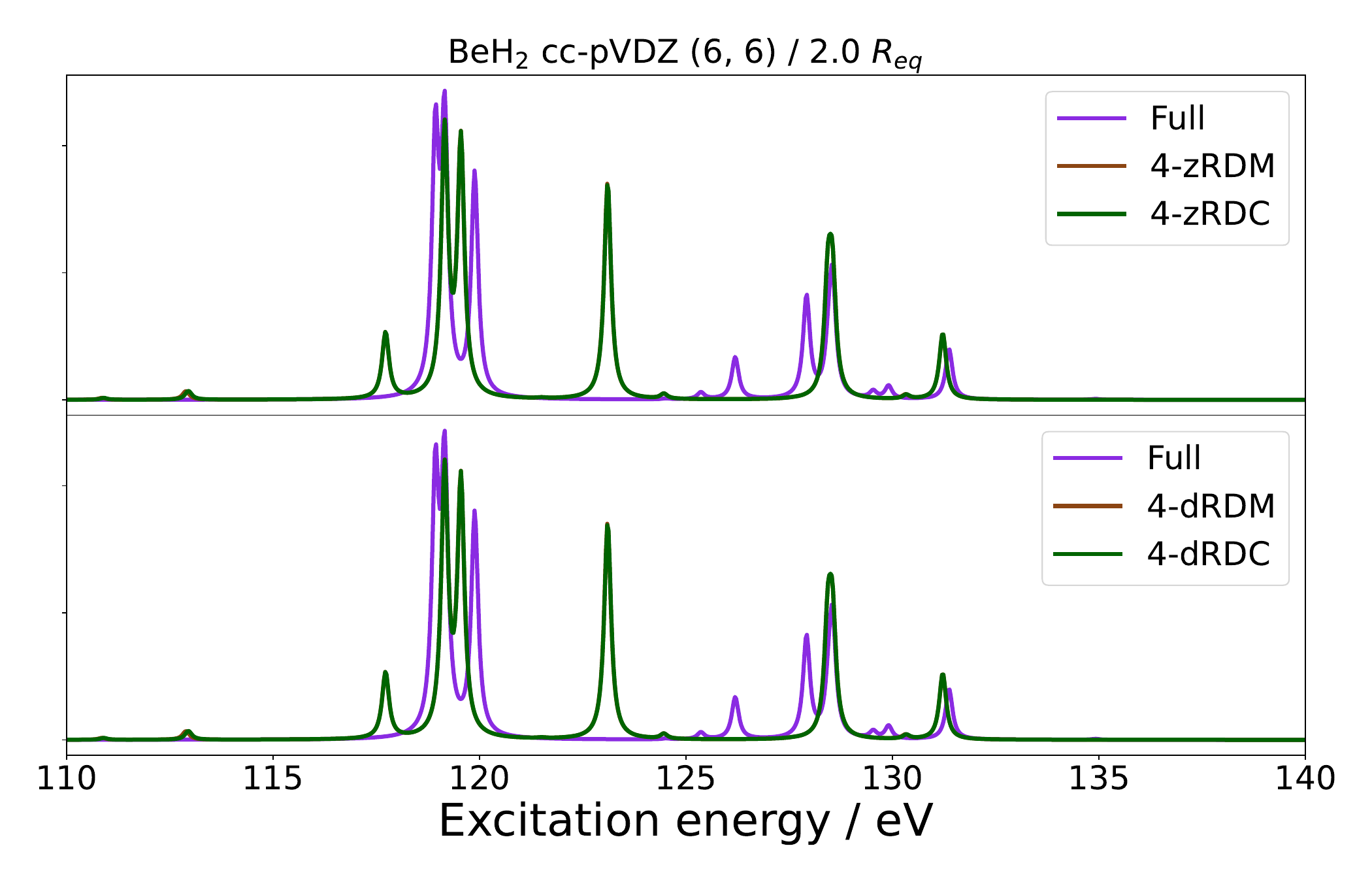}
    \caption{Absorption spectra of BeH$_2$ in a (6, 6) active space with the cc-pVDZ basis set at differing symmetric Be-H stretches in the valence [left] and core [right] excitation regions. Each figure contains two panels comparing the naive qLRSD absorption spectrum with no approximation to the absorption spectrum of naive qLRSD using the 4-zRDM and 4-zRDC approximations (first panel) and the absorption spectrum of naive qLRSD using the 4-dRDM and 4-dRDC approximations (second panel).}
    \label{fig:stretched_BeH2_valence_6_6}
\end{figure}

We have seen that for molecules in their equilibrium geometry the 4-RDM does not contribute significantly to the qLR algorithm. Since the 4-RDM is important for the qLR Hessian element with double excitations, the lack of relevance of the 4-RDM could be due to a dominance of single excitations at the equilibrium geometry. In order to test our approximations further, we turn our attention to more strongly correlated systems. Thus, in Figs.~\ref{fig:stretched_water_valence_dz} and~\ref{fig:stretched_BeH2_valence_6_6} we show the valence absorption spectra of H$_2$O and BeH$_2$ for symmetric stretches of, respectively, the O-H and Be-H bonds to 1.0, 1.5 and 2.0 times the equilibrium bond length ($R_{\text{eq}}$) of 0.92 \AA\ and 1.35 \AA. For both systems we used a (6, 6) active space and the cc-pVDZ basis set. The valence excitation energies have been restricted from 8 to 16 eV for H$_2$O and from 5 to 20 eV for BeH$_2$. Further examples can be found in the SI, namely for the K-edge of oxygen and beryllium with different active space combinations as well as results for the 3-RDM and 3-RDC approximations, see Figs.~\ref{SI-fig:stretched_water_core_dz} --~\ref{SI-fig:stretched_beh2_core_6_6} and tables~\ref{SI-tab:H2O_valence} --~\ref{SI-tab:BeH2_core}.

We continue to see the trend that RDC approximations perform on par with the RDM approximations at the equilibrium structure. At the same time, the 4-z and 4-d approximations perform the same within the RDM and RDC approximations while the 3- \& 4- approximations continue to have large errors for the valence excitation energies. With an elongation of the bond, the error increases and becomes significant even for the 4-RDM approximations. As expected, this is due to the strongly correlated systems' reliance on double excitation contributions expressed by the 4-RDMs.

The core excitation energies paint a different picture. For H$_2$O in a (6, 6) active space, the errors are very small for all bond lengths. This can be attributed to the core excitation energies being in the inactive space and thus being described by single excitations that are in turn dominated by lower order RDMs. In the case of H$_2$O this would allow for the core excitation energies to be calculated without the 3- and 4-RDM (as long as they are single excitation dominated), as seen in Fig.~\ref{SI-fig:stretched_water_core_dz}. A similar trend can be observed for BeH$_2$(4, 4) (see Fig.~\ref{SI-fig:stretched_beh2_core_4_4}). However, if the core electrons are placed in the active space, as done for BeH$_2$ (6, 6) in Fig.~\ref{SI-fig:stretched_beh2_core_6_6}, this is no longer the case and the errors of the core excitation energies at large bond distances become comparable to the errors of the valence excitation energies.

\section{Summary}
\label{sec:summary}

We have investigated eight approximations to our previously derived reduced density matrix formulation of naive orbital-optimized quantum linear response~\cite{von2024reduced} in order to reduce both the classical and quantum computational demands. The approximations are differentiated by firstly being applied to either only the 4-RDM or both 3- and 4-RDM and secondly by (I) keeping only the diagonal of an RDM, (II) discarding the entire RDM, (III) reconstructing the entire RDM from exact lower order RDCs, and (IV) keeping the exact diagonal elements of the RDM and reconstructing the off-diagonal elements from exact lower order RDCs. 

We start out by highlighting measurement costs of evaluating the RDMs explicitly in various mapping schemes. As the 4-RDM scales $N_A^8$ it clearly dominates the measurement cost and approximations to it can drastically reduce the computational workload. For example, for a system with an (8, 8) active space the 4-RDM cost dominate the overall costs by over a factor of five. 

Applying our approximations to the excitation energies and absorption spectra of the H$_2$ ladder model system, we concluded that entirely removing the 4-RDM only resulted in slight errors while any approximation to the 3-RDM resulted in huge errors. When including the diagonal of the 4-RDM no changes were observed, however, including the diagonal of the 3-RDM improved the performance compared to completely removing the entire 3-RDM. All errors increased with the size of the active space. Going to  chemically stable molecules confirmed the trends of the model system conclusion. In fact, the 4-RDM approximations had slightly smaller errors than in the H$_2$ ladders and the 3-RDM approximations continued to give huge errors making them unusable. 
We still note that the diagonal 3-RDC approximations perform the best out of all 3-RDM and 3-RDC approximations.

For strongly correlated systems, here studied via stretched H$_2$O and BeH$_2$, we show that the 4-RDM cannot be ignored for high quality results as double excitations become important. Only core-excitation energies with the core orbitals in the inactive space (i.e.\ single excitations) are insensitive to our approximations regardless of bond length. In fact, in these spectra there were little to no errors caused by disregarding the 3-RDM as well. However, when the core orbitals are included in the active space, as in BeH$_2$ (6, 6), any approximations to the RDMs lead to large errors in the presence of strong correlation.

In summary, we showed that qLRSD can produce good results using approximations to the 4-RDM or 4-RDC for equilibrium systems or core-excitations, but struggles when either approximations to the 3-RDM or 3-RDC are introduced or when the system exhibits strong correlation, which limits the potential applicability to quantum computing.

\section*{Acknowledgments}
We acknowledge support from the Novo Nordisk Foundation (NNF) for the focused research project \textit{Hybrid Quantum Chemistry on Hybrid Quantum Computers} (HQC)$^2$ (grant number: NNFSA220080996). 

\section*{Supporting Information}

This information is available free of charge via the Internet at http://pubs.acs.org

\section*{Data Availability}

The data that support the findings of this study are available from the corresponding authors upon reasonable request.

\newpage
\bibliography{literature}

\end{document}


\date{August 10, 2025}

\maketitle

\clearpage

\section{Tables and figures}
\label{sec:TabFig}

\begin{figure}[p]
    \centering
    \includegraphics[width=.47\linewidth]{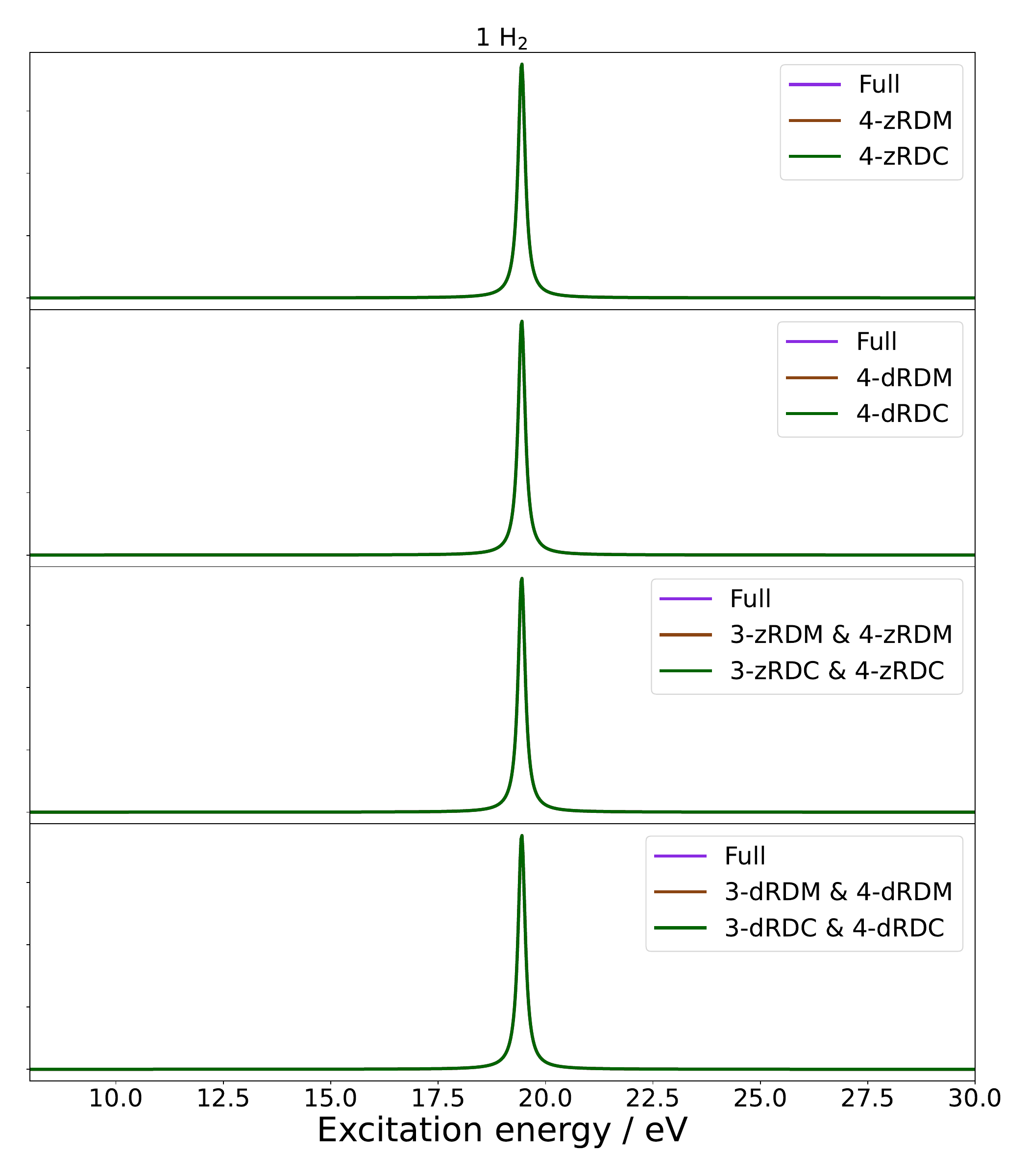}
    \includegraphics[width=.47\linewidth]{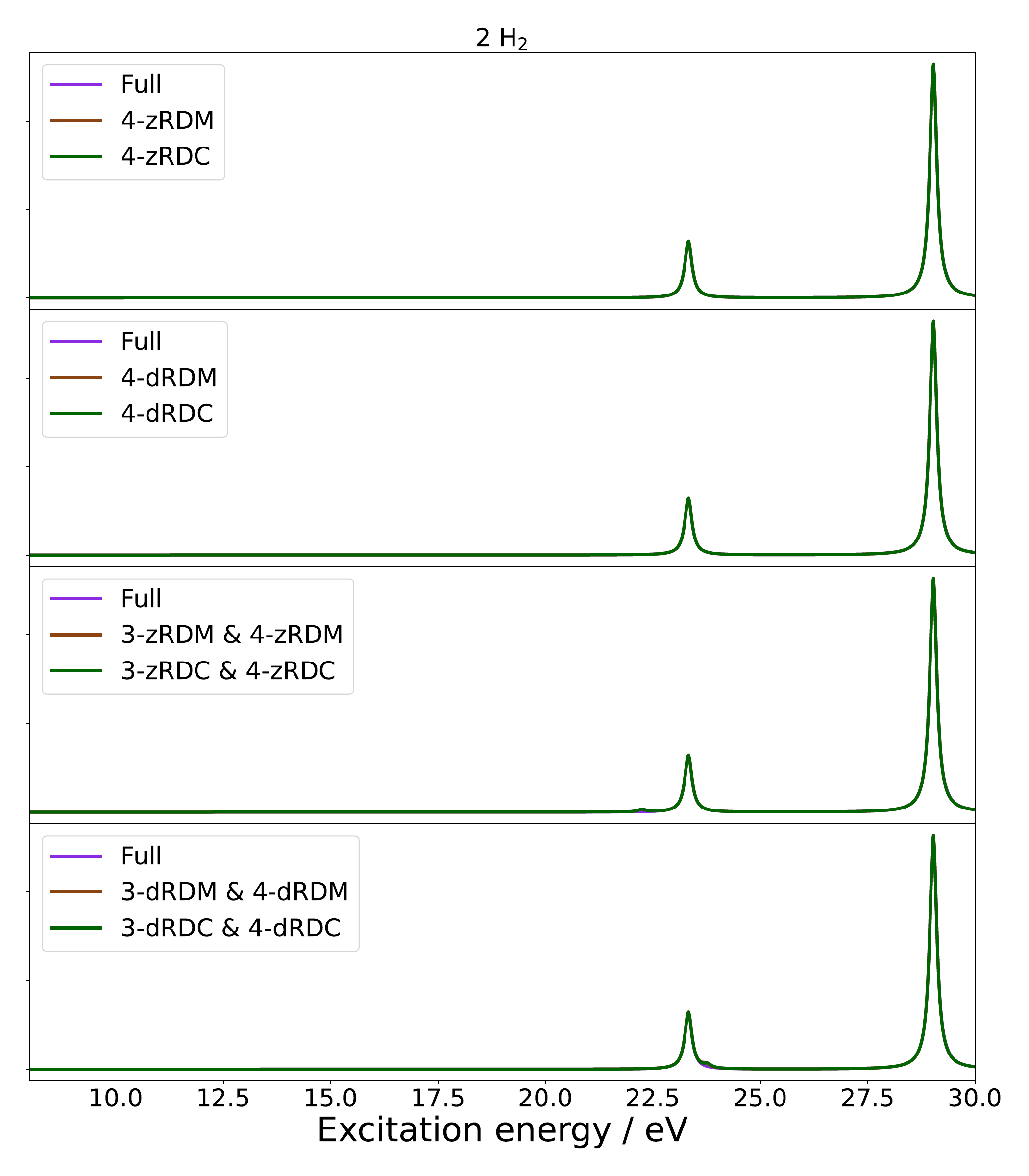}
    \includegraphics[width=.47\linewidth]{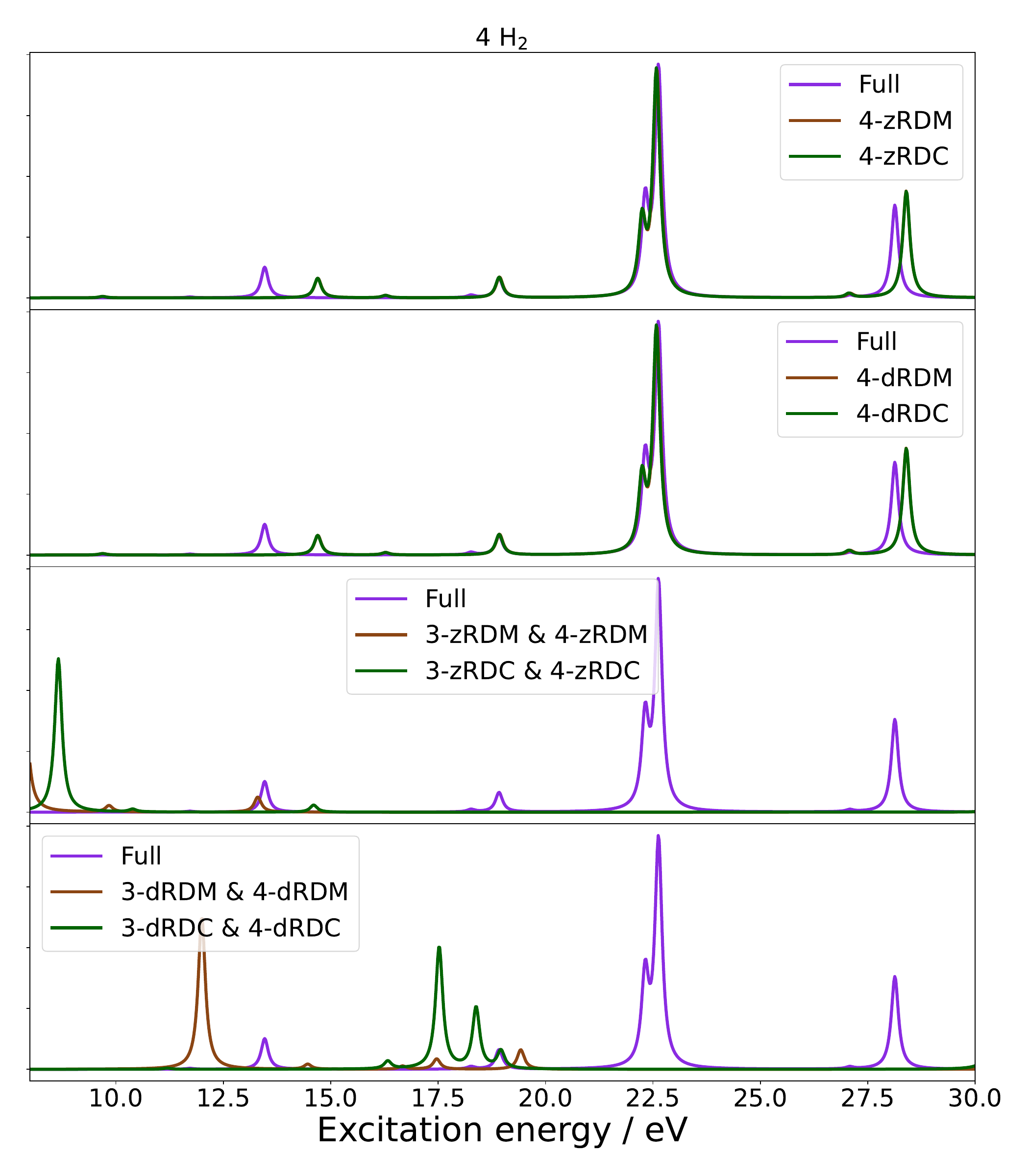}
    \includegraphics[width=.47\linewidth]{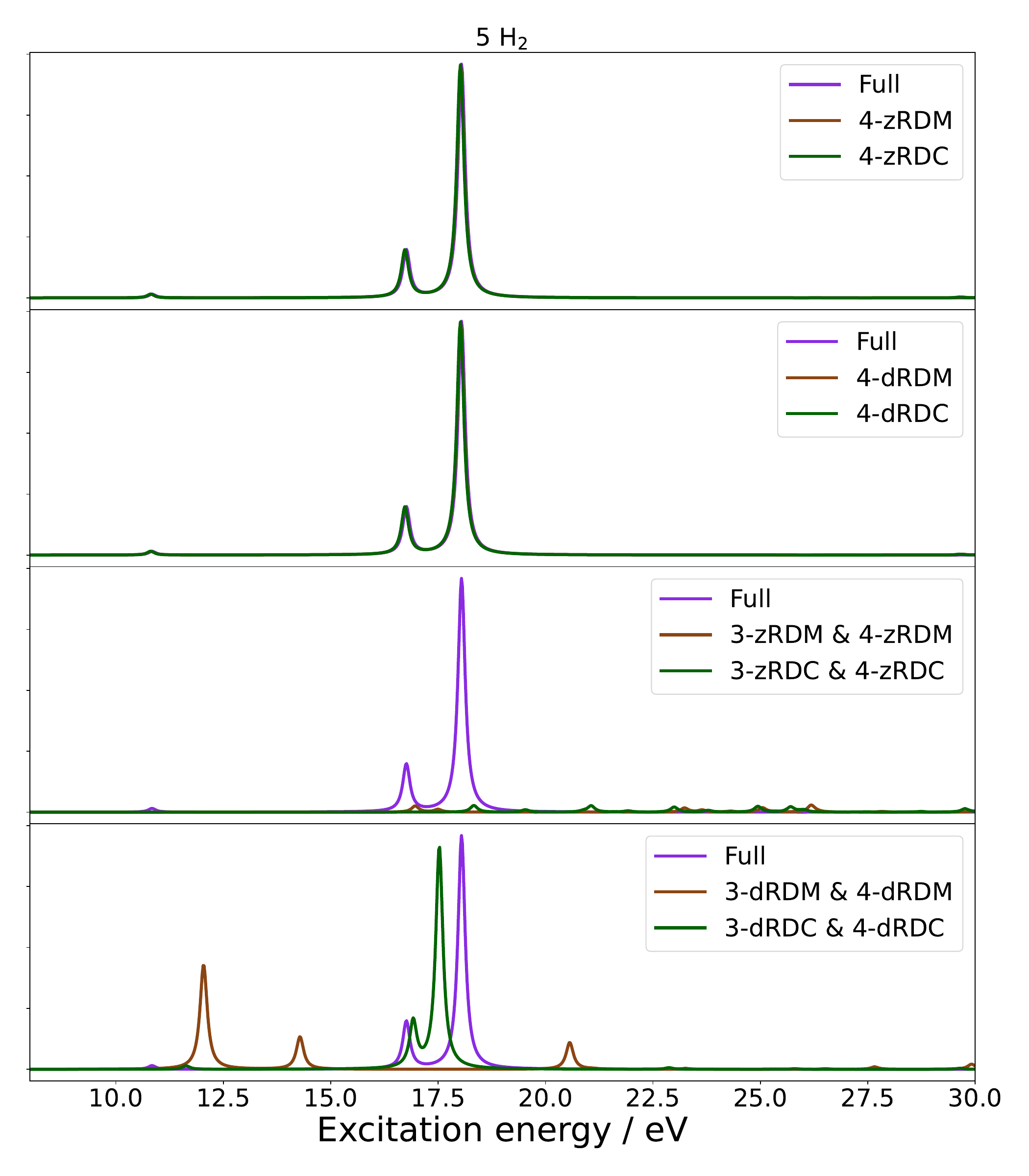}
    \caption{Absorption spectra of H$_2$ ladders containing one link (top left), two links (top right), four links (bottom left), and five links (bottom right) using the naive qLRSD method on top of a FCI wave function. The four panels in each subfigure contain: (first panel) the absorption spectrum of naive qLRSD using the 4-zRDM and 4-zRDC approximations, (second panel) the absorption spectrum of naive qLRSD using the 4-dRDM and 4-dRDC approximations, (third panel) absorption spectrum of naive qLRSD using the 3-zRDM \& 4-zRDM and 3-zRDC \& 4-zRDC approximations, and (fourth panel) absorption spectrum of naive qLRSD using the 3-dRDM \& 4-dRDM and 3-dRDC \& 4-dRDC approximations.}
    \label{fig:H2}
\end{figure}

\begin{table}[p]
    \centering
    \begin{tabular}{lc|cccc|cccc}
        \hline\hline & & \multicolumn{4}{c|}{RDM approximations} & \multicolumn{4}{c}{RDC approximations} \\
        System & Errors & 4-z & 3- \& 4-z & 4-d & 3- \& 4-d & 4-z & 3- \& 4-z & 4-d & 3- \& 4-d  \\
        \hline
        \multirow{5}{2em}{1 H$_2$} & MAE$_\text{exc}$ & 0.0 & 0.0 & 0.0 & 0.0 & 0.0 & 0.0 & 0.0 & 0.0 \\
         & $\sigma_\text{exc}$ & 0.0 & 0.0 & 0.0 & 0.0 & 0.0 & 0.0 & 0.0 & 0.0 \\
         & MAE$_\text{osc}$ & 0.0 & 0.0 & 0.0 & 0.0 & 0.0 & 0.0 & 0.0 & 0.0 \\
         & $\sigma_\text{osc}$ & 0.0 & 0.0 & 0.0 & 0.0 & 0.0 & 0.0 & 0.0 & 0.0 \\
         & \# zero & 0 & 0 & 0 & 0 & 0 & 0 & 0 & 0 \\
        \hline
        \multirow{5}{2em}{2 H$_2$} & MAE$_\text{exc}$ & 0.2999 & 4.2836 & 0.2999 & 2.2334 & 0.2999 & 4.2835 & 0.2999 & 2.2291 \\
         & $\sigma_\text{exc}$ & 0.3265 & 4.2002 & 0.3265 & 2.2884 & 0.3265 & 4.2001 & 0.3265 & 2.2674 \\
         & MAE$_\text{osc}$ & 0.0001 & 0.1901 & 0.0001 & 0.1893 & 0.0001 & 0.1901 & 0.0001 & 0.1893 \\
         & $\sigma_\text{osc}$ & 0.0005 & 0.4058 & 0.0005 & 0.4557 & 0.0005 & 0.4058 & 0.0005 & 0.4557 \\
         & \# zero & 0 & 0 & 0 & 0 & 0 & 0 & 0 & 0 \\
        \hline
        \multirow{5}{2em}{3 H$_2$} & MAE$_\text{exc}$ & 0.4273 & 7.4679 & 0.4273 & 1.6097 & 0.4414 & 7.0933 & 0.4296 & 1.7333 \\
         & $\sigma_\text{exc}$ & 0.3198 & 3.9506 & 0.3198 & 1.5681 & 0.3148 & 4.0119 & 0.3187 & 1.5794 \\
         & MAE$_\text{osc}$ & 0.0002 & 0.0960 & 0.0002 & 0.0855 & 0.0002 & 0.0941 & 0.0002 & 0.0849 \\
         & $\sigma_\text{osc}$ & 0.0005 & 0.3606 & 0.0005 & 0.4005 & 0.0005 & 0.3698 & 0.0005 & 0.4122 \\
         & \# zero & 0 & 0 & 0 & 0 & 0 & 0 & 0 & 0 \\
        \hline
        \multirow{5}{2em}{4 H$_2$} & MAE$_\text{exc}$ & 0.8669 & 9.6414 & 0.8669 & 4.9432 & 0.8443 & 8.2002 & 0.8443 & 2.0725 \\
         & $\sigma_\text{exc}$ & 0.7796 & 2.5250 & 0.7796 & 1.9843 & 0.7905 & 2.1759 & 0.7905 & 1.1005 \\
         & MAE$_\text{osc}$ & 0.0088 & 0.1303 & 0.0088 & 0.1307 & 0.0087 & 0.1217 & 0.0087 & 0.1741 \\
         & $\sigma_\text{osc}$ & 0.0338 & 0.3322 & 0.0338 & 0.2883 & 0.0339 & 0.3033 & 0.0339 & 0.5127 \\
         & \# zero & 0 & 5 & 0 & 5 & 0 & 5 & 0 & 7 \\
        \hline
        \multirow{5}{2em}{5 H$_2$} & MAE$_\text{exc}$ & 0.6302 & 12.4120 & 0.6302 & 4.8132 & 0.6972 & 10.4634 & 0.6972 & 2.7017 \\
         & $\sigma_\text{exc}$ & 0.2620 & 0.7847 & 0.2620 & 0.7685 & 0.2865 & 0.7042 & 0.2865 & 0.9576 \\
         & MAE$_\text{osc}$ & 0.0017 & 0.0556 & 0.0017 & 0.0884 & 0.0020 & 0.0721 & 0.0020 & 0.0959 \\
         & $\sigma_\text{osc}$ & 0.0047 & 0.3885 & 0.0047 & 0.4265 & 0.0049 & 0.3959 & 0.0049 & 0.5321 \\
         & \# zero & 0 & 0 & 0 & 0 & 0 & 0 & 0 & 0 \\
        \hline
        \multirow{5}{2em}{6 H$_2$} & MAE$_\text{exc}$ & 0.8699 & 15.593 & 0.8699 & 8.0726 & 0.8781 & 14.0385 & 0.8781 & 3.1770 \\
         & $\sigma_\text{exc}$ & 0.3287 & 2.0410 & 0.3287 & 1.0246 & 0.3317 & 1.5616 & 0.3317 & 0.7908 \\
         & MAE$_\text{osc}$ & 0.1177 & 0.0709 & 0.1177 & 0.0971 & 0.1177 & 0.0762 & 0.1177 & 0.0883 \\
         & $\sigma_\text{osc}$ & 0.6318 & 0.4516 & 0.6318 & 0.4524 & 0.6318 & 0.4359 & 0.6318 & 0.4612 \\
         & \# zero & 0 & 3 & 0 & 0 & 0 & 0 & 0 & 0 \\
        \hline\hline
    \end{tabular}
    \caption{Mean absolute errors (MAE) and standard deviations ($\sigma$) for the eight RDM and RDC approximations for a maximum of 100 non-zero excitation energies and their corresponding oscillator strengths. Errors and standard deviations for excitation energies are given in eV. The number of zero excitation energies for the molecules and approximations are also given.}
    \label{tab:H2}
\end{table}

\begin{figure}[p]
    \centering
    \includegraphics[width=.47\linewidth]{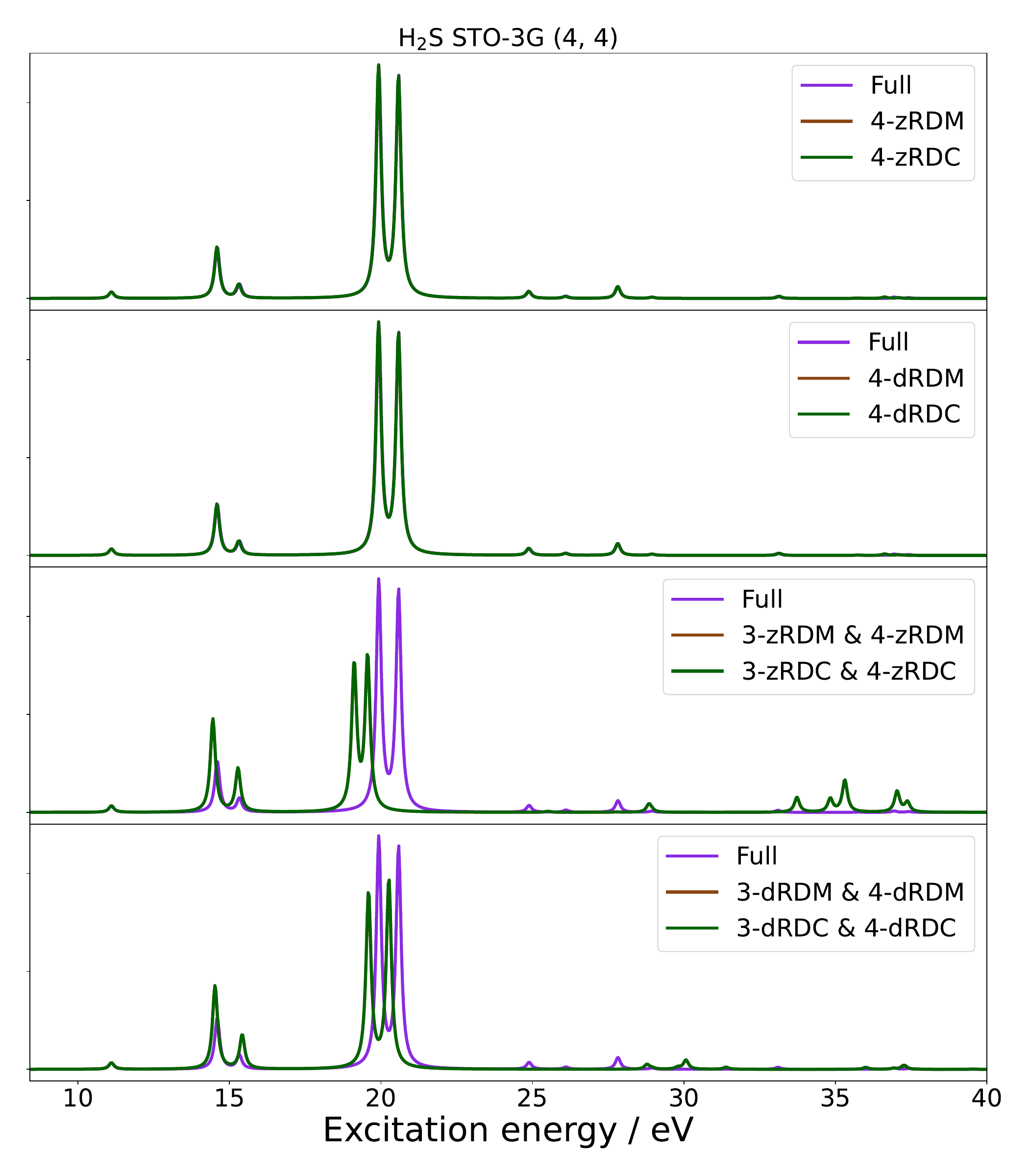}
    \includegraphics[width=.47\linewidth]{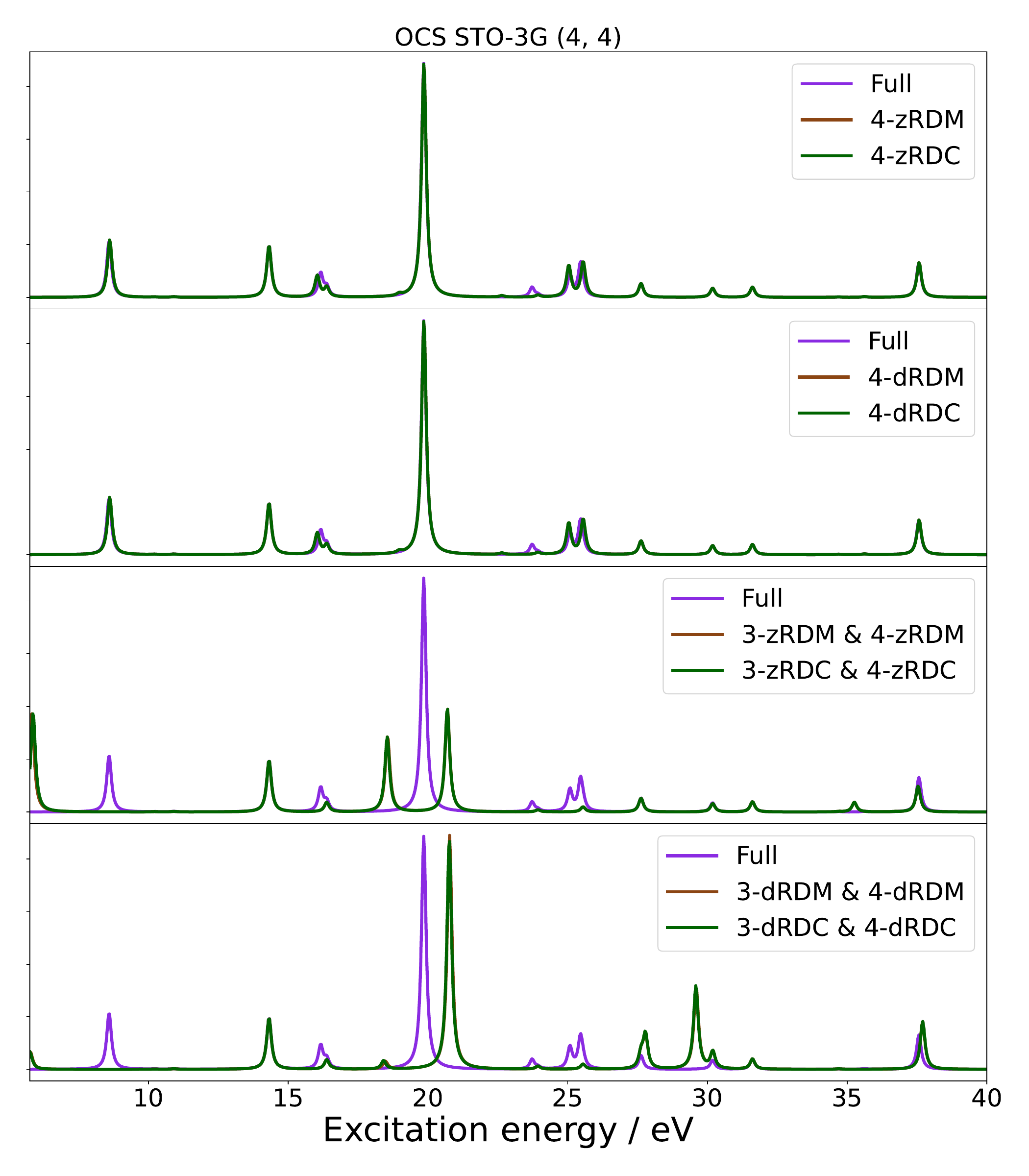}
    \includegraphics[width=.47\linewidth]{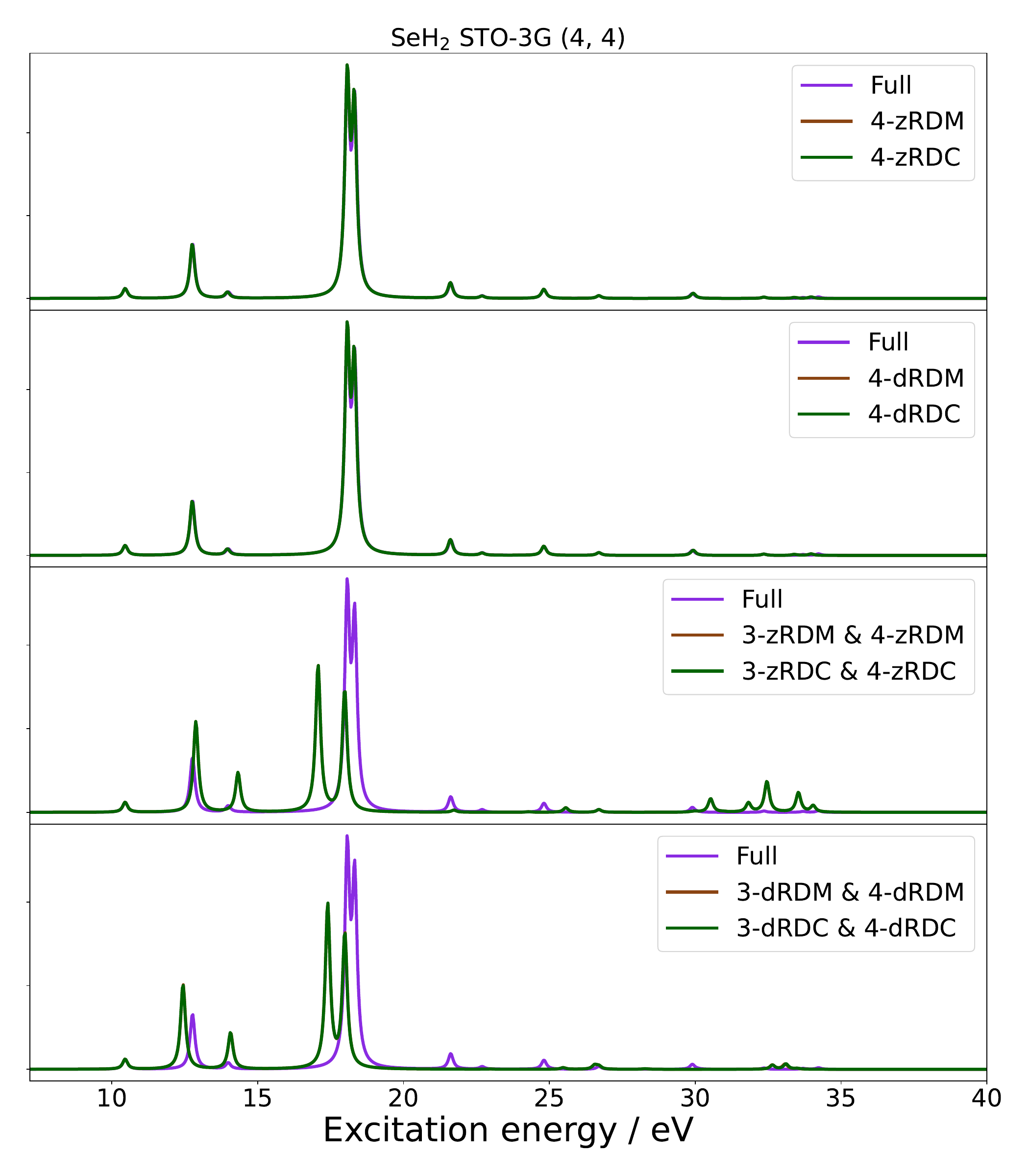}
    \caption{Absorption spectra of H$_2$S (4, 4) (top left), OCS (4, 4) (top right), and SeH$_2$ (4, 4) (bottom) using the naive qLRSD method on top of a FCI wave function. The four panels contain: (first panel) the absorption spectrum of naive qLRSD using the 4-zRDM and 4-zRDC approximations; (second panel) the absorption spectrum of naive qLRSD using the 4-dRDM and 4-dRDC approximations; (third panel) absorption spectrum of naive qLRSD using the 3-zRDM \& 4-zRDM and 3-zRDC \& 4-zRDC approximations; (fourth panel) absorption spectrum of naive qLRSD using the 3-dRDM \& 4-dRDM and 3-dRDC \& 4-dRDC approximations.}
    \label{fig:other_molecules_4-4}
\end{figure}

\begin{figure}[p]
    \centering
    \includegraphics[width=0.47\linewidth]{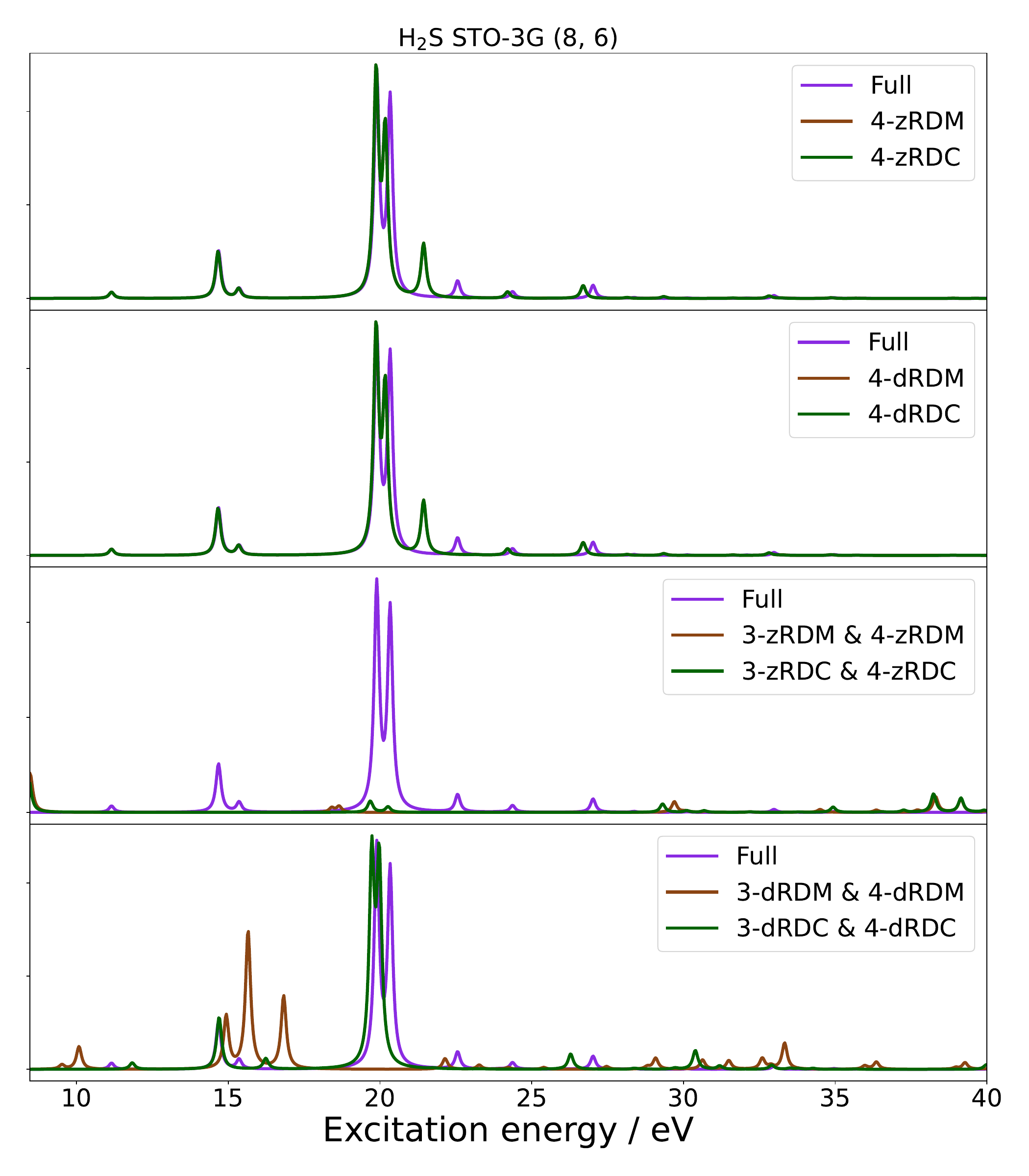}
    \includegraphics[width=0.47\linewidth]{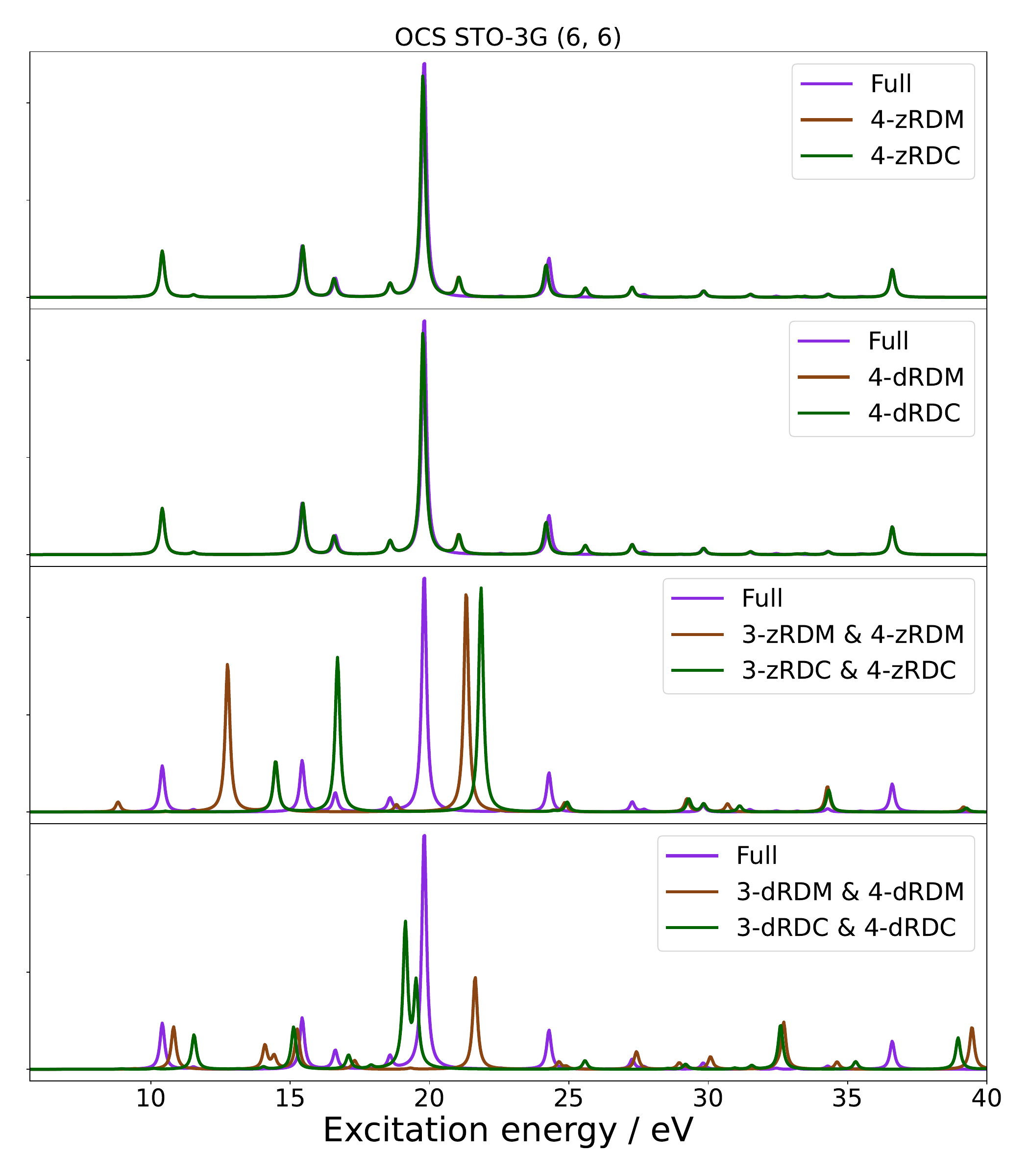}
    \includegraphics[width=0.47\linewidth]{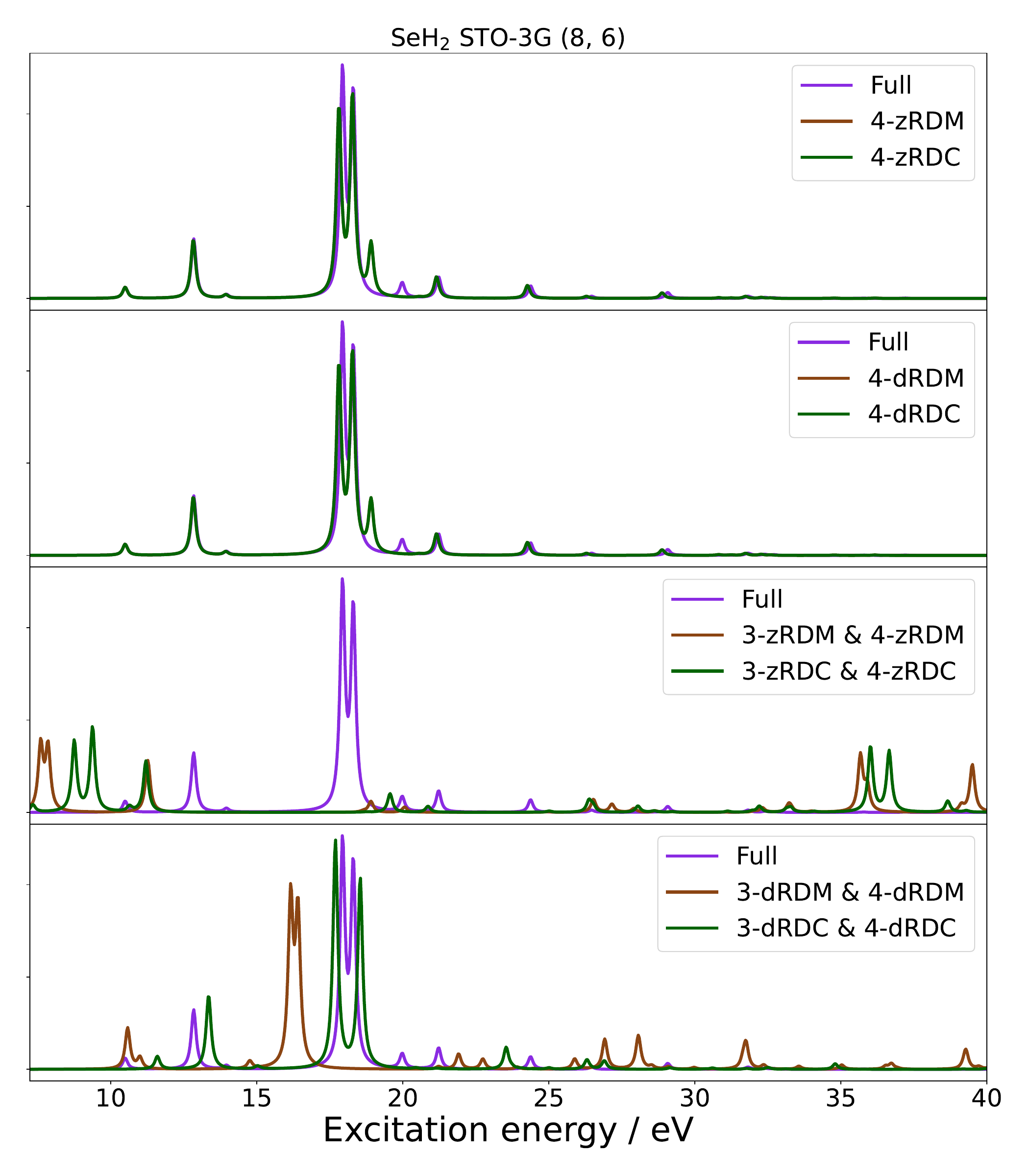}
    \caption{Absorption spectrum of H$_2$S (8, 6) (top left), OCS (6, 6) (top right), and SeH$_2$ (bottom). The four panels contain: (first panel) the absorption spectrum of naive qLRSD using the 4-zRDM and 4-zRDC approximations; (second panel) the absorption spectrum of naive qLRSD using the 4-dRDM and 4-dRDC approximations; (third panel) absorption spectrum of naive qLRSD using the 3-zRDM \& 4-zRDM and 3-zRDC \& 4-zRDC approximations; (fourth panel) absorption spectrum of naive qLRSD using the 3-dRDM \& 4-dRDM and 3-dRDC \& 4-dRDC approximations.}
    \label{fig:molecules_large_active_space}
\end{figure}

\clearpage

\begin{longtable}{lc|cccc|cccc}
    \caption{Mean absolute errors (MAE) and standard deviations ($\sigma$) for the eight RDM and RDC approximations for a maximum of 100 non-zero excitation energies and their corresponding oscillator strengths. Errors and standard deviations for excitation energies are given in eV. The number of zero excitation energies for the molecules and approximations are also given.} \\
    \hhline{==========} & & \multicolumn{4}{c|}{RDM approximations} & \multicolumn{4}{c}{RDC approximations} \\
    System & Errors & 4-z & 3- \& 4-z & 4-d & 3- \& 4-d & 4-z & 3- \& 4-z & 4-d & 3- \& 4-d  \\\hline
    \multirow{5}{3em}{\centering H$_2$S \\ (4, 4)} & MAE$_\text{exc}$ & 0.0748 & 1.0614 & 0.0748 & 0.6222 & 0.0748 & 1.0614 & 0.0748 & 0.6227 \\
     & $\sigma_\text{exc}$ & 0.1958 & 2.1579 & 0.1958 & 1.1400 & 0.1958 & 2.1579 & 0.1958 & 1.1411 \\
     & MAE$_\text{osc}$ & 0.0008 & 0.0416 & 0.0008 & 0.0255 & 0.0008 & 0.0416 & 0.0008 & 0.0255 \\
     & $\sigma_\text{osc}$ & 0.0023 & 0.0911 & 0.0023 & 0.0544 & 0.0023 & 0.0911 & 0.0023 & 0.0544 \\
     & \# zero & 0 & 0 & 0 & 0 & 0 & 0 & 0 & 0 \\\hline
    \multirow{5}{3em}{\centering H$_2$S \\ (8, 6)} & MAE$_\text{exc}$ & 0.3286 & 1.4537 & 0.3286 & 1.1384 & 0.3295 & 1.3105 & 0.3295 & 1.1497 \\
     & $\sigma_\text{exc}$ & 0.4135 & 2.8625 & 0.4135 & 1.4253 & 0.4121 & 2.4769 & 0.4121 & 1.5255 \\
     & MAE$_\text{osc}$ & 0.0079 & 0.0677 & 0.0079 & 0.0377 & 0.0079 & 0.0619 & 0.0079 & 0.0087 \\
     & $\sigma_\text{osc}$ & 0.0331 & 0.1868 & 0.0331 & 0.0976 & 0.0331 & 0.1817 & 0.0331 & 0.0214 \\
     & \# zero & 0 & 0 & 0 & 0 & 0 & 0 & 0 & 0 \\\hline\pagebreak\hline
    \multirow{5}{3em}{\centering OCS \\ (4, 4)} & MAE$_\text{exc}$ & 0.0835 & 26.5346 & 0.0835 & 25.3310 & 0.0835 & 26.5340 & 0.0835 & 25.3311 \\
     & $\sigma_\text{exc}$ & 0.2598 & 189.5359 & 0.2598 & 189.6802 & 0.2598 & 189.5360 & 0.2598 & 189.6802 \\
     & MAE$_\text{osc}$ & 0.0540 & 0.0808 & 0.0540 & 0.0936 & 0.0540 & 0.0808 & 0.0540 & 0.0935 \\
     & $\sigma_\text{osc}$ & 0.3121 & 0.2680 & 0.3121 & 0.3256 & 0.3121 & 0.2681 & 0.3121 & 0.3240 \\
     & \# zero & 0 & 1 & 0 & 1 & 0 & 1 & 0 & 1 \\\hline
    \multirow{5}{3em}{\centering OCS \\ (6, 6)} & MAE$_\text{exc}$ & 0.5964 & 10.1088 & 0.5964 & 5.9769 & 0.5934 & 13.3942 & 0.5934 & 3.8050 \\
     & $\sigma_\text{exc}$ & 0.6293 & 6.3799 & 0.6293 & 12.3586 & 0.6276 & 15.6987 & 0.6276 & 8.4392 \\
     & MAE$_\text{osc}$ & 0.0232 & 0.1065 & 0.0232 & 0.0951 & 0.0231 & 0.1088 & 0.0231 & 0.0934 \\
     & $\sigma_\text{osc}$ & 0.0717 & 0.3532 & 0.0717 & 0.2404 & 0.0716 & 0.3727 & 0.0716 & 0.3005 \\
     & \# zero & 0 & 0 & 0 & 2 & 0 & 3 & 0 & 1 \\\hline
    \multirow{5}{3em}{\centering SeH$_2$ \\ (4, 4)} & MAE$_\text{exc}$ & 0.0368 & 0.4996 & 0.0368 & 0.3318 & 0.0368 & 0.4995 & 0.0368 & 0.3318 \\
     & $\sigma_\text{exc}$ & 0.1378 & 1.4655 & 0.1378 & 0.7644 & 0.1378 & 1.4655 & 0.1378 & 0.7646 \\
     & MAE$_\text{osc}$ & 0.0010 & 0.0235 & 0.0010 & 0.0193 & 0.0010 & 0.0235 & 0.0010 & 0.0193 \\
     & $\sigma_\text{osc}$ & 0.0046 & 0.0702 & 0.0046 & 0.0543 & 0.0046 & 0.0702 & 0.0046 & 0.0543 \\
     & \# zero & 0 & 0 & 0 & 0 & 0 & 0 & 0 & 0 \\\hline
    \multirow{5}{3em}{\centering SeH$_2$ \\ (8, 6)} & MAE$_\text{exc}$ & 0.2327 & 1.1153 & 0.2327 & 0.6882 & 0.2341 & 1.0362 & 0.2341 & 0.8265 \\
     & $\sigma_\text{exc}$ & 0.3609 & 1.5935 & 0.3609 & 1.0650 & 0.3612 & 1.4677 & 0.3612 & 1.2087 \\
     & MAE$_\text{osc}$ & 0.0068 & 0.0467 & 0.0068 & 0.0231 & 0.0068 & 0.0452 & 0.0068 & 0.0085 \\
     & $\sigma_\text{osc}$ & 0.0313 & 0.1547 & 0.0313 & 0.0567 & 0.0313 & 0.1554 & 0.0313 & 0.0212 \\
     & \# zero & 0 & 0 & 0 & 0 & 0 & 0 & 0 & 0 \\
    \hhline{==========}
    \label{tab:other_molecules}
\end{longtable}

\clearpage

\begin{figure}[p]
    \centering
    \includegraphics[width=.47\linewidth]{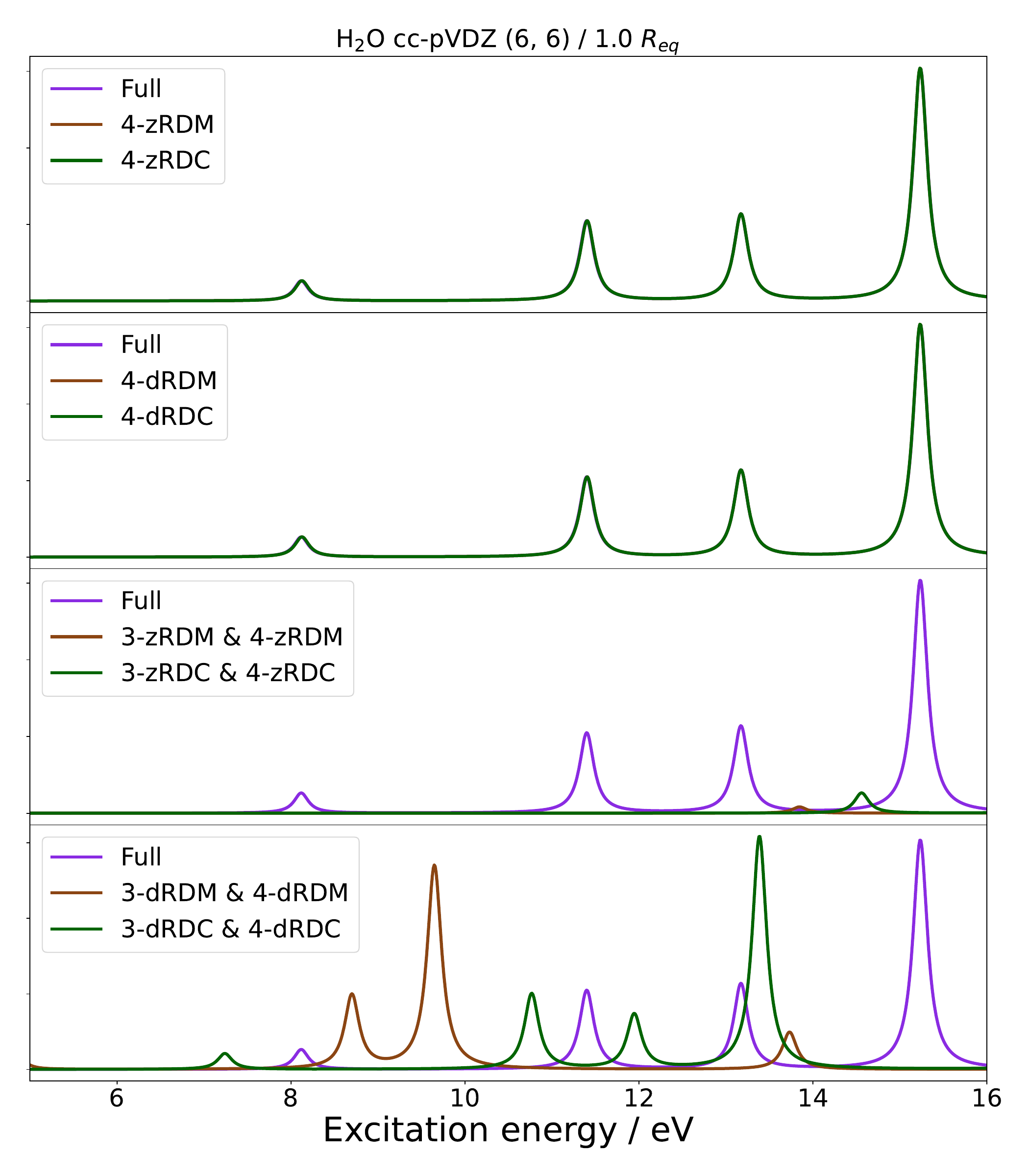}
    \includegraphics[width=.47\linewidth]{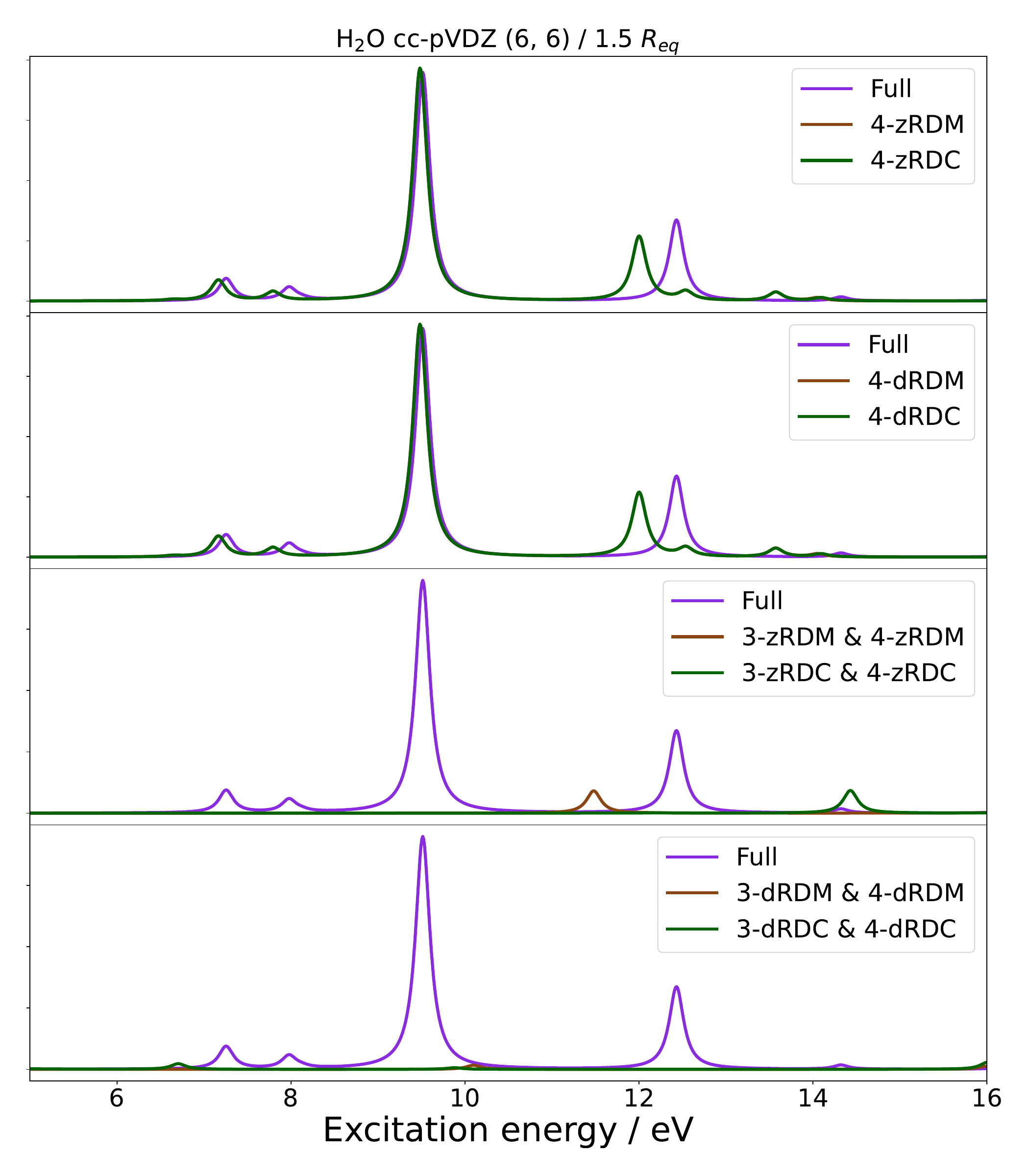}
    \includegraphics[width=.47\linewidth]{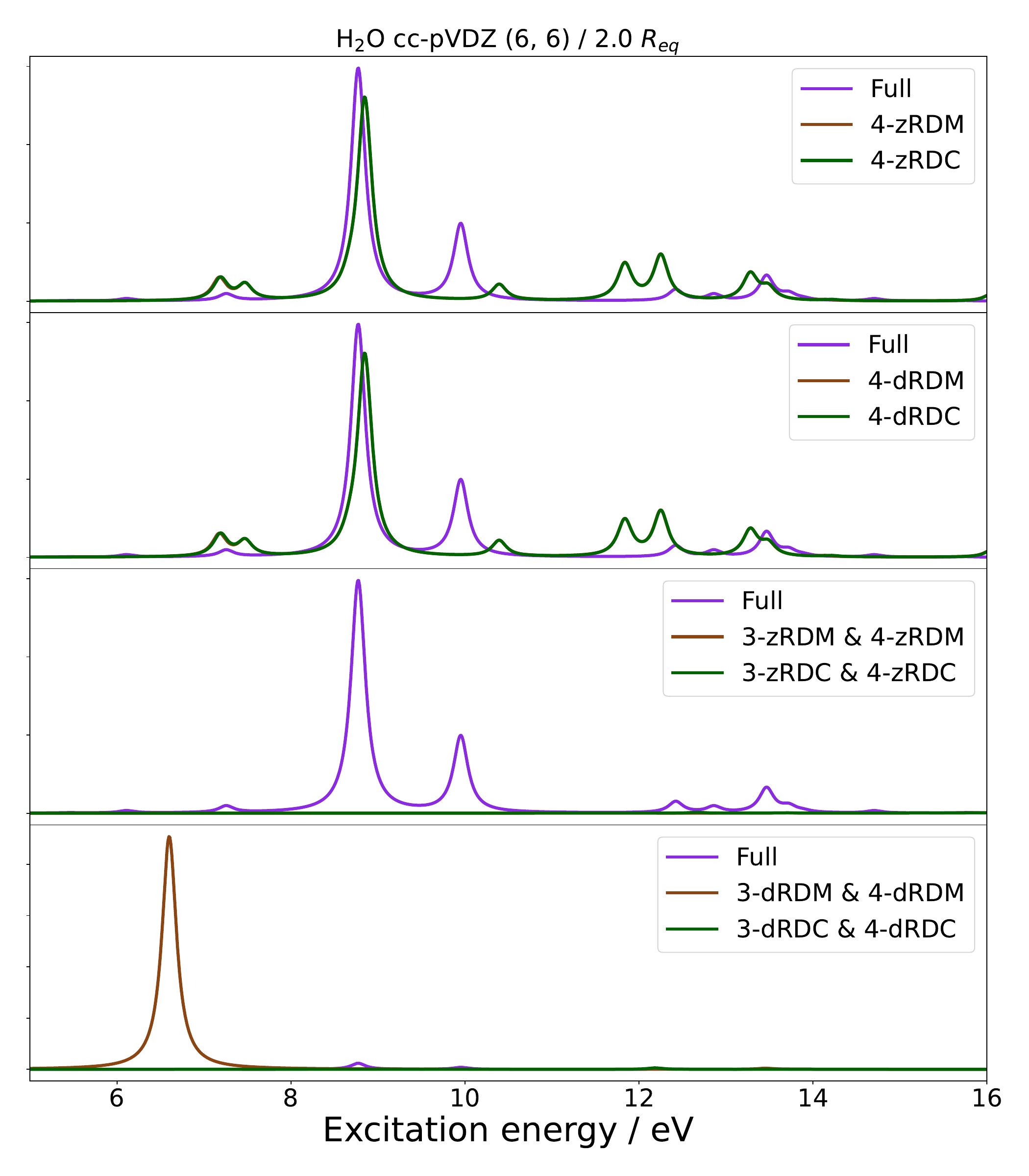}
    \caption{Absorption spectra in the valence excitation region of H$_2$O in a (6, 6) active space with the cc-pVDZ basis set at differing symmetric O-H stretches. Each figure contains four panels comparing the naive qLRSD absorption spectrum with no approximation to the absorption spectrum of naive qLRSD using the 4-zRDM and 4-zRDC approximations (first panel), the absorption spectrum of naive qLRSD using the 4-dRDM and 4-dRDC approximations (second panel), absorption spectrum of naive qLRSD using the 3-zRDM \& 4-zRDM and 3-zRDC \& 4-zRDC approximations, (third panel) and absorption spectrum of naive qLRSD using the 3-dRDM \& 4-dRDM and 3-dRDC \& 4-dRDC approximations (fourth panel).}
    \label{fig:stretched_water_valence_dz}
\end{figure}

\begin{figure}[p]
    \centering
    \includegraphics[width=.47\linewidth]{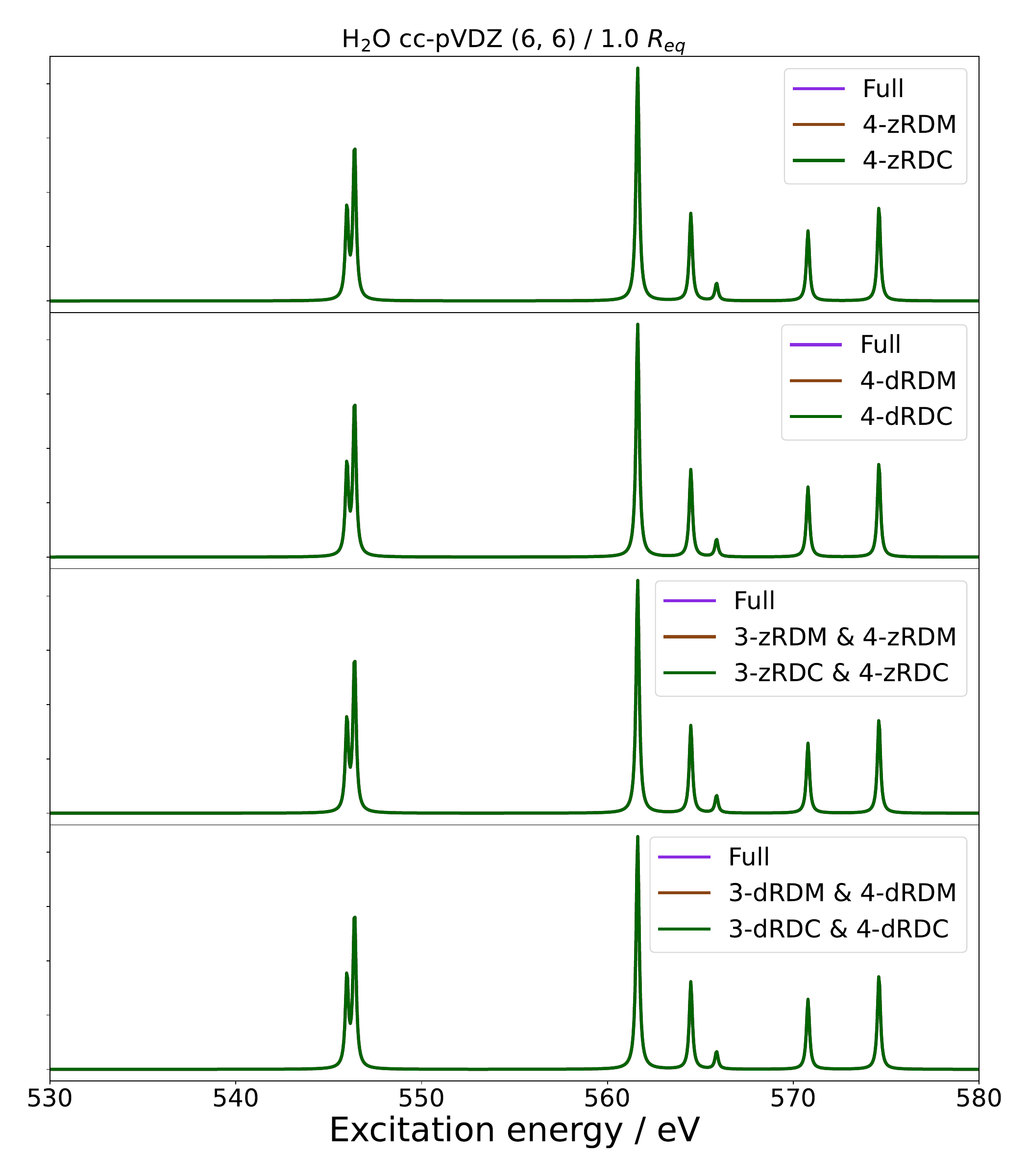}
    \includegraphics[width=.47\linewidth]{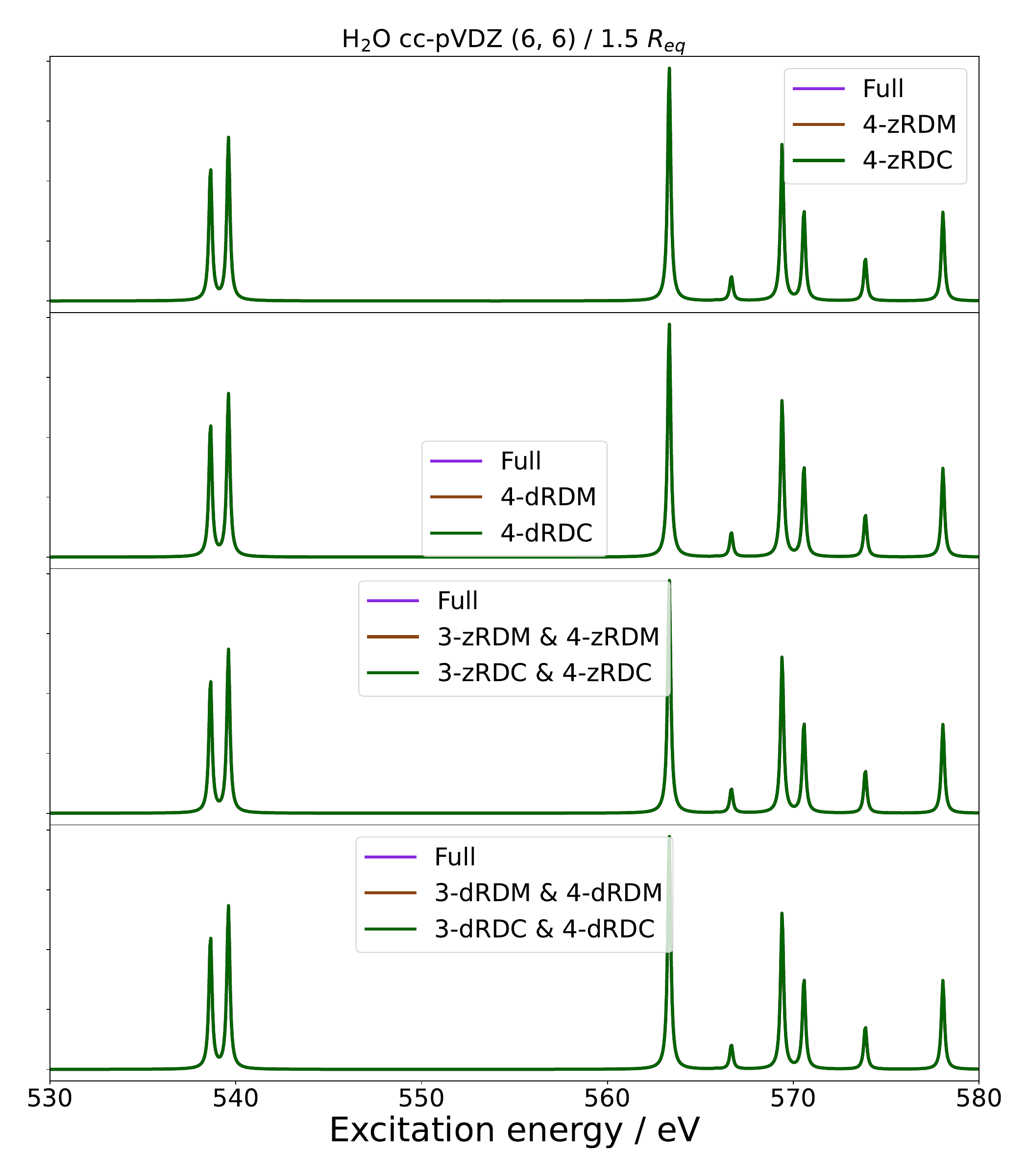}
    \includegraphics[width=.47\linewidth]{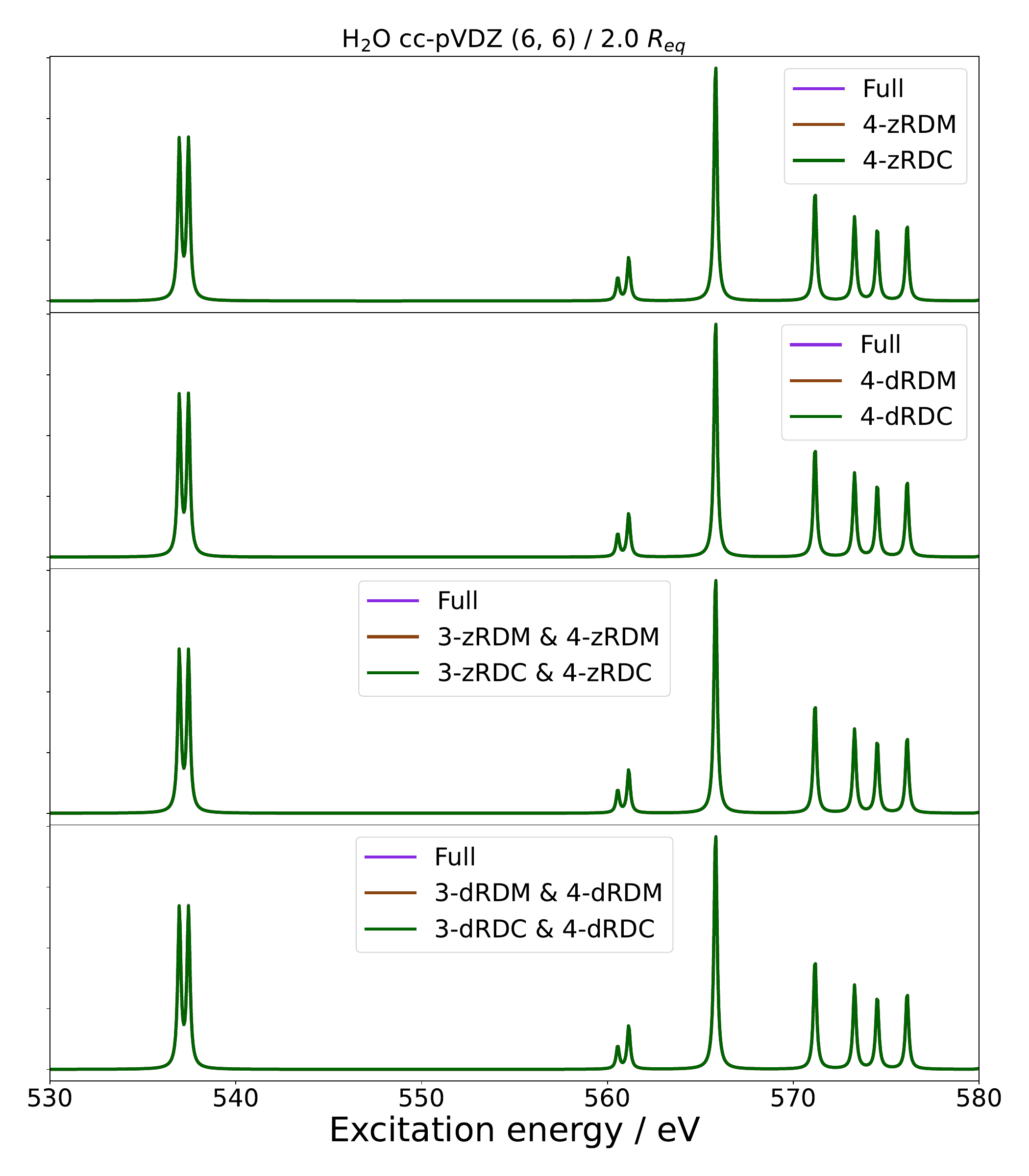}
    \caption{Oxygen K-edge absorption spectra of H$_2$O in a (6, 6) active space with the cc-pVDZ basis set at differing symmetric O-H stretches. 
    Each figure contains four panels comparing the naive qLRSD absorption spectrum with no approximation to the absorption spectrum of naive qLRSD using the 4-zRDM and 4-zRDC approximations (first panel), the absorption spectrum of naive qLRSD using the 4-dRDM and 4-dRDC approximations (second panel), absorption spectrum of naive qLRSD using the 3-zRDM \& 4-zRDM and 3-zRDC \& 4-zRDC approximations, (third panel) and absorption spectrum of naive qLRSD using the 3-dRDM \& 4-dRDM and 3-dRDC \& 4-dRDC approximations (fourth panel).}   \label{fig:stretched_water_core_dz}
\end{figure}

\begin{figure}[p]
    \centering
    \includegraphics[width=.47\linewidth]{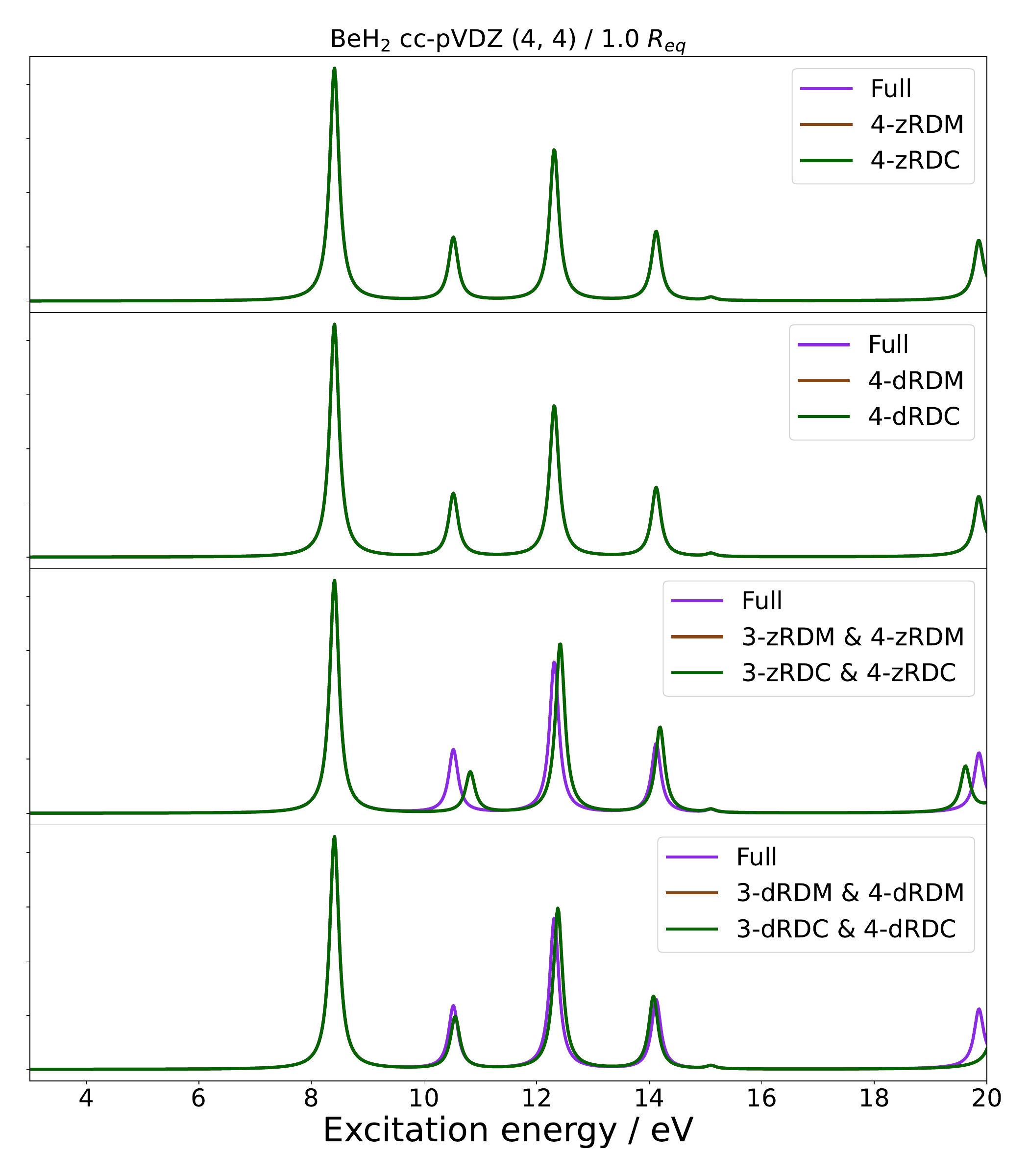}
    \includegraphics[width=.47\linewidth]{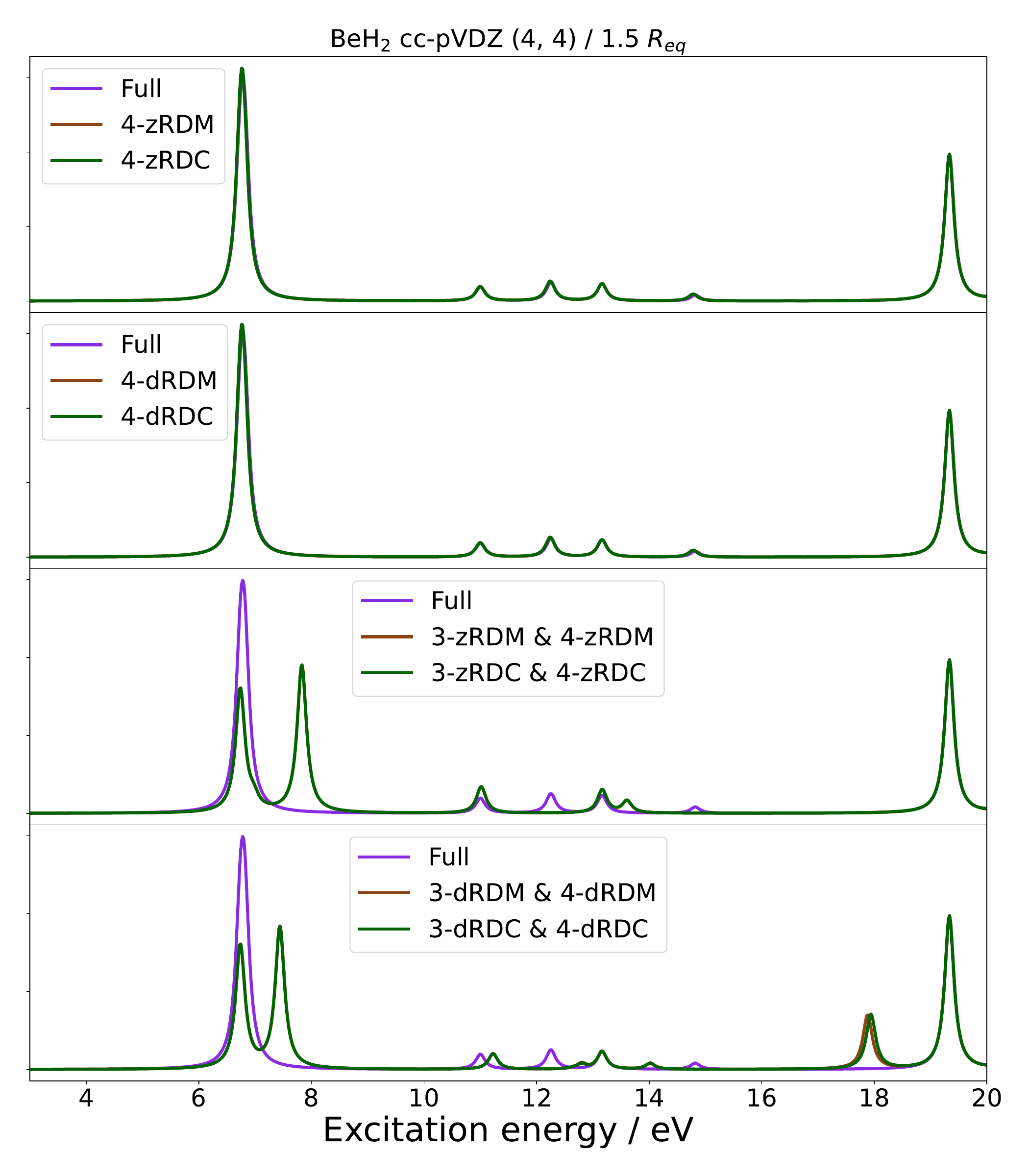}
    \includegraphics[width=.47\linewidth]{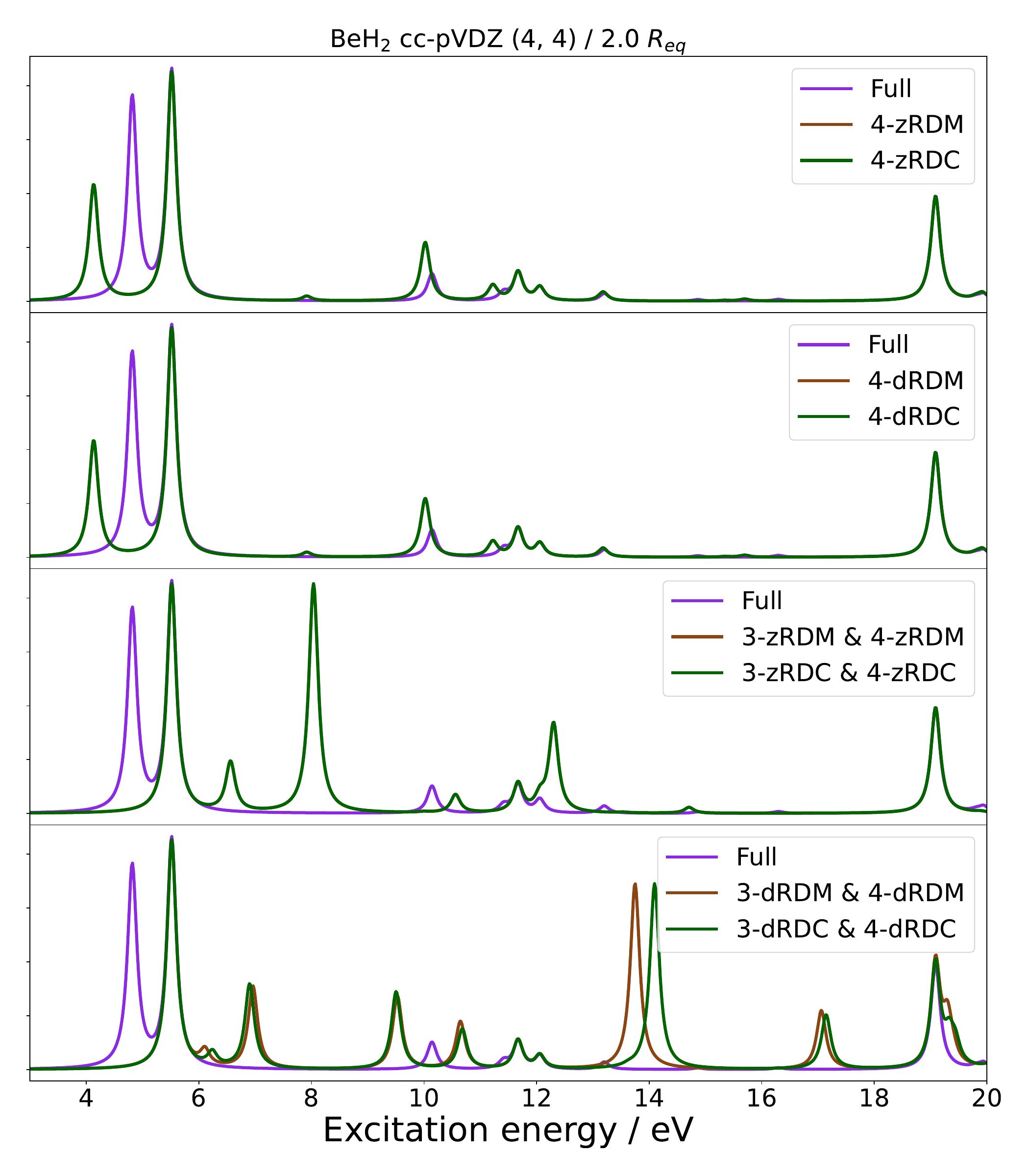}
    \caption{Absorption spectra in the valence excitation region of BeH$_2$ in a (4, 4) active space with the cc-pVDZ basis set at differing symmetric Be-H stretches. Each figure contains four panels comparing the naive qLRSD absorption spectrum with no approximation to the absorption spectrum of naive qLRSD using the 4-zRDM and 4-zRDC approximations (first panel), the absorption spectrum of naive qLRSD using the 4-dRDM and 4-dRDC approximations (second panel), absorption spectrum of naive qLRSD using the 3-zRDM \& 4-zRDM and 3-zRDC \& 4-zRDC approximations, (third panel) and absorption spectrum of naive qLRSD using the 3-dRDM \& 4-dRDM and 3-dRDC \& 4-dRDC approximations (fourth panel).}\label{fig:stretched_beh2_valence_4_4}
\end{figure}

\begin{figure}[p]
    \centering
    \includegraphics[width=.47\linewidth]{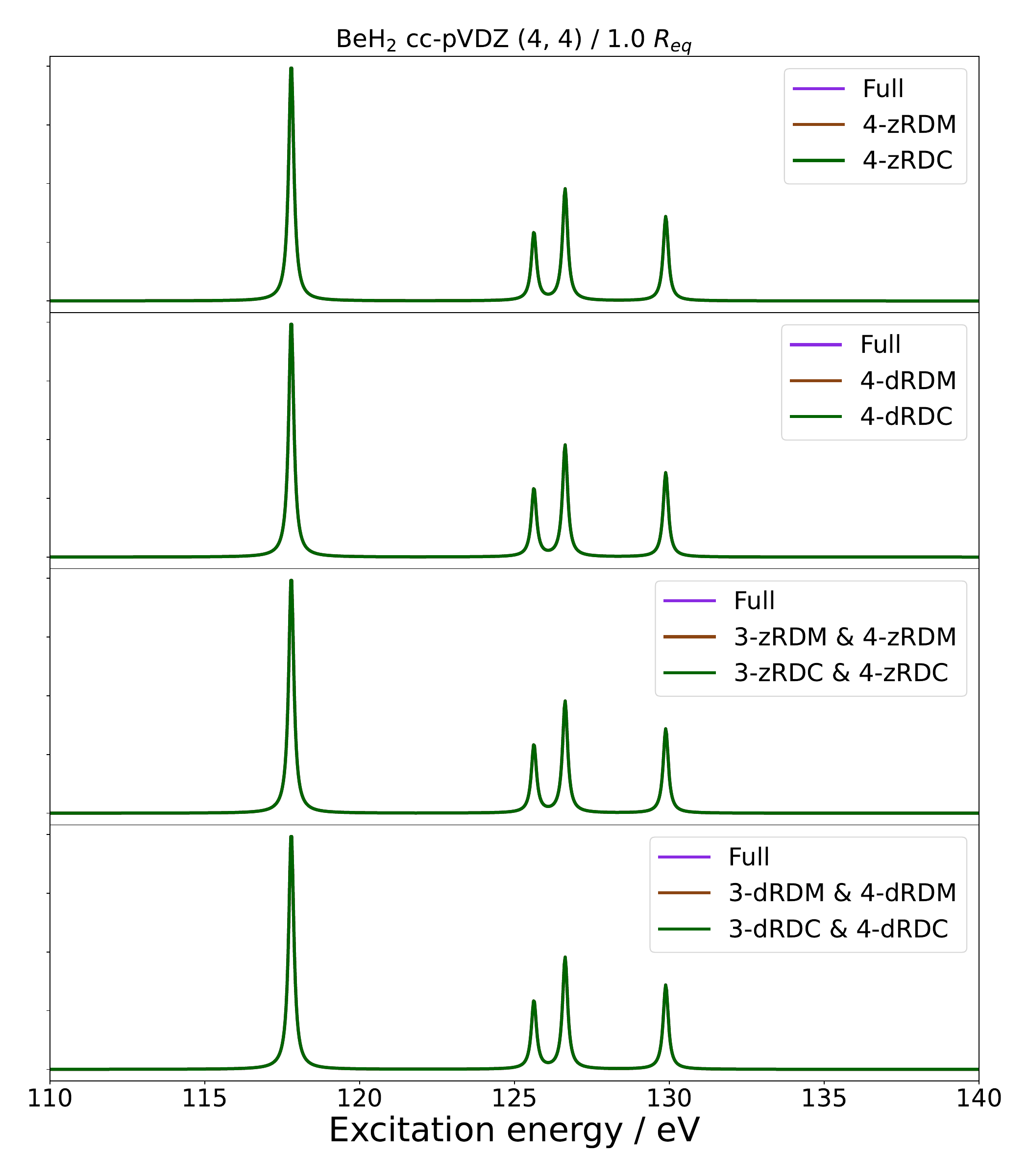}
    \includegraphics[width=.47\linewidth]{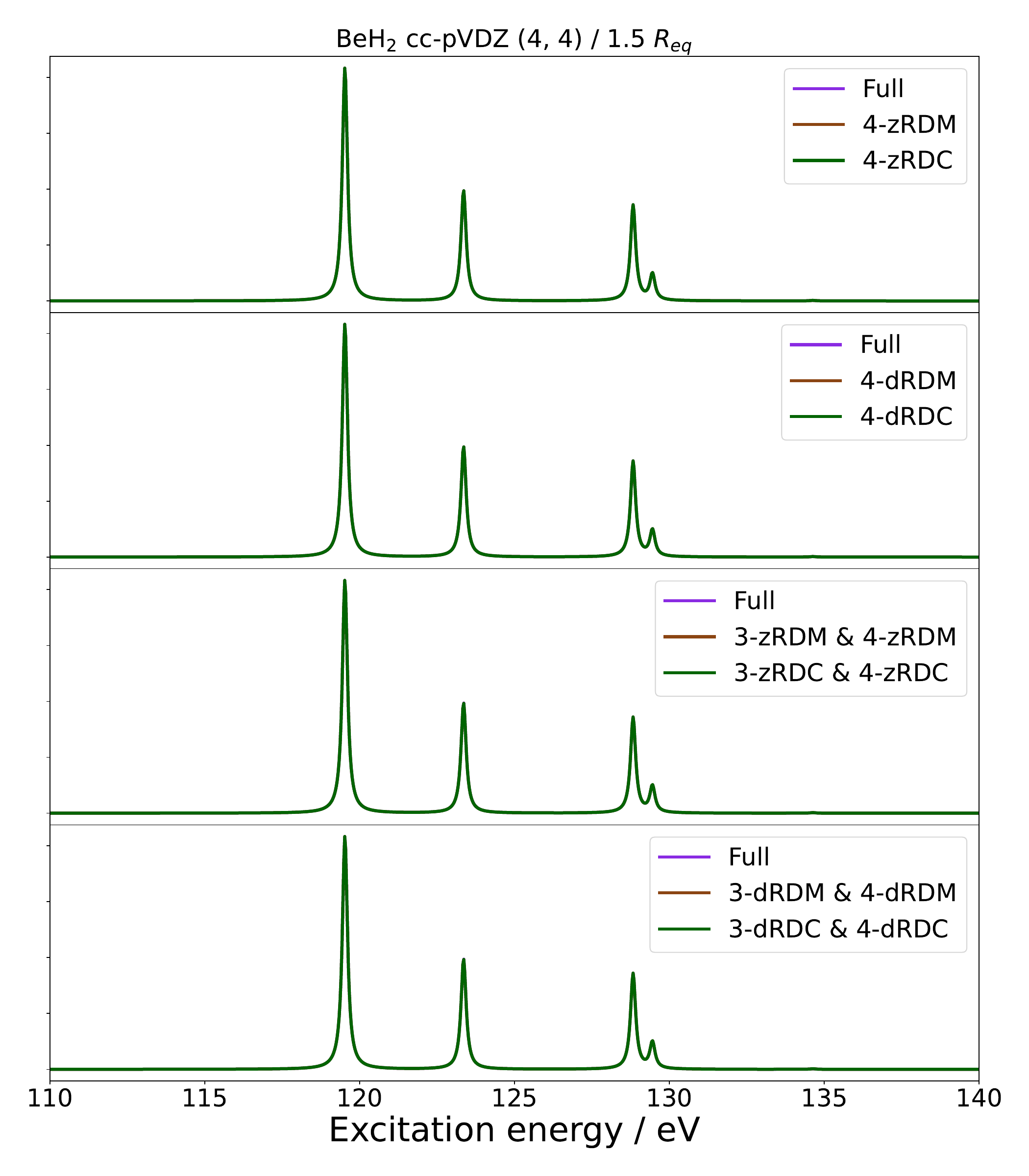}
    \includegraphics[width=.47\linewidth]{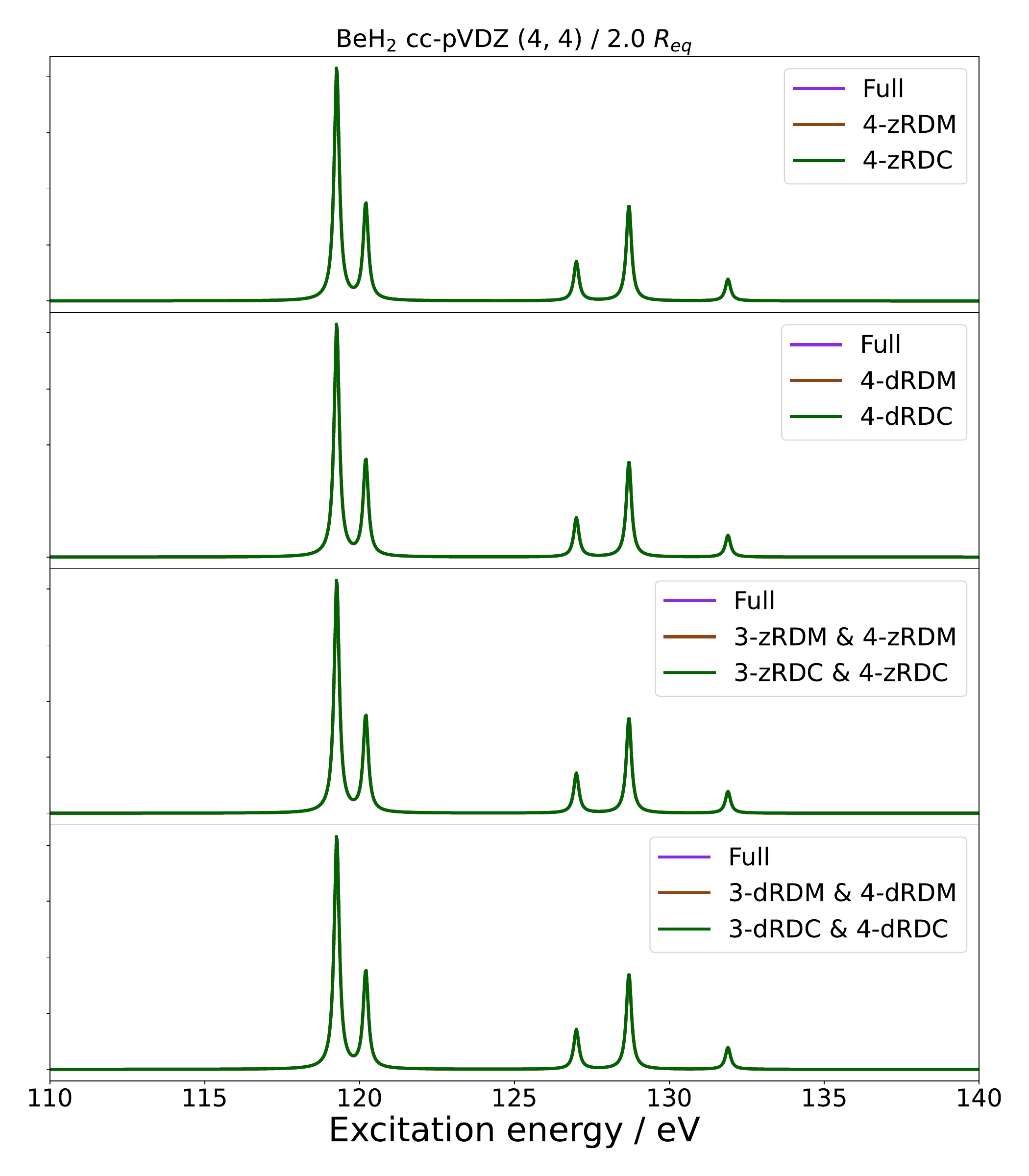}
    \caption{Beryllium K-edge absorption spectra of BeH$_2$ in a (4, 4) active space with the cc-pVDZ basis set at differing symmetric Be-H stretches. 
    Each figure contains four panels comparing the naive qLRSD absorption spectrum with no approximation to the absorption spectrum of naive qLRSD using the 4-zRDM and 4-zRDC approximations (first panel), the absorption spectrum of naive qLRSD using the 4-dRDM and 4-dRDC approximations (second panel), absorption spectrum of naive qLRSD using the 3-zRDM \& 4-zRDM and 3-zRDC \& 4-zRDC approximations, (third panel) and absorption spectrum of naive qLRSD using the 3-dRDM \& 4-dRDM and 3-dRDC \& 4-dRDC approximations (fourth panel).}\label{fig:stretched_beh2_core_4_4}
\end{figure}

\begin{figure}[p]
    \centering
    \includegraphics[width=.47\linewidth]{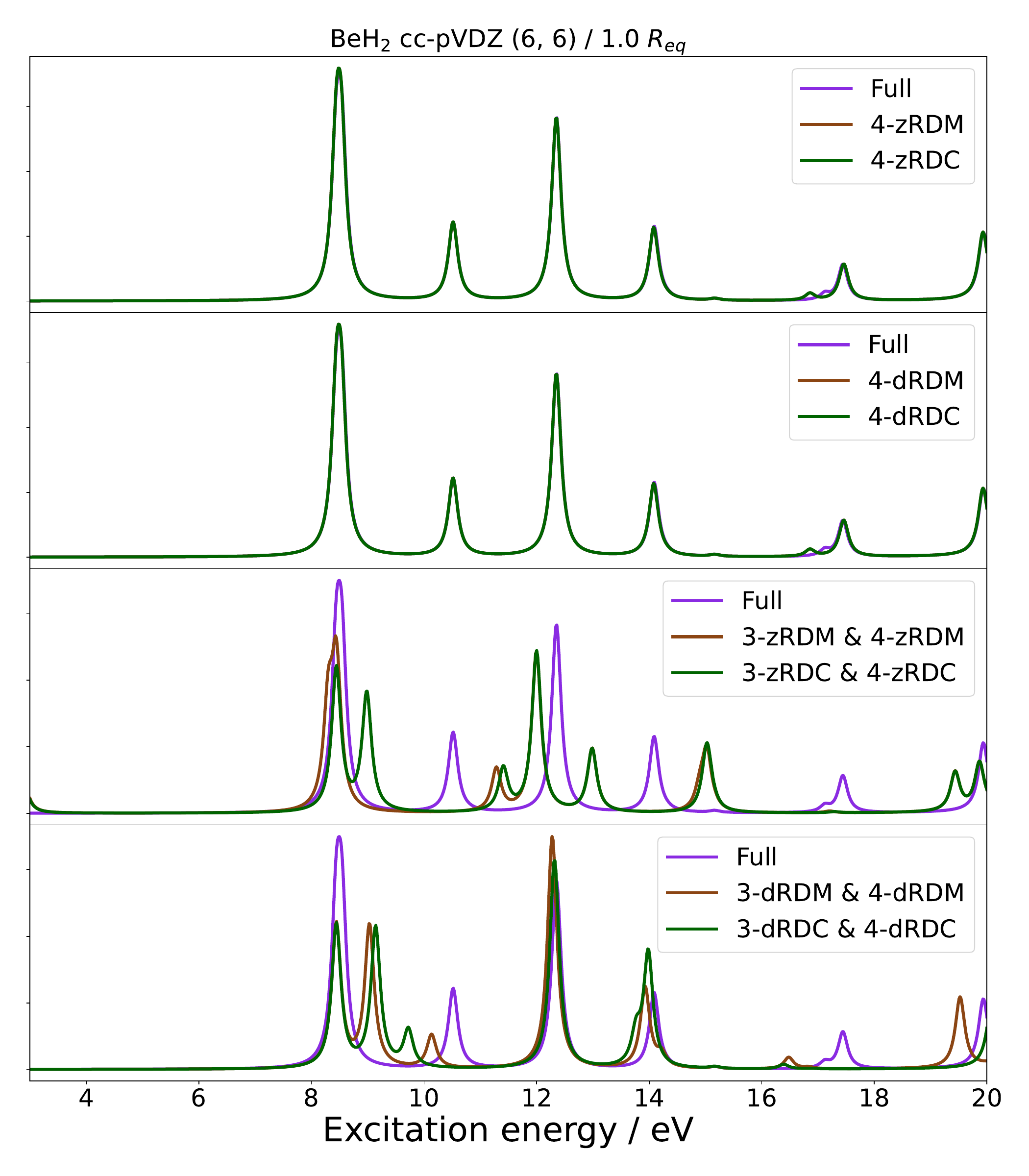}
    \includegraphics[width=.47\linewidth]{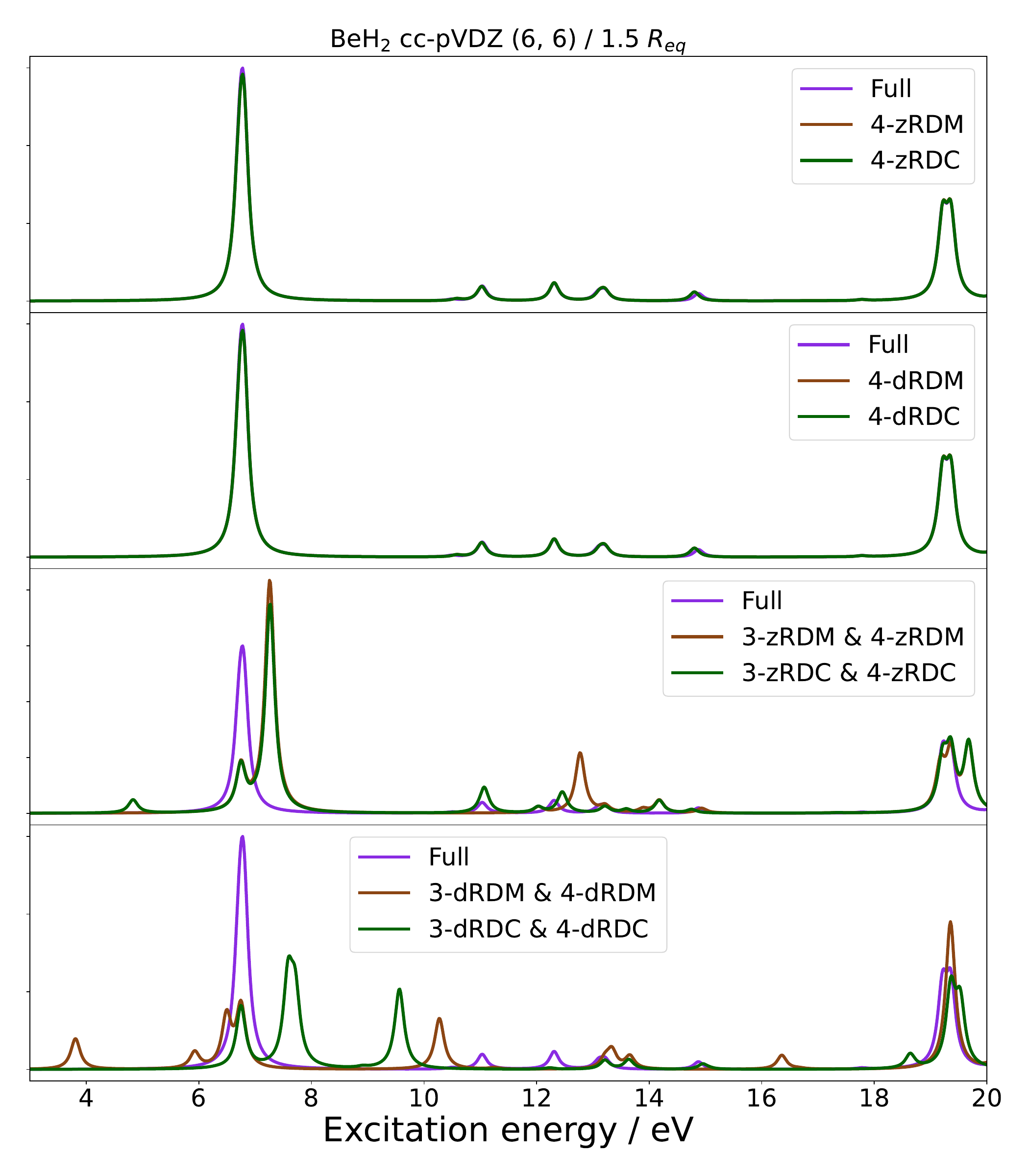}
    \includegraphics[width=.47\linewidth]{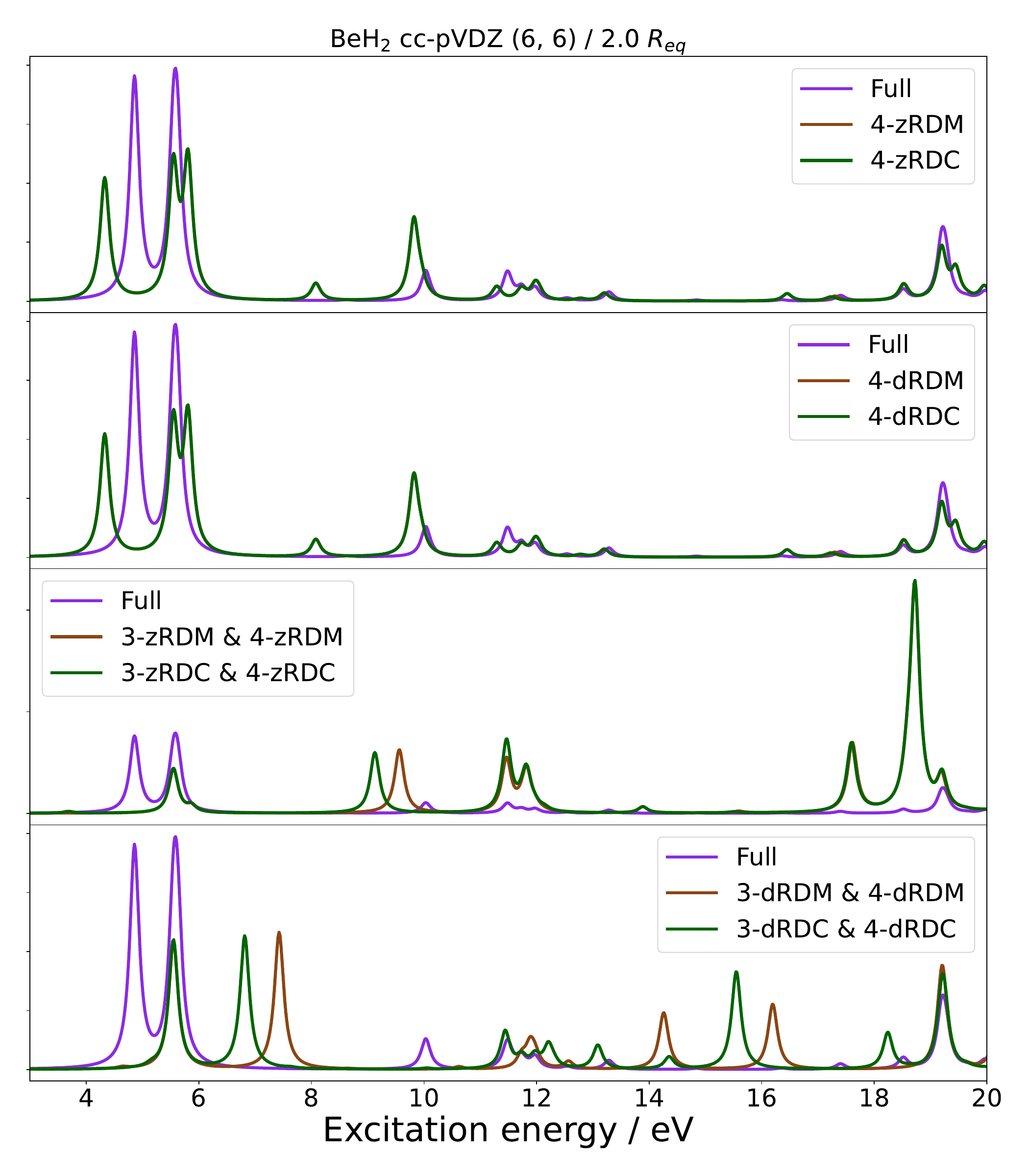}
    \caption{Absorption spectra in the valence excitation region of BeH$_2$ in a (6, 6) active space with the cc-pVDZ basis set at differing symmetric Be-H stretches. Each figure contains four panels comparing the naive qLRSD absorption spectrum with no approximation to the absorption spectrum of naive qLRSD using the 4-zRDM and 4-zRDC approximations (first panel), the absorption spectrum of naive qLRSD using the 4-dRDM and 4-dRDC approximations (second panel), absorption spectrum of naive qLRSD using the 3-zRDM \& 4-zRDM and 3-zRDC \& 4-zRDC approximations, (third panel) and absorption spectrum of naive qLRSD using the 3-dRDM \& 4-dRDM and 3-dRDC \& 4-dRDC approximations (fourth panel).}
    \label{fig:stretched_beh2_valence_6_6}
\end{figure}

\begin{figure}[p]
    \centering
    \includegraphics[width=.47\linewidth]{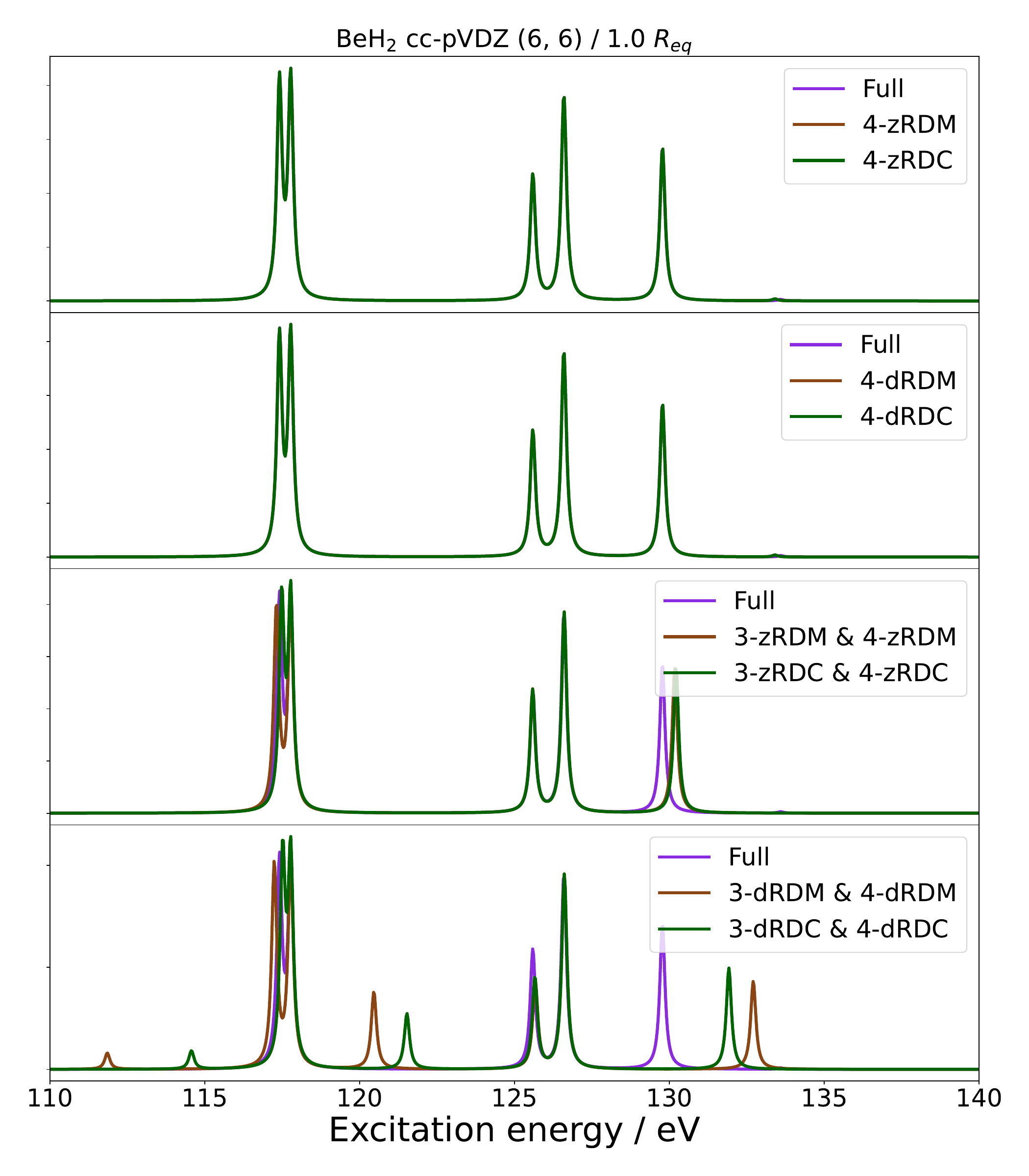}
    \includegraphics[width=.47\linewidth]{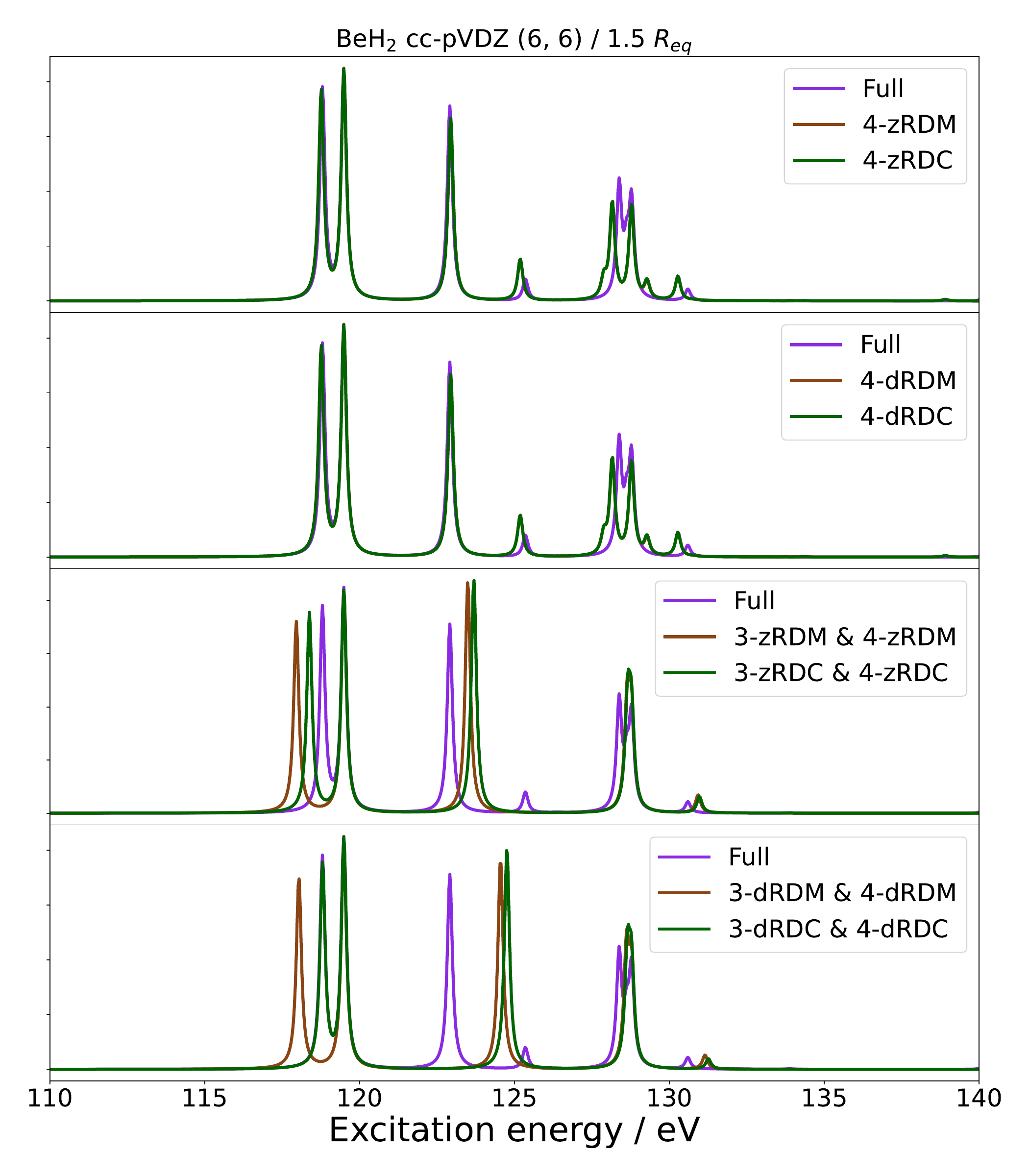}
    \includegraphics[width=.47\linewidth]{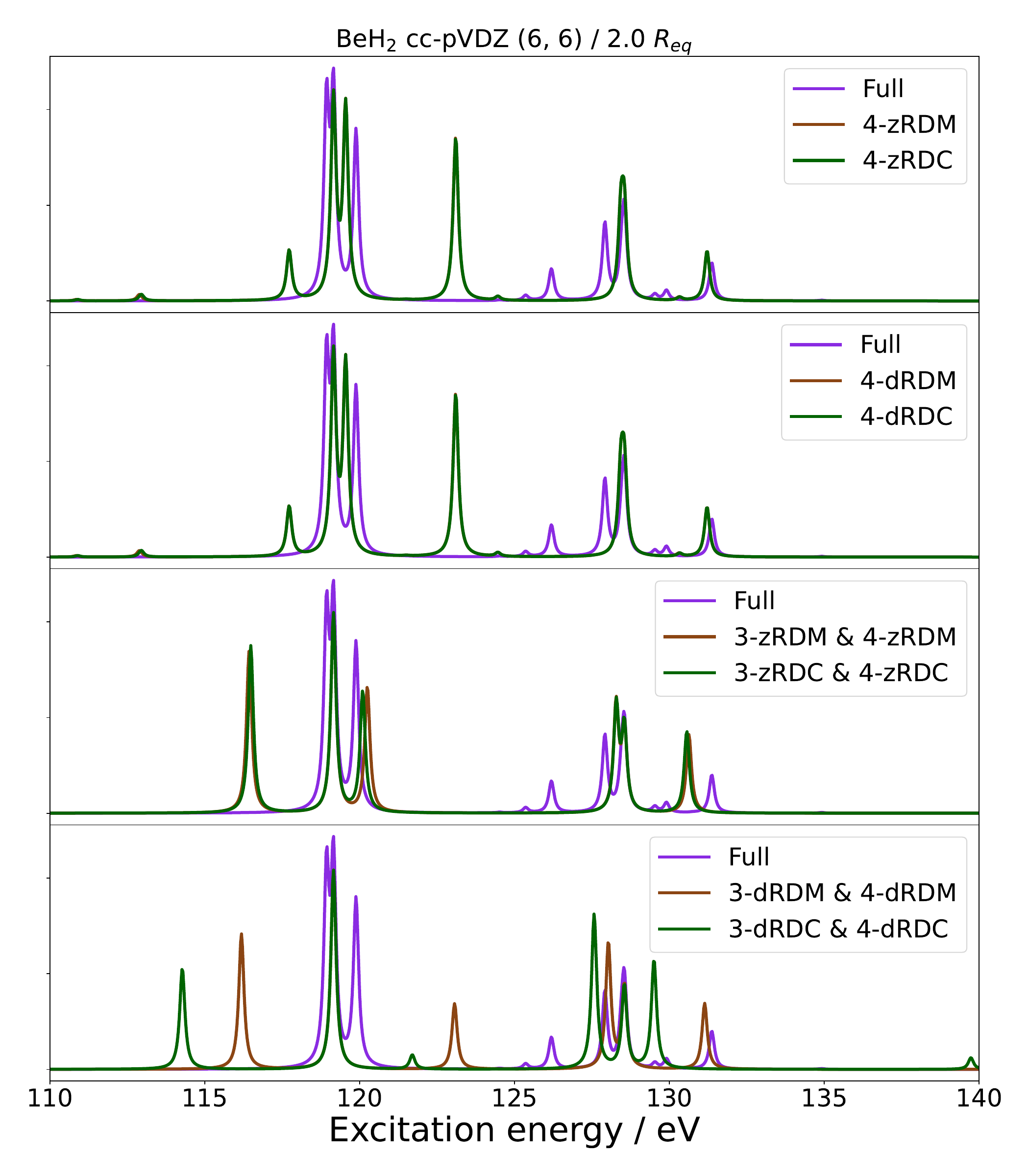}
    \caption{Beryllium K-edge absorption spectra of BeH$_2$ in a (6, 6) active space with the cc-pVDZ basis set at differing symmetric Be-H stretches. 
    Each figure contains four panels comparing the naive qLRSD absorption spectrum with no approximation to the absorption spectrum of naive qLRSD using the 4-zRDM and 4-zRDC approximations (first panel), the absorption spectrum of naive qLRSD using the 4-dRDM and 4-dRDC approximations (second panel), absorption spectrum of naive qLRSD using the 3-zRDM \& 4-zRDM and 3-zRDC \& 4-zRDC approximations, (third panel) and absorption spectrum of naive qLRSD using the 3-dRDM \& 4-dRDM and 3-dRDC \& 4-dRDC approximations (fourth panel).}
    \label{fig:stretched_beh2_core_6_6}
\end{figure}

\clearpage

\begin{longtable}{cc|cccc|cccc}
    \caption{Mean absolute errors (MAE) and standard deviations ($\sigma$) for the eight RDM and RDC approximations for the valence excitation energies of H$_2$O between 8 and 16 eV and their corresponding oscillator strengths. Errors and standard deviations for excitation energies are given in eV.} \\
    \hhline{==========} & & \multicolumn{4}{c|}{RDM approximations} & \multicolumn{4}{c}{RDC approximations} \\
    System & Errors & 4-z & 3- \& 4-z & 4-d & 3- \& 4-d & 4-z & 3- \& 4-z & 4-d & 3- \& 4-d  \\\hhline{==========}
    \multirow{4}{4.5em}{\centering H$_2$O \\ 1.0 R$_\text{eq}$ \\ (6, 6) \\ cc-pVDZ}
     & MAE$_\text{exc}$ & 0.0039 & 9.3353 & 0.0039 & 3.9895 & 0.0039 & 9.7281 & 0.0039 & 1.1694 \\
     & $\sigma_\text{exc}$ & 0.0027 & 0.8164 & 0.0027 & 1.0031 & 0.0027 & 1.8239 & 0.0027 & 0.4114 \\
     & MAE$_\text{osc}$ & 0.0001 & 0.1095 & 0.0001 & 0.0339 & 0.0001 & 0.116 & 0.0001 & 0.0111 \\
     & $\sigma_\text{exc}$ & 0.0001 & 0.1038 & 0.0001 & 0.041 & 0.0001 & 0.1005 & 0.0001 & 0.015 \\\hline
    \multirow{4}{4.5em}{\centering H$_2$O \\ 1.5 R$_\text{eq}$ \\ (6, 6) \\ cc-pVDZ}
     & MAE$_\text{exc}$ & 1.1432 & 4.0005 & 1.1432 & 5.4876 & 1.1374 & 2.3629 & 1.1374 & 4.1683 \\
     & $\sigma_\text{exc}$ & 1.1593 & 1.9593 & 1.1593 & 1.1696 & 1.1558 & 1.8578 & 1.1558 & 1.9807 \\
     & MAE$_\text{osc}$ & 0.027 & 0.2002 & 0.027 & 0.2065 & 0.0271 & 0.1941 & 0.0271 & 0.1741 \\
     & $\sigma_\text{exc}$ & 0.0151 & 0.2838 & 0.0151 & 0.2948 & 0.0151 & 0.278 & 0.0151 & 0.2054 \\\hline
    \multirow{4}{4.5em}{\centering H$_2$O \\ 2.0 R$_\text{eq}$ \\ (6, 6) \\ cc-pVDZ}
     & MAE$_\text{exc}$ & 4.5834 & 6.477 & 4.5834 & 7.2408 & 4.5718 & 6.0023 & 4.5718 & 8.2661 \\
     & $\sigma_\text{exc}$ & 1.1791 & 1.1023 & 1.1791 & 0.9927 & 1.1811 & 0.8666 & 1.1811 & 1.409 \\
     & MAE$_\text{osc}$ & 0.1198 & 0.1055 & 0.1198 & 1.8201 & 0.1197 & 0.095 & 0.1197 & 0.0711 \\
     & $\sigma_\text{exc}$ & 0.1885 & 0.1583 & 0.1885 & 4.0758 & 0.1885 & 0.1614 & 0.1885 & 0.1598 \\\hline
    \hhline{==========}
    \label{tab:H2O_valence}
\end{longtable}

\clearpage

\begin{longtable}{cc|cccc|cccc}
    \caption{Mean absolute errors (MAE) and standard deviations ($\sigma$) for the eight RDM and RDC approximations for the oxygen K-edge excitation energies of H$_2$O between 540 and 600 eV and their corresponding oscillator strengths. Errors and standard deviations for excitation energies are given in eV.} \\
    \hhline{==========} & & \multicolumn{4}{c|}{RDM approximations} & \multicolumn{4}{c}{RDC approximations} \\
    System & Errors & 4-z & 3- \& 4-z & 4-d & 3- \& 4-d & 4-z & 3- \& 4-z & 4-d & 3- \& 4-d  \\\hhline{==========}
    \multirow{4}{4.5em}{\centering H$_2$O \\ 1.0 R$_\text{eq}$ \\ (6, 6) \\ cc-pVDZ}
     & MAE$_\text{exc}$ & 0.0 & 0.0017 & 0.0 & 0.0018 & 0.0 & 0.0015 & 0.0 & 0.0015 \\
     & $\sigma_\text{exc}$ & 0.0 & 0.0039 & 0.0 & 0.004 & 0.0 & 0.0033 & 0.0 & 0.0034 \\
     & MAE$_\text{osc}$ & 0.0 & 0.0001 & 0.0 & 0.0001 & 0.0 & 0.0 & 0.0 & 0.0001 \\
     & $\sigma_\text{exc}$ & 0.0 & 0.0001 & 0.0 & 0.0001 & 0.0 & 0.0001 & 0.0 & 0.0001 \\\hline
    \multirow{4}{4.5em}{\centering H$_2$O \\ 1.5 R$_\text{eq}$ \\ (6, 6) \\ cc-pVDZ}
     & MAE$_\text{exc}$ & 0.0 & 0.0002 & 0.0 & 0.0002 & 0.0 & 0.0002 & 0.0 & 0.0002 \\
     & $\sigma_\text{exc}$ & 0.0 & 0.0005 & 0.0 & 0.0004 & 0.0 & 0.0005 & 0.0 & 0.0005 \\
     & MAE$_\text{osc}$ & 0.0 & 0.0001 & 0.0 & 0.0001 & 0.0 & 0.0 & 0.0 & 0.0 \\
     & $\sigma_\text{exc}$ & 0.0 & 0.0001 & 0.0 & 0.0001 & 0.0 & 0.0 & 0.0 & 0.0 \\\hline
    \multirow{4}{4.5em}{\centering H$_2$O \\ 2.0 R$_\text{eq}$ \\ (6, 6) \\ cc-pVDZ}
     & MAE$_\text{exc}$ & 0.0 & 0.0003 & 0.0 & 0.0003 & 0.0 & 0.0003 & 0.0 & 0.0003 \\
     & $\sigma_\text{exc}$ & 0.0 & 0.0009 & 0.0 & 0.001 & 0.0 & 0.0009 & 0.0 & 0.001 \\
     & MAE$_\text{osc}$ & 0.0 & 0.0 & 0.0 & 0.0 & 0.0 & 0.0 & 0.0 & 0.0 \\
     & $\sigma_\text{exc}$ & 0.0 & 0.0 & 0.0 & 0.0 & 0.0 & 0.0 & 0.0 & 0.0 \\\hline
    \hhline{==========}
    \label{tab:H2O_core}
\end{longtable}

\clearpage

\begin{longtable}{cc|cccc|cccc}
    \caption{Mean absolute errors (MAE) and standard deviations ($\sigma$) for the eight RDM and RDC approximations for the valence excitation energies of BeH$_2$ between 5 and 20 eV and their corresponding oscillator strengths. Errors and standard deviations for excitation energies are given in eV.} \\
    \hhline{==========} & & \multicolumn{4}{c|}{RDM approximations} & \multicolumn{4}{c}{RDC approximations} \\
    System & Errors & 4-z & 3- \& 4-z & 4-d & 3- \& 4-d & 4-z & 3- \& 4-z & 4-d & 3- \& 4-d  \\\hhline{==========}
    \multirow{4}{4.5em}{\centering BeH$_2$ \\ 1.0 R$_\text{eq}$ \\ (4, 4) \\ cc-pVDZ}
     & MAE$_\text{exc}$ & 0.001 & 0.0929 & 0.001 & 0.0532 & 0.001 & 0.0929 & 0.001 & 0.0529 \\
     & $\sigma_\text{exc}$ & 0.0016 & 0.1254 & 0.0016 & 0.0802 & 0.0016 & 0.1255 & 0.0016 & 0.0798 \\
     & MAE$_\text{osc}$ & 0.0001 & 0.0157 & 0.0001 & 0.0074 & 0.0001 & 0.0157 & 0.0001 & 0.0074 \\
     & $\sigma_\text{exc}$ & 0.0003 & 0.0283 & 0.0003 & 0.014 & 0.0003 & 0.0283 & 0.0003 & 0.0139 \\\hline
    \multirow{4}{4.5em}{\centering BeH$_2$ \\ 1.5 R$_\text{eq}$ \\ (4, 4) \\ cc-pVDZ}
     & MAE$_\text{exc}$ & 0.0207 & 0.8594 & 0.0207 & 0.4011 & 0.0207 & 0.8591 & 0.0207 & 0.4031 \\
     & $\sigma_\text{exc}$ & 0.0492 & 0.9442 & 0.0492 & 0.6175 & 0.0492 & 0.9437 & 0.0492 & 0.6201 \\
     & MAE$_\text{osc}$ & 0.0046 & 0.1332 & 0.0046 & 0.0622 & 0.0046 & 0.1332 & 0.0046 & 0.0629 \\
     & $\sigma_\text{exc}$ & 0.0121 & 0.263 & 0.0121 & 0.1453 & 0.0121 & 0.263 & 0.0121 & 0.1455 \\\hline
    \multirow{4}{4.5em}{\centering BeH$_2$ \\ 2.0 R$_\text{eq}$ \\ (4, 4) \\ cc-pVDZ}
     & MAE$_\text{exc}$ & 0.1378 & 1.2946 & 0.1378 & 0.4514 & 0.1378 & 1.2943 & 0.1378 & 0.3923 \\
     & $\sigma_\text{exc}$ & 0.305 & 1.1786 & 0.305 & 0.5616 & 0.305 & 1.1814 & 0.305 & 0.4925 \\
     & MAE$_\text{osc}$ & 0.0061 & 0.0646 & 0.0061 & 0.0645 & 0.0061 & 0.0661 & 0.0061 & 0.0632 \\
     & $\sigma_\text{exc}$ & 0.0198 & 0.1387 & 0.0198 & 0.1304 & 0.0198 & 0.1383 & 0.0198 & 0.1289 \\\hline
    \multirow{4}{4.5em}{\centering BeH$_2$ \\ 1.0 R$_\text{eq}$ \\ (6, 6) \\ cc-pVDZ}
     & MAE$_\text{exc}$ & 0.0185 & 5.3726 & 0.0185 & 4.0771 & 0.0185 & 5.2981 & 0.0185 & 4.1017 \\
     & $\sigma_\text{exc}$ & 0.0566 & 1.5667 & 0.0566 & 1.8491 & 0.0566 & 1.586 & 0.0566 & 1.8524 \\
     & MAE$_\text{osc}$ & 0.0005 & 0.1919 & 0.0005 & 0.1622 & 0.0005 & 0.2011 & 0.0005 & 0.2011 \\
     & $\sigma_\text{exc}$ & 0.001 & 0.1823 & 0.001 & 0.1944 & 0.001 & 0.1818 & 0.001 & 0.2077 \\\hline
    \multirow{4}{4.5em}{\centering BeH$_2$ \\ 1.5 R$_\text{eq}$ \\ (6, 6) \\ cc-pVDZ}
     & MAE$_\text{exc}$ & 0.0288 & 4.4654 & 0.0288 & 4.0373 & 0.0307 & 4.2543 & 0.0307 & 4.1191 \\
     & $\sigma_\text{exc}$ & 0.0434 & 0.9591 & 0.0434 & 1.5287 & 0.045 & 1.1347 & 0.045 & 0.8232 \\
     & MAE$_\text{osc}$ & 0.0018 & 0.1853 & 0.0018 & 0.1302 & 0.0335 & 0.1799 & 0.0335 & 0.1574 \\
     & $\sigma_\text{exc}$ & 0.0039 & 0.2996 & 0.0039 & 0.188 & 0.1205 & 0.2749 & 0.1205 & 0.2329 \\\hline\pagebreak\hline
    \multirow{4}{4.5em}{\centering BeH$_2$ \\ 2.0 R$_\text{eq}$ \\ (6, 6) \\ cc-pVDZ}
     & MAE$_\text{exc}$ & 0.1863 & 2.315 & 0.1863 & 3.0703 & 0.1792 & 2.5989 & 0.1792 & 3.5342 \\
     & $\sigma_\text{exc}$ & 0.2593 & 1.5148 & 0.2593 & 2.2088 & 0.2383 & 1.655 & 0.2383 & 1.7659 \\
     & MAE$_\text{osc}$ & 0.0235 & 0.1517 & 0.0235 & 0.0753 & 0.0254 & 0.153 & 0.0254 & 0.0773 \\
     & $\sigma_\text{exc}$ & 0.0536 & 0.3559 & 0.0536 & 0.1248 & 0.0545 & 0.353 & 0.0545 & 0.1326 \\\hline
    \hhline{==========}
    \label{tab:BeH2_valence}
\end{longtable}

\clearpage

\begin{longtable}{cc|cccc|cccc}
    \caption{Mean absolute errors (MAE) and standard deviations ($\sigma$) for the eight RDM and RDC approximations for the beryllium K-edge excitation energies of BeH$_2$ between 110 and 140 eV and their corresponding oscillator strengths. Errors and standard deviations for excitation energies are given in eV.} \\
    \hhline{==========} & & \multicolumn{4}{c|}{RDM approximations} & \multicolumn{4}{c}{RDC approximations} \\
    System & Errors & 4-z & 3- \& 4-z & 4-d & 3- \& 4-d & 4-z & 3- \& 4-z & 4-d & 3- \& 4-d  \\\hhline{==========}
    \multirow{4}{4.5em}{\centering BeH$_2$ \\ 1.0 R$_\text{eq}$ \\ (4, 4) \\ cc-pVDZ}
     & MAE$_\text{exc}$ & 0.0 & 0.0 & 0.0 & 0.0 & 0.0 & 0.0 & 0.0 & 0.0 \\
     & $\sigma_\text{exc}$ & 0.0 & 0.0 & 0.0 & 0.0 & 0.0 & 0.0 & 0.0 & 0.0 \\
     & MAE$_\text{osc}$ & 0.0 & 0.0 & 0.0 & 0.0 & 0.0 & 0.0 & 0.0 & 0.0 \\
     & $\sigma_\text{exc}$ & 0.0 & 0.0 & 0.0 & 0.0 & 0.0 & 0.0 & 0.0 & 0.0 \\\hline
    \multirow{4}{4.5em}{\centering BeH$_2$ \\ 1.5 R$_\text{eq}$ \\ (4, 4) \\ cc-pVDZ}
     & MAE$_\text{exc}$ & 0.0 & 0.0003 & 0.0 & 0.0002 & 0.0 & 0.0003 & 0.0 & 0.0002 \\
     & $\sigma_\text{exc}$ & 0.0 & 0.0005 & 0.0 & 0.0004 & 0.0 & 0.0005 & 0.0 & 0.0004 \\
     & MAE$_\text{osc}$ & 0.0 & 0.0 & 0.0 & 0.0 & 0.0 & 0.0 & 0.0 & 0.0 \\
     & $\sigma_\text{exc}$ & 0.0 & 0.0001 & 0.0 & 0.0001 & 0.0 & 0.0001 & 0.0 & 0.0001 \\\hline
    \multirow{4}{4.5em}{\centering BeH$_2$ \\ 2.0 R$_\text{eq}$ \\ (4, 4) \\ cc-pVDZ}
     & MAE$_\text{exc}$ & 0.0 & 0.0002 & 0.0 & 0.0002 & 0.0 & 0.0002 & 0.0 & 0.0002 \\
     & $\sigma_\text{exc}$ & 0.0 & 0.0003 & 0.0 & 0.0003 & 0.0 & 0.0003 & 0.0 & 0.0003 \\
     & MAE$_\text{osc}$ & 0.0 & 0.0001 & 0.0 & 0.0001 & 0.0 & 0.0001 & 0.0 & 0.0001 \\
     & $\sigma_\text{exc}$ & 0.0 & 0.0002 & 0.0 & 0.0002 & 0.0 & 0.0002 & 0.0 & 0.0002 \\\hline
    \multirow{4}{4.5em}{\centering BeH$_2$ \\ 1.0 R$_\text{eq}$ \\ (6, 6) \\ cc-pVDZ}
     & MAE$_\text{exc}$ & 0.0542 & 4.8742 & 0.0542 & 22.1473 & 0.0539 & 4.8856 & 0.0539 & 20.5622 \\
     & $\sigma_\text{exc}$ & 0.1126 & 8.6498 & 0.1126 & 7.9299 & 0.1122 & 8.6587 & 0.1122 & 6.928 \\
     & MAE$_\text{osc}$ & 0.0001 & 0.0014 & 0.0001 & 0.0456 & 0.0001 & 0.0013 & 0.0001 & 0.0454 \\
     & $\sigma_\text{exc}$ & 0.0003 & 0.0023 & 0.0003 & 0.0363 & 0.0003 & 0.0023 & 0.0003 & 0.0369 \\\hline
    \multirow{4}{4.5em}{\centering BeH$_2$ \\ 1.5 R$_\text{eq}$ \\ (6, 6) \\ cc-pVDZ}
     & MAE$_\text{exc}$ & 0.6947 & 20.8381 & 0.6947 & 19.8295 & 0.693 & 20.746 & 0.693 & 18.8591 \\
     & $\sigma_\text{exc}$ & 0.7636 & 16.9839 & 0.7636 & 17.6396 & 0.7654 & 16.8733 & 0.7654 & 16.4454 \\
     & MAE$_\text{osc}$ & 0.0051 & 0.0096 & 0.0051 & 0.0304 & 0.0051 & 0.0094 & 0.0051 & 0.0312 \\
     & $\sigma_\text{exc}$ & 0.0121 & 0.0156 & 0.0121 & 0.0372 & 0.0112 & 0.0154 & 0.0112 & 0.0382 \\\hline\pagebreak\hline
    \multirow{4}{4.5em}{\centering BeH$_2$ \\ 2.0 R$_\text{eq}$ \\ (6, 6) \\ cc-pVDZ}
     & MAE$_\text{exc}$ & 3.319 & 22.6587 & 3.319 & 20.1038 & 3.3008 & 22.5023 & 3.3008 & 18.3609 \\
     & $\sigma_\text{exc}$ & 2.9365 & 16.1129 & 2.9365 & 19.3291 & 2.9325 & 15.9754 & 2.9325 & 19.2645 \\
     & MAE$_\text{osc}$ & 0.0274 & 0.0105 & 0.0274 & 0.0398 & 0.0274 & 0.0106 & 0.0274 & 0.0503 \\
     & $\sigma_\text{exc}$ & 0.0364 & 0.0164 & 0.0364 & 0.0804 & 0.0363 & 0.0164 & 0.0363 & 0.1088 \\\hline
    \hhline{==========}
    \label{tab:BeH2_core}
\end{longtable}

\clearpage